\documentclass[12pt]{report}

\usepackage[a4paper,top=2cm,bottom=2cm,left=3cm,right=3cm,marginparwidth=1.75cm]{geometry}
\usepackage[english]{babel}
\usepackage[utf8]{inputenc}

\usepackage[numbers]{natbib}
\usepackage{url}

\usepackage{amsmath}

\usepackage[normalem]{ulem}
\DeclareMathOperator{\arccosh}{arccosh}
\DeclareMathOperator{\arcsinh}{arcsinh}

\usepackage{etoolbox}
\usepackage{tikz}
\usepackage[svgnames]{xcolor}
\usepackage{empheq}
\usepackage{graphicx}
\usepackage[colorlinks=true, allcolors=blue]{hyperref}
\usepackage{lipsum}
\usepackage{float}
\usepackage{epsfig}
\usepackage{graphicx}
\usepackage{xcolor}
\usepackage{color}

\usepackage{amssymb}
\usepackage{amsfonts}
\usepackage{amsmath}
\usepackage{amsthm}
\usepackage{epsfig}
\usepackage{subcaption}
\usepackage{cancel}

\usepackage{float}
\usepackage{placeins}
\usepackage{threeparttable}

\usepackage[normalem]{ulem}

\usepackage{setspace}
\date{\today}

\usepackage[most]{tcolorbox}
\usepackage{listings}
\usepackage{xcolor}
\lstdefinestyle{meucodigo}{
    language=Python,
    basicstyle=\ttfamily\small,
    keywordstyle=\color{blue!70!black},
    commentstyle=\color{green!40!black},
    stringstyle=\color{orange!60!black},
    numberstyle=\tiny\color{gray},
    numbers=left,
    stepnumber=1,
    numbersep=8pt,
    showstringspaces=false,
    breaklines=true,
    breakatwhitespace=true,
    frame=none,
    tabsize=4
}

\usepackage{fancyhdr}
\usepackage[font=small,labelfont=bf,labelsep=colon]{caption}

\begin{document}
\begin{titlepage}
\begin{center}

    \vspace*{1.5cm}
    \Huge
    \text{University of Brasília} \\ \text{Institute of Physics}
    
     \vspace*{3cm}
    \Huge
    \textbf{Doctoral Thesis}\\
    
    \vspace*{1cm}
    \Huge
    \textbf{Scale invariance, fractal dynamics, and critical exponents at the phase transition.}
    \vspace*{-0.5cm}

    \vspace*{3cm}
    \Large
    \textbf{}\\ 
    \vspace*{0.2cm}
    \textbf{Henrique Alves de Lima}\\
    \vspace*{0.2cm}

     \text{Brasília} \\ \text{\today}
\end{center}
\date{\today}
\end{titlepage}

\newpage
 %blank page

\begin{titlepage}
\begin{center}

     \vspace*{3cm}
     \Large
     \textbf{Henrique Alves de Lima}\\
     \vspace*{2cm}

    \vspace*{0.5cm}
    \Huge
    %\textbf{Fractals in the Ising Model}
    \textbf{Scale invariance, fractal dynamics, and critical exponents at the phase transition.}
    \vspace*{2cm}
    \Large
    \textbf{}\\ 
    \vspace*{0.2cm}
    \vspace*{0.2cm}
    \textbf{Advisor: Prof. Dr. Fernando A. Oliveira}\\
    \textbf{Co-Advisor: Prof. Dr. Ismael S. S. Carrasco}\\
    \textbf{Co-Advisor: Prof. Dr. Jairo Almeida}\\
    
    \vspace*{0.2cm}
    \textbf{}

    \large
    \vfill
    Doctoral thesis \\\
    \vspace*{1.0cm}

    \text{University of Brasília} \\ \text{Institute of Physics}

     \text{Brasília} \\ \text{\today}
\end{center}
\date{\today}
\end{titlepage}

\newpage
\renewcommand{\abstractname}{Abstract}
\begin{abstract}
\paragraph{} In this work we propose that, at criticality, the dynamics of equilibrium systems do not effectively unfold throughout the entire Euclidean space, but rather within a fractal subspace. Starting from the fact that correlation functions constitute a central tool for describing the behavior of physical variables in space and time, we investigate the geometric structure underlying the correlation function in systems near a phase transition, returning to Fisher's formulation of the order-parameter autocorrelation function. We argue that the usual critical correction associated with the exponent $\eta$ can be given a geometric interpretation through a correlation fractal dimension associated with this effective subspace. This fractal subspace is distinct from that associated with the order parameter and acts as a defining property of criticality, restoring the behavior of the correlation function at the critical point and explaining the ``failure'' of the fluctuation-dissipation theorem. Based on this hypothesis, we investigate the physics and mathematical structure of critical correlations, emphasizing the connection between scaling behavior, critical exponents, and fractal geometry. We show that the application of modern fractional calculus tools makes it possible to write an expression for the correlation function that correctly recovers the critical exponents below the upper critical dimension. From this formulation, we obtain an explicit relation for Fisher's critical exponent $\eta$ in terms of a fractal dimension $d_R$, associated with the Riesz fractional derivative, while also examining the Rushbrooke scaling relation, whose validity has been questioned in some magnetic systems. We further analyze the behavior of physical quantities in non-integer-dimensional systems and in disordered systems, controlled by a disorder parameter $\sigma$, seeking to evaluate the limits within which our hypothesis remains valid. %\nocite{fisher1964, Mandelbrot82, Kardar07, Lima24, Lima25}

\end{abstract}
\newpage
\tableofcontents
\newpage

\listoftables

\listoffigures

\chapter*{List of Symbols}
\begin{table}[h]
    \begin{tabular}{r|l}
    Symbol & Description \\ \hline \\
        $T$ & Temperature  \\
        $V$ & Temperature  \\
        $P$ & Temperature  \\
        $n$ & Number of \textit{moles} in the system  \\
        $R$ & Universal gas constant   \\
        $L$ & Latent heat of the transition  \\
        $Z$ & Partition function \\
        $N_{+}$ & Number of \textit{spin-up} states  \\
        $N_{-}$ & Number of \textit{spin-down} states  \\
        $g(T, H = 0)$ & Gibbs free energy per particle \\
        E & Energy \\
        $\varrho^2$ & Normalization factor of the pair correlation function\\
        $\beta_t$ & Constant equivalent to $1/k_B T$  \\
        $k_B$ & Boltzmann constant  \\
        $T_O$ & Onsager critical temperature  \\
        $T_c$ & Critical temperature in units of $T_O$ \\
        $J_{ij}$ & Coupling between spins $i$ and $j$\\
        $J$ & Mean coupling between spins\\
        $\sigma_i$ & \textit{Spin} variable with index $i$  \\
        H & System Hamiltonian \\
        N & Number of bodies in the system \\
        M & System magnetization value \\
        U & System energy\\
        $C_v$ & Heat capacity at constant volume\\
        $\chi$ & Magnetic susceptibility of the system\\
        G(r) & Correlation function \\
        $\rho$ & Correlation length\\
        $\eta$ & Fisher exponent\\
        $\sigma$ & Standard deviation of the mean coupling value \textit{J}\\
        $\beta$ & Critical exponent of the magnetization  \\
        $\alpha$ & Critical exponent of the specific heat\\
        $\gamma$ & Critical exponent of the magnetic susceptibility\\
        $\nu$ & Critical exponent of the correlation length\\
        $d_R$ & Riesz fractal dimension\\
        $d_f$ & Fractal dimension of the largest cluster\\
 
    \end{tabular}
    
\end{table}

\cleardoublepage
\thispagestyle{empty}

\vspace*{\fill}

\begin{center}
\textbf{Dedication}
\end{center}

\vspace{1cm}

To my mother, Dona Rosinha, who, with an incomplete elementary education and the sweat of her brow, fought so that her son could be illuminated by the light of knowledge. Thank you for fighting, for fleeing the drought, for living in other people's homes, for having worked as a domestic worker, for taking care of me when the man who fathered me abandoned me. For every sadness we went through together, every laugh and every hug, my eternal teacher, this achievement is yours, Mom.\\

To my dear mentor and father, Prof. Fernando Albuquerque de Oliveira, the most exuberant mind I have ever known, without whom none of this would have happened, thank you for being a light when I needed it most. Thank you very much.\\

To my fiancée, Cristina, who taught me to feel and appreciate the taste of life again, who brought color to my limited and monochromatic existence, who shared with me even what she did not have so that there could be something of her within me, thank you.\\

To my parents-in-law, Cristiane and Raimundo, for accepting me and welcoming me with so much affection into your lives. Family is everyone with whom the heart identifies, and my heart identifies with you. Thank you.\\

To all my friends, who shared with me the moments of this journey. You know everything I feel at this moment as I write; you did not arrive here today and you will not leave anytime soon. Thank you.\\

To the professors Ismael Carrasco, Leonardo Castro, Fábio Aarão Reis, Pedro Pinto, Jairo Almeida, Martin Oettel, Paulo Narciso, Jales Dantas, Márcio Sampaio, among the countless others who were fundamental to my education, thank you very much.\\

To all the employees of the University of Brasília---custodians, security staff, librarians, university restaurant staff, secretaries, coordinators, and professors---who, in different ways, were also part of this journey.\\

\noindent I dedicate this work to all of you.

\vspace*{\fill}

\cleardoublepage
\markboth{}{}
\thispagestyle{fancy}

\vspace*{3cm}

\subsection*{Acknowledgments}

To the University of Brasília and the Institute of Physics;\\
To CAPES for funding this research;\\
To the Complex Systems Research Group;\\
To my dear advisor Prof. Dr. Fernando Albuquerque de Oliveira;\\
To Professor Márcio Sampaio and LNCC, for supporting the computational part of this work by providing access to the SDumont system;\\
In advance, to all the members of my evaluation committee;\\
To everyone who contributed to the completion of this stage;\\

\noindent Thank you very much.

\vspace{80mm}
\begin{flushright}
\itshape
``There are decades when nothing happens and there are weeks when decades happen'',\\
Lenin.
\end{flushright}

\chapter{Introduction} \label{Introdução}
%\addcontentsline{toc}{chapter}{Introduction}
\paragraph{} The understanding of the phases of matter has come a long way from the earliest empirical observations that materials transform under certain external conditions to the formulation of theories capable of quantitatively describing such changes. Throughout this process, phase transitions ceased to be viewed merely as familiar macroscopic phenomena (such as transitions among the solid, liquid, and gaseous states) and became a central topic in statistical physics, condensed matter physics, and, more broadly, the study of complex systems. This change in perspective made it possible to understand that phase transitions represent not only changes of state, but also collective reorganizations of matter, frequently associated with the emergence of new properties.\\

Several historical milestones contributed to this consolidation. Among them are the works of R. Boyle, in the seventeenth century, in establishing quantitative laws for the behavior of gases \cite{Boyle1662}, and of J. Black, in the eighteenth century, in introducing the concept of latent heat and distinguishing heat from temperature \cite{Black1761}. In the nineteenth century, Thomas Andrews showed, in his studies of carbon dioxide, that liquid and gas can become indistinguishable above a certain temperature, revealing the existence of the critical point \cite{Andrews1869}. Shortly thereafter, J. D. van der Waals proposed his celebrated equation of state, incorporating the role of intermolecular interactions and the finite volume of molecules into the description of the liquid--vapor transition \cite{vdW1873}. These advances were fundamental in transforming the notion of a phase change into a quantitative problem, amenable to mathematical description and deeper physical interpretation.\\

However, it was in the twentieth century that the problem acquired its modern formulation. The development of statistical mechanics provided the framework needed to connect microscopic properties to macroscopic behaviors, making it possible to interpret phase transitions in terms of cooperation among a very large number of degrees of freedom. In this context, the exact solution of the two-dimensional Ising model by Lars Onsager, in 1944, represented a turning point by explicitly revealing the existence of thermodynamic singularities and the central role of long-range correlations near the critical temperature. From then on, it became clear that criticality is not merely a singular point in a phase diagram, but a regime in which the system exhibits universal properties, independent of many microscopic details.\\

Among these properties, scale invariance occupies a prominent position. Near the critical point, the system ceases to exhibit a dominant characteristic length, and fluctuations on very different scales begin to coexist. This absence of a privileged scale manifests itself in the power-law behavior of several physical observables and is directly related to the appearance of the so-called critical exponents, which characterize how quantities such as magnetization, susceptibility, specific heat, and correlation length behave near the transition. These exponents constitute some of the most important elements of the modern theory of phase transitions, as they allow very different systems to be classified in terms of common universality classes.\\

In this setting, the exponent associated with the behavior of the correlation function at criticality, introduced by Fisher \cite{fisher1964}, occupies a particularly important position. Its presence is essential for describing the asymptotic form of spatial correlations in critical systems and, consequently, for establishing the connection between the microscopic structure of fluctuations and the macroscopic response of the system. Despite its phenomenological success and broad use in different contexts, the deeper physical origin of this exponent is still not completely transparent. In much of the literature, it is introduced effectively: it is an indispensable parameter for the consistency of scaling relations and for agreement with experimental and numerical results, although its geometric or spatial interpretation is not always made explicit in a satisfactory manner.\\

It is precisely at this point that the central motivation of this thesis arises. The hypothesis guiding this work is that a deeper understanding of critical behavior requires a revision of the very notion of the space effectively accessible to the dynamics of the system near the phase transition. In other words, we propose that part of the properties observed at criticality arise not only from thermal fluctuations in a conventional Euclidean space, but also from a more subtle geometric reorganization associated with a non-integer effective dimensionality. Such a hypothesis becomes particularly natural when one recalls that scale invariance, a hallmark of criticality, is also one of the fundamental properties of fractal objects.\\

From this perspective, it becomes plausible to interpret the collective dynamics of critical \textit{spins} as dynamics of a fractal nature \cite{Lima24,Lima25,Carrasco26,lima26}. In this approach, the critical exponents cease to be viewed merely as constants fitted through scaling arguments and instead admit a more direct geometric interpretation associated with the reduction of the effective space in which the correlations of the system develop. Rather than thinking of criticality only as a regime in which fluctuations become intense, it is also regarded as a regime in which the relevant geometry of the problem becomes less trivial, possibly requiring descriptions based on non-integer dimensions. \\

This change in perspective is not only conceptually attractive, but also methodologically challenging. Unlike ordinary differential calculus, which has been firmly established over centuries, methods associated with the description of fractal structures and non-integer generalizations of geometric and differential operators are still under development. In recent decades, important contributions have sought to establish consistent extensions for this type of formalism \cite{Muslih10,Muslih10b}, but it remains a field under construction, especially when one intends to apply it to real physical systems and criticality problems. Thus, investigating the relationship between phase transitions and fractal geometry implies dealing simultaneously with conceptual, analytical, and numerical issues.\\

From the computational point of view, the difficulties are also significant. The best-known methods for estimating fractal dimensions, such as the \textit{box counting} method, can become extremely costly when applied to large systems, especially near criticality, where the structures of interest are distributed across multiple scales. The computational cost grows rapidly with system size and with the need to obtain reliable statistics, making this type of analysis particularly demanding. In this context, the use of high-performance computing techniques ceases to be merely convenient and becomes essential. Among the tools available to address this problem, CUDA technology \cite{CudaSite} stands out, allowing the implementation of parallel algorithms on graphics processing units. The use of this type of architecture enables the treatment of larger systems, more extensive sampling, and the execution of numerical procedures that would be prohibitively slow on conventional architectures. Thus, parallel computing appears in this work not merely as an auxiliary technical resource, but as an integral part of the methodological strategy employed to explore the proposed geometric hypothesis.\\

In light of these considerations, this thesis was organized around analyzing the extent to which behaviors observed at criticality, usually described by means of scaling relations and critical exponents, can be reinterpreted from a more refined geometric notion of space. In particular, we seek to investigate whether the apparent tension between certain classical results, such as the formulation of the fluctuation-dissipation theorem in critical systems, and the singular behavior of correlations can be understood more naturally when one admits that the relevant dynamics of the system occurs in an effective space of fractal nature.\\

This thesis is organized so as to lead progressively from the classical formulation of the problem of phase transitions to the central hypothesis defended here and its main consequences. Chapter 2 presents the theoretical foundations supporting the discussion, emphasizing the description of phase transitions in simple systems, Landau phenomenology, the exact solution of the two-dimensional Ising model, and Fisher's proposed generalization for the critical behavior of correlations. Chapter 3 then develops the conceptual core of the work, in which a geometric interpretation of critical exponents is proposed based on the notion of fractal dynamics, combining theoretical arguments, experimental results, and numerical analyses. Chapter 4 extends this discussion to structures of non-integer dimension, examining the implications of the proposed hypothesis in more general geometric contexts. In Chapter 5, the investigation is extended to disordered systems, with the aim of analyzing the effects of disorder on the critical exponents and on the validity of the geometric interpretation developed throughout the thesis. Finally, Chapter 6 presents the general conclusions of the work. The appendices, in turn, gather the methodological procedures supporting the analyses performed, including the \textit{box counting} method, the calculation of the spatial correlation function by spectral amplitude, and the \textit{checkerboard} update in the parallel Monte Carlo method.\\ 

Thus, the objective of this work is to articulate critical phenomena, fractal geometry, and high-performance computing within a cohesive framework capable of offering a more fundamental physical interpretation of the origin and meaning of critical exponents.
\chapter{Phase transitions} \label{chap2}

\section{Introduction to phase transitions}

\paragraph{} The distinction among phases of matter is one of society's oldest empirical observations. It is well known that when we heat a quantity of water in the liquid state, after sufficient time it begins to boil and the matter passes into a gaseous state. Analogously, if we sufficiently reduce the temperature of that quantity of water, it enters a solid phase, ice. The transition process from the liquid to the gaseous state is called \textbf{boiling}, whereas the reverse process, gas to liquid, is called \textbf{condensation}. Between the solid and liquid states we have \textbf{melting}, ice to liquid water, and \textbf{solidification} for the change from liquid water to its solid state. Note that the critical behavior of water, even in low dimensions, is very complicated \cite{Barbosa11,Silva15,Braz25,Habitzreuter25}, so that it is not as simple a liquid as it may seem at first sight. Even the pressure-driven transition of the simplest element, hydrogen~\cite{belitz94,Garavelli91,Garavelli92,Yukalov98,Patil01,Penna09,Diniz08,Penna10}, is not so simple.\\

\begin{figure}[h!]
    \centering
    \includegraphics[width=0.9\linewidth]{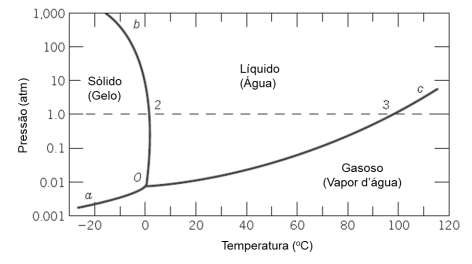}
    \caption[Phase diagram of water.]{Phase diagram of water. Retrieved from: https://www.researchgate.net/figure/Figura-1-Diagrama-de-fases-da-agua\_fig1\_351462066. Accessed on 19/07/2025.}
    \label{dfa1}
\end{figure}

It would be impossible to describe each interaction and physical quantity of the particles of a material individually; therefore, we analyze the properties of materials using macroscopic parameters of the system. Figure \ref{dfa1} presents the phase diagram of water as a function of temperature, in degrees Celsius, and pressure in units of atm. Also in \ref{dfa1}, we see how the system varies as a function of these parameters and can also observe the curves that delimit the boundaries between phases. In 1873, J. D. Van der Waals proposed a successful theory to explain the liquid and gaseous transitions of water, as well as presenting the equation of state for gases

\begin{equation}
    (P+\frac{n^{2}a}{V^2})(V-nb) = nRT, 
\end{equation}

\noindent where $a$ and $b$ are parameters that vary depending on the material under study, $P$ represents the system pressure, $V$ the volume occupied by the gas, $n$ the number of moles, $R$ the universal gas constant, and $T$ the absolute temperature. The parameter $a$ is associated with attractive interactions between particles, while $b$ accounts for the finite volume occupied by them. Thus, this equation incorporates corrections to the ideal gas model, allowing the behavior of simple systems to be described more realistically. For $H_{2}O$ molecules, for example, the values of $a$ and $b$ are respectively $5.47\times10^{-1}$ and $30.52\times10^{-6}$. The van der Waals equation shows that even a still-simple description of matter is already capable of incorporating essential aspects absent from the ideal gas model. In particular, the presence of the parameters $a$ and $b$ makes it possible to account, albeit approximately, for both the attractive interactions between particles and the finite volume they occupy. As a result, simple systems described by this equation already exhibit, at least qualitatively, fundamental features such as phase coexistence and critical behavior.\\

In addition, curves that delimit the stability regions of each phase also have a well-defined thermodynamic interpretation. A classical relation that describes these boundaries is given by the Clausius--Clapeyron equation,
\begin{equation}
    \frac{dP}{dT}=\frac{L}{T\Delta V},
\end{equation}
where $L$ is the latent heat of the transition and $\Delta V$ represents the volume change between phases in equilibrium. This equation establishes that the slope of the coexistence curves in the phase diagram depends directly on the physical properties of the material, constituting one of the most important relations in the description of simple systems.\\

The study of these fundamental systems provides the conceptual basis necessary for understanding more abstract systems, such as spin models, in which phase transitions also emerge as collective phenomena.

\subsection{Ferromagnetic System}

\begin{figure}[h!]
    \centering
    \includegraphics[width=0.6\linewidth]{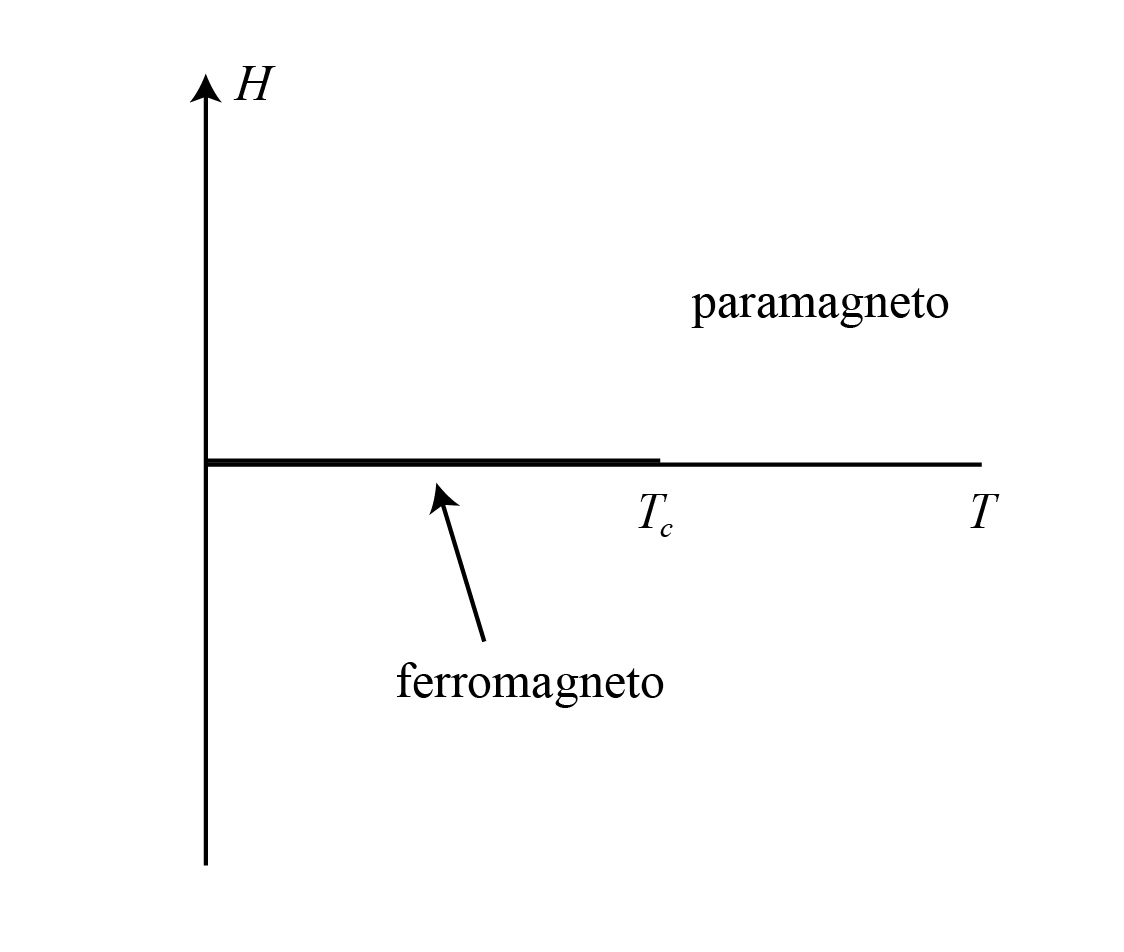}
    \caption[Phase diagram of a one-dimensional ferromagnet as a function of temperature $T$ and the external magnetic field $H$.]{Phase diagram of a one-dimensional ferromagnet as a function of temperature $T$ and the external magnetic field $H$. The bold region represents the region in which the system is in a ferromagnetic phase.}
    \label{dfm1}
\end{figure}

\paragraph{} In the case of magnetic systems, the analysis is conducted using other macroscopic parameters of interest, seeking to understand how different phases manifest themselves as these quantities are varied. Let us then use the system temperature $T$ as one of the parameters and, considering that the particles composing our material may interact with an external magnetic field, let us also consider the parameter $H$ referring to the external field applied to the material. Considering a one-dimensional ferromagnetic material, the phase diagram as a function of $T$ and $H$ is easily described by Figure \ref{dfm1}.\\

The line in 2.2 defined by $H=0$ and $T<T_c$ is called the coexistence curve, because in this region states with opposite magnetization coexist and have the same value of the Gibbs free energy per particle $g(T,H=0)$, corresponding to the phases usually called ferro-1 and ferro-2 \cite{salinas}. The total magnetization $M$ of the system, given by the difference between the number of \textit{spin-up} and \textit{spin-down} states, $M = N_{+}-N_{-}$, is the most appropriate parameter for understanding how the phases of these simple magnetic systems are distinguished. The critical point $T_{c}$ determines the temperature at which the system, for a field $H = 0$, leaves a ferromagnetic phase and enters a paramagnetic one. These phases can be described as:

\begin{itemize}
    \item Ferromagnetic phase: For zero external field, the magnetization of the system is different from $0$ for values of $T$ lower than a $T_{c}$, the temperature that describes the critical point at which a phase transition occurs;
    
    \item Paramagnetic phase: Considering the field $H=0$, the magnetization of the system is $M = 0$ for temperatures above the critical temperature $T_c$.
\end{itemize}

\begin{figure}[h!]
    \centering
    \includegraphics[width=1.05\linewidth]{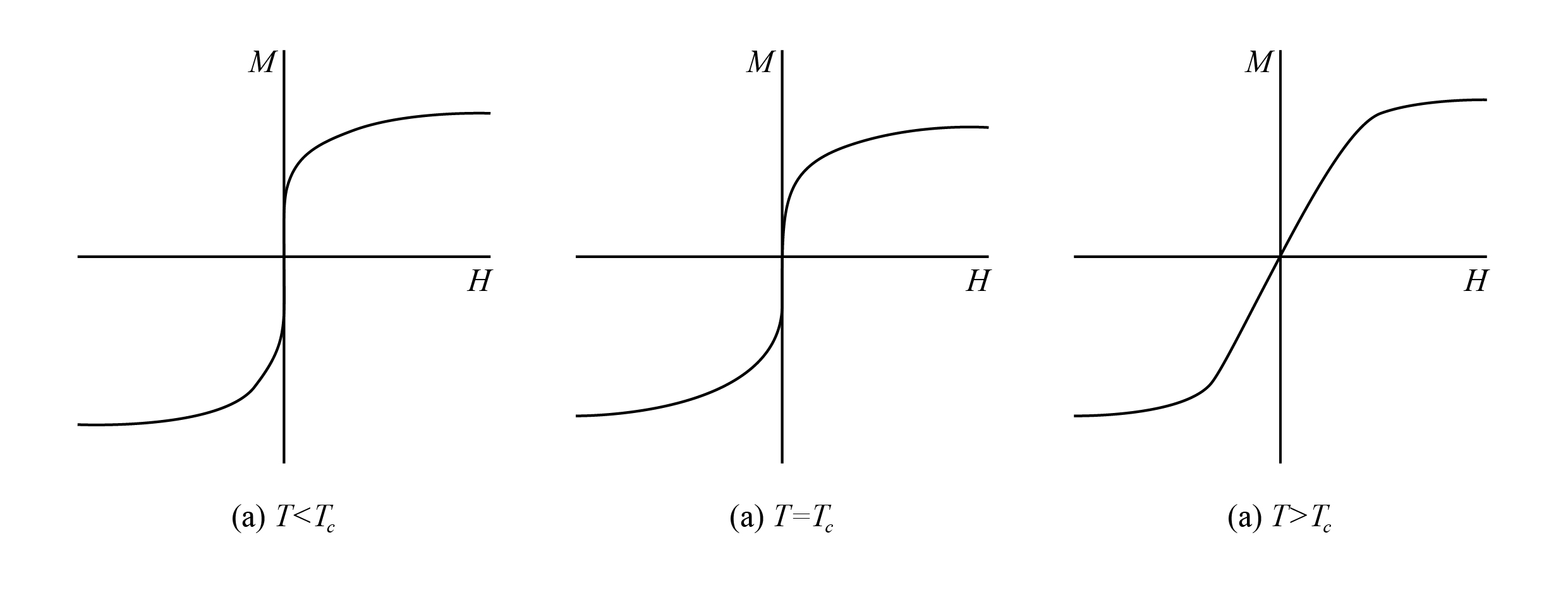}
    \caption[Typical behavior of the magnetization $M$ as a function of the external field $H$ for different temperature regimes.]{Typical behavior of the magnetization $M$ as a function of the external field $H$ for different temperature regimes: below, in the vicinity of, and above the critical temperature $T_c$.}
    \label{fig:enter-label}
\end{figure}

\begin{figure}[h!]
    \centering
    \includegraphics[width=0.6\linewidth]{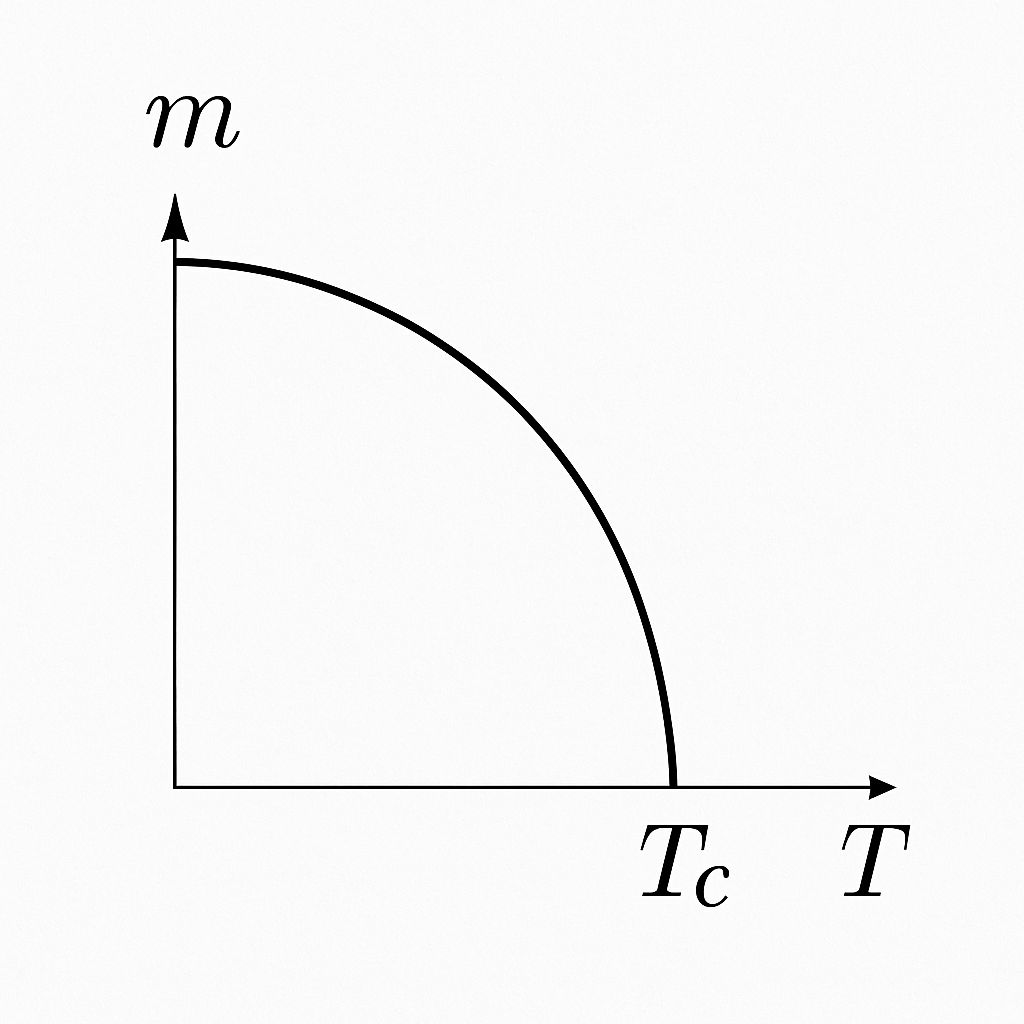}
    \caption[Typical behavior of the order parameter $m$ as a function of temperature $T$.]{Typical behavior of the order parameter $m$ as a function of temperature $T$, showing the disappearance of spontaneous magnetization when $T$ reaches the critical temperature $T_c$.}
    \label{op1}
\end{figure}

Advancing the concept of magnetization $M$, we can analyze the spontaneous magnetization per \textit{spin} of the system. Spontaneous magnetization corresponds to the value of the system magnetization even in the absence of an external magnetic field. This behavior arises below the critical temperature, when the \textit{spins} tend to align collectively due to the internal interactions of the system, characterizing the ordered ferromagnetic phase. The spontaneous magnetization of the system is given by

\begin{equation} 
    m = \frac{1}{N}\sum_{i=1}^{N}\sigma_{i},
\label{mag_def}
\end{equation}

\noindent where $\sigma_{i}$ represents a \textit{spin} of the system and $N$ is the total number of \textit{spins}. The magnetization per \textit{spin} $m$ of the system is responsible for describing the order parameter proposed by Lev Landau in the 1930s \cite{landau1937theory}. The quantity $m(T)$ is useful for describing, using the notion of spontaneous magnetization of the system \textit{spins}, the transition between the ferromagnetic and paramagnetic phases. The ordered ferromagnetic phase, represented by the bold region in Figure \ref{dfm1}, is primarily explained by the alignment of the \textit{spins} in a direction pointing upward (\textit{spin up}) or downward (\textit{spin down}). The paramagnetic phase can be explained by the existence of a state symmetric under inversion of the value of the \textit{spins}. If we observe the change in the parameter $m$, present in \ref{mag_def}, as a function of the system temperature $T$, we obtain behavior similar to that shown in Figure \ref{op1}. Figure \ref{op1} presents the magnetization-per-\textit{spin} curve as a function of the system temperature $T$, a curve that will be used as the order parameter that describes the phase of our magnetic system.

Near the critical point $T_{c}$ and with the magnetic field $H \sim 0$, the order-parameter curve $m(T)$ behaves as

\begin{equation}
    m(T) \approx A(-t)^{\beta}
\label{mag1}
\end{equation}

\noindent where

\begin{equation}
    t = \frac{T_{c}-T}{T_{c}}.
\end{equation}

Experimentally, the critical exponent $\beta$ in equation \ref{mag1} is calculated to be approximately 1/3 for the case of a simple uniaxial ferromagnet. In addition to the spontaneous magnetization $m(T)$, other physical quantities related to the ferromagnetic model exhibit behavior related to critical exponents. The magnetic susceptibility, concerning how the order parameter $m(T)$ behaves upon exposure to the magnetic field, is given by

\begin{equation}
    \chi(T,H) = \left(\frac{\partial m}{\partial H}\right)_T ,
    \label{suscept1}
\end{equation}

\noindent behaves as

\begin{equation}
\chi(T, H=0) \sim
\begin{cases}
    a_{+}t^{-\gamma}; & T > T_{c}, \\[6pt]
    a_{-}(-t)^{-\gamma}; & T < T_{c},
\end{cases}  
\end{equation}

\noindent with $\gamma$ experimentally measured at values ranging between $1.2$ and $1.4$.\\

Taking the magnetic field $H$ to be zero, the specific heat $C_v$ of the system, a quantity that expresses how the system temperature varies when its thermal energy is modified, exhibits the behavior

\begin{equation}
C_v(T, H=0) \sim
\begin{cases}
    b_{+}t^{-\alpha}; & T > T_{c}, \\[6pt]
    b_{-}(-t)^{-\alpha}; & T < T_{c},
\end{cases}  
\end{equation}

\noindent with an experimentally calculated value of approximately $\alpha \sim 0$.

\subsection{Landau phenomenology}

\paragraph{} Seeking an efficient analytical description to explain continuous phase transitions, Lev D. Landau presented, in the 1930s, a technically simple proposal for understanding these phenomena. Using the concept of the order parameter $m(T)$ and knowing that this parameter is zero for $T>T_c$, where the system is symmetric under the transformation $m\to -m$, the free energy $f(T,m)$ is expanded around the point $m=0$. For the ferromagnetic case, this expansion takes the form
\begin{equation}
    f(T,m)=f_0 + A(T)m^2 + B(T)m^4 + \ldots
    \label{expf}
\end{equation}
in which the odd terms are discarded by virtue of the symmetry of the system.\\

In the absence of an external field, the equilibrium state is obtained by minimizing this free energy with respect to the order parameter. However, when an external magnetic field $H$ is applied, the coupling between the field and the magnetization must be taken into account. In this case, the relevant thermodynamic potential is no longer only $f(T,m)$ and is instead given by
\begin{equation}
    g(T,H,m)=f(T,m)-mH.
\end{equation}
The equilibrium magnetization is then determined by the value of $m$ that minimizes this new function. In other words, the Gibbs free energy per particle can be written as
\begin{equation}
    g(T,H)=\min_m \left\{ f(T,m)-mH \right\}.
\end{equation}
Substituting the Landau expansion into this expression, one obtains
\begin{equation}
    g(T,H,m)=f_0(T)-Hm + A(T)m^2 + B(T)m^4 + \ldots
\label{expg}
\end{equation}

Considering the external magnetic field $H$ to be zero, we have the relation $g(T,0, m) = f(T,m)$. Near the critical point $T_{c}$, the value of the magnetization $m$ is sufficiently small that we can discard terms of order higher than 4. Truncating expression \ref{expf} at the $m^{4}$ term and minimizing the expansion as a function of the order parameter $m$, we obtain

\begin{equation}
    \frac{\partial f}{\partial m} = 0 = 2A(T)m + 4B(T)m^{3}
\end{equation}
\noindent and
\begin{equation}
    m = \pm(2A(T)/B(T))^{1/2}.
\end{equation}

\noindent Since the coefficient $A(T) \sim a(T-T_{C})$ and $B(T) = b$ with $b > 0$, we have

\begin{equation}
    m \sim c(T-T_c)^{1/2}
    \label{maglandau}
\end{equation}

\noindent where $c=2a/b$ is a constant and consequently the exponent value is $\beta = 1/2$. Minimizing equation \ref{expg} for small values of $m$, we obtain the relation

\begin{equation}
    \frac{\partial g}{\partial m} =  2A(T)m-H = 0,
\end{equation}

\noindent thus,

\begin{equation}
    m = \frac{H}{2A(T)} \sim \frac{H}{2a(T-T_{c})}.
    \label{mdeh}
\end{equation}

\noindent Using equations \ref{suscept1} and \ref{mdeh}, we obtain the magnetic susceptibility defined by

\begin{equation}
    \chi(T) =  \left(\frac{\partial m}{\partial H}\right) = \frac{1}{2A(T)} \sim (T-T_{c})^{-\gamma} \sim (T-T_{c})^{-1},
    \label{suscept2}
\end{equation}

\noindent where the exponent is $\gamma = 1$.  \textcolor{black}{  $\alpha=0$, $\beta=1/2$ and $\gamma = 1$ are the mean-field exponents. They are independent of dimension.}

\subsection{Onsager's exact solution for the two-dimensional Ising model}

\paragraph{} The phenomenological descriptions of the phases of matter span more than half a century through a series of names such as Boltzmann, van der Waals, L. Landau, P. Curie, among others. However, the phenomenologies presented up to the beginning of the twentieth century had an evident limitation: they dealt only with mean-field systems. The immediate consequences of this methodological choice are that these approaches qualitatively explain the existence of a critical point; however, the anomalies associated with the existence of this point lack precision in their proper analyses. Moreover, analyses for low dimensions are inefficient. In addition, by discarding the microscopic details of these systems, we lose the important contribution of fluctuations to an accurate theoretical description. It was then in \textbf{1944} that the physicist Lars Onsager presented an exact solution for the two-dimensional Ising model\cite{onsager44}. For a system with energy given by

\begin{equation}
    \label{ising_2d}
    \mathcal{H} = -J\sum_{<i,j>}s_{i}s_{j}-h\sum_{i}s_{i},
\end{equation}

\noindent where $s_{i}$ and $s_{j}$ are the system \textit{spin} variables with values $\pm 1$, $J$ is the coupling variable of these \textit{spins}, and $h$ represents the value of the external magnetic field. In his work, Onsager arrives at an exact expression for a fundamental thermodynamic quantity that opens the door to a complete understanding of the system, the free energy per \textit{spin} $f(T)$. The expression for the quantity $f(T)$ is presented by Onsager as

\begin{equation}
    \label{onsager_f(T)}
    -\beta f(T) = \ln \lambda_{\infty}
= \frac{1}{2}\ln\!\big(2 \sinh 2H_{}\big)
+ \frac{1}{2\pi} 
  \int_{0}^{\pi} 
  \gamma(w),
\end{equation}

\noindent where $H$ is defined by

\begin{equation}
    H = \frac{J}{k_{B}T},
\end{equation}

\noindent and

\begin{equation}
    \cosh\gamma(w) = \cosh (2H')\cosh(2H^{*})-\sinh(2H')\sinh(2H^{*})\cos(w).
\end{equation}

\noindent Using the relation

\begin{equation}
    \arccosh{x} = \ln(x+\sqrt{x^{2}+1}),
\end{equation}

\noindent we have

\begin{equation}
    \label{gamma(w)}
    \gamma(w)  = \ln (a-b\cos(w)+\sqrt{(a-b\cos(w))^{2}-1} ),
\end{equation}

\noindent where

\begin{equation}
    a = \cosh (2H')\cosh(2H^{*}),
\end{equation}

\noindent and

\begin{equation}
    b = \sinh(2H')\sinh(2H^{*}).
\end{equation}

\noindent Making the assumption that $H'=H^{*}$, which means assuming that the system is isotropic, that is, there are no differences among the couplings $J$ of the \textit{spins}, we can simplify the expression by considering $H'=H^{*}= K$. We then have

\begin{equation}
    \begin{aligned}
        a = \cosh^2(2K); \\
        \\b = \sinh^{2}(2K), 
    \end{aligned}
\end{equation}

\noindent where equation \ref{gamma(w)} becomes

\begin{equation}
    \label{gamma(w)_2}
    \gamma(w) = \ln(\cosh^{2}(2K)-\sinh^{2}(2K)\cos(w)\sqrt{(cosh^{2}(2K)-\sinh^{2}(2K)\cos(w))^{2}-1}.
\end{equation}

\noindent To simplify the previous equation, we can use the relations

\begin{equation}
    \begin{aligned}
        \cosh^{2}(2K) = 1 +\sinh^{2}(2K),\\
        \\ 1-\cos(w)= 2\sin^{2}(w/2), 
    \end{aligned}
\end{equation}

\noindent thereby obtaining

\begin{equation}
    \label{gamma(w)_3}
    \gamma(w)=\ln \left(\cosh(2K)\left[ \frac{1}{2}\left(1+\sqrt{1-\kappa^{2}\sin^{2}(\phi)}\right) \right]\right),
\end{equation}

\noindent where the change of variable $\phi = \frac{w}{2}$ was made and

\begin{equation}
    \kappa = \frac{2 \sinh(2K)}{\cosh^{2}(2K)}.
\end{equation}

Applying equation \ref{gamma(w)_3} to \ref{onsager_f(T)} and performing the appropriate algebraic manipulations, we arrive at a compact expression for the free energy per \textit{spin} $f(T)$ given by

\begin{equation} 
    \label{onsager_f(T)_2}
-\beta f(K)
= \ln\!\big(2\cosh 2K\big)
+ \frac{1}{2\pi}\int_{0}^{\pi}
\ln\!\left[
\tfrac{1}{2}\Big(1+\sqrt{\,1-\kappa^2\sin^2(\phi)\,}\Big)
\right] d\phi.
\end{equation}

\noindent Equation \ref{onsager_f(T)_2} is an exact result for an important physical quantity and allows us to proceed in the search for an analytical expression for the critical temperature $T_{c}$. Analyzing the integrand in \ref{onsager_f(T)_2}, we can see that the element inside the square root reaches the value $0$ when

\begin{equation}
    \label{1-kappa}
    1-\kappa^{2}\sin^{2}(\phi)=0 \iff \kappa |\sin(\phi)| = 1,
\end{equation}

\noindent which would be the first point at which analyticity is lost within the function $f(T)$ and its respective derivatives. Assuming that $\kappa(T_{c}) = 1$ and writing $\kappa$ from the change of variables $q \equiv \sinh(2K)$ together with the relation $\cosh^{2}(2K)= 1+\sinh^{2}(2K) = 1+q^{2}$, we obtain the expression

\begin{equation}
    \label{kappa(q)}
    \kappa = \frac{2q}{1+q^{2}}, 
\end{equation}

\noindent and using the condition $\kappa(T_{c})= 1$,

\begin{equation}
    \label{kappa(q)_2}
    \frac{2q}{1+q^{2}} = 1 \Rightarrow 2q = 1+q^{2} \Rightarrow (q-1)^{2} = 0 \iff q = 1 \Rightarrow \sinh(2K_{c}) = 1
\end{equation}

\noindent Using the relation $2K_{c} = \arcsinh(1) = \ln(1+\sqrt{2})$, we have

\begin{equation}
    \label{kappa(q)_3}
    K_{c} = \frac{\ln(1+\sqrt{2})}{2} \Rightarrow k_{B}T_{c} = \frac{J}{K_{c}} = \frac{2J}{\ln(1+\sqrt{2})},
\end{equation}

\noindent thus,

\begin{equation}
    \boxed{T_{c}= \frac{2J}{k_{b}\ln(1+\sqrt{2})}}
\end{equation}

\noindent which is the exact solution for the critical temperature $T_{c}$ of the two-dimensional Ising model. In \textbf{1952}, C. N. Yang presented an exact solution for the spontaneous magnetization $m(T)$. This solution is given by the expression

\begin{equation}
    m(T) =
    \begin{cases}
    \left[\,1 - \sinh^{-4}(2K)\,\right]^{1/8}, & T < T_c,\\[6pt]
    0, & T \geq T_c,
    \end{cases}
\end{equation}

\noindent which has the exponent $\beta = 1/8$, different from the value $\beta = 1/2$ of mean-field theories. The solutions initially proposed by Onsager and later by Yang constitute some of the most emblematic solutions in all of statistical physics.  

\subsection{Limits of classical theory: The collapse at the critical point.}

\paragraph{} In his \textbf{1964} paper, M. Fisher~\cite{fisher1964} sought to obtain \textit{"insights"} into the nature of matter in critical regions. M. S. Green had already pointed out problems in the definition of $G(r) \sim 1/r$, for systems of dimension $d=2$, at $T = T_{c}$, a fact noted by Fisher himself in \textbf{1964}. To this end, Fisher began the analysis from a many-particle distribution function (\textit{many-particle distribution function})\cite{fisher1964}. The function used is the pair correlation function, determined by

\begin{equation}
    \label{mpdf}
    g(r_{12}) = n_{2}(r_{1}, r_{2})/n_{1}(r_{1})n_{2}(r_{2}) = n_{2}(r_{1}, r_{2})/\varrho^{2}
\end{equation}

\noindent where $n_{2}(r_{1}, r_{2})$ is the probability of finding one particle at $r_{1}$ and another at $r_{2}$ at the same time, and $\varrho^{2}$ refers to a normalization factor that makes this quantity dimensionless, with $\varrho$ being the particle density. When $r \rightarrow \infty$, function \ref{mpdf} tends to the value $1$; thus, Fisher uses the net correlation function (\textit{net correlation function}), given by

\begin{equation}
    \label{ncf}
    G(r)=g(r)-1
\end{equation}

\noindent which exhibits the behavior $G(r)\rightarrow0$ when $r\rightarrow\infty$. Function \ref{ncf} can be interpreted as the measure of one particle on another at a distance $r$. This measure is also related to the calculation of physical quantities, an important result for describing how correlations within systems produce measurable responses of these quantities. The example used here is the fluctuation theorem for isothermal compressibility, expressed by

\begin{equation}
    \label{fluctuation_teorem}
    k_{B}T\left( \frac{\partial \varrho}{\partial p}\right)_{T} = k_{B}T\varrho K_{T} = 1 +\varrho\int G(r)dr.
\end{equation}
where $k_B$ is the Boltzmann constant, $T$ is the absolute temperature, $\varrho$ is the particle density, $p$ is the pressure, $K_T$ is the isothermal compressibility, $G(r)$ is the total pair correlation function for a separation $r$, and $\left(\partial \varrho/\partial p\right)_T$ represents the variation of density with pressure at constant temperature. Equation \ref{fluctuation_teorem} is a classic example of the fluctuation-dissipation theorem, in which fluctuations of the particle correlations $G(r)$ are related to a response of the system, which in this case is given by the isothermal compressibility $K_{T}$. It is known that $K_{T}$ diverges at the temperature $T = T_{c}$, $K_{T_{c}} \rightarrow \infty$, which, upon analyzing equation \ref{fluctuation_teorem}, is possible only if the integral of $G(r)$ becomes divergent at the upper limits. This statement is analogous to saying that, at the critical point, the correlations among the particles of the system become long ranged. In a systematic review of the theories presented mainly by Ornstein and Zernike for the effects of criticality, Fisher proposes some \textit{ad hoc} corrections, that is, corrections "imposed by hand" to avoid failures in the theory.\\

Starting from the Ornstein and Zernike proposal for a function $G(r)$ whose Fourier transform is given by

\begin{equation}
    \label{g_de_r_1}
    \tilde{G}(k) = \frac{1}{k^{2}+\kappa^{2}}.
\end{equation}

\noindent Applying the inverse Fourier transform to \ref{g_de_r_1}, we obtain

\begin{equation}
    \label{g_de_r_2}
    \mathcal{F}^{-1}[(k^{2}+\kappa^{2})\hat{G}(k)] = \mathcal{F}^{-1}[1],
\end{equation}

\noindent resulting in an equation of the form

\begin{equation}
    \label{g_de_r_3}
    (-\nabla^{2}+\kappa^{2})G(r) = \delta(r),
\end{equation}

\noindent where $\kappa = \rho^{-1}$ is the inverse correlation length $\rho$ and equation \ref{g_de_r_3} has solutions of the form

\begin{equation}
    \label{g_de_r_4}
    G(r) \sim \frac{e^{-r/\rho}}{r^{d-2}}.
\end{equation}
At the phase transition, where the correlation length $\rho$ tends to infinity, the correlation function \ref{g_de_r_4} exhibits the nonphysical behavior $G(r) \sim r^{2-d}$, which is a correlation function independent of the distance $r$ for dimension $d=2$. We then have a transition in which the correlation lengths diverge as

\begin{equation}
\rho \propto |T-T_c|^{-\nu},
\label{rhodivergence}
\end{equation}
the correlation function

\begin{equation}
    \label{g_de_r_5}
    G(r) \sim \frac{1}{r^{2-d}}.
\end{equation}
However, it was already known from Kaufman and Onsager \cite{kaufman_1949} that at the phase transition the correlation function behavior for the $2d$ Ising model behaves as $G(r) \sim r^{1/4}$ and, in addition, for $d=3$, X-ray scattering experiments in fluids exhibit a deviation from the exponent $2-d$. Equations \ref{g_de_r_4} and \ref{g_de_r_5} are results expected from Ornstein--Zernike theory; however, Fisher analyzes the failures of classical theory and proposes that, for the critical region of a $d$-dimensional system, an equation of the form be used

\begin{equation}
    \label{g_de_r_6}
    \boxed{ G(r) \sim \frac{1}{r^{d-2+\eta}} }
\end{equation}

\noindent where $\eta$ measures the deviation from the result predicted by Ornstein and Zernike~\cite{fisher1964, ornstein_zernike_1914}. Outside the critical region, where the correlation lengths $\rho$ are smaller than the distances $r$, the correlation function $G(r)$ is given by \ref{g_de_r_4}. Thus, the correlation function $G(r)$ has the behavior given by

\begin{equation}
\label{G2}
G(r) \propto
\begin{cases}
r^{2-d} \exp(-r/\rho) , &\text{ if~~ } r>\rho,\\
r^{2-d-\eta}. &\text{ if~~ }  r \ll\rho.\\
\end{cases}
\end{equation}

Thus, the introduction of the exponent $\eta$ represents more than a simple formal correction to classical theory; it signals that, at criticality, the spatial structure of the system correlations can no longer be adequately described by what was proposed by Ornstein and Zernike. The replacement of the predicted asymptotic dependence by a modified power law shows that the critical regime requires a more refined description capable of incorporating the absence of a characteristic scale of the phase transition. In this sense, Fisher's proposal not only restores consistency between theory and observed behavior, but also suggests that criticality involves a deeper reorganization of the effective geometry of the system. This issue will be central to the discussions developed in the following chapters, in which we will seek to interpret the exponent $\eta$ from a geometric perspective.

\chapter{A new fractal analysis of mean-field theory at the phase transition}

\paragraph{} The result proposed by Fisher for the correlation function $G(r)$, equation \ref{g_de_r_6}, presents an \textit{ad hoc} solution to the failure of the mean-field result at the phase transition. The exponent $\eta$ initially appears without a physical context, as a device to explain the deviation from the theoretical results expected for the critical phenomenon; thus, to better understand the problem, we begin by looking at the entire context in which this phenomenon is embedded.\\

Phases of matter, symmetry, and spontaneous symmetry breaking are key ideas for describing nature and many physical phenomena. Phase transitions presuppose the existence of parameters or characteristics capable of unambiguously distinguishing one phase from another. Near criticality, for second-order phase transitions, scale invariance commonly arises, \textcolor{black}{ characterized by (\ref{rhodivergence})} and, in some cases, this appears as self-affinity, a geometric characteristic. Therefore, analyzing the geometry of these configurations helps explain an important part of the behavior of the systems.\\

Our main objective is to establish a new response theory in fractal space. To this end, magnetic materials provide a particularly clear environment for exploring these ideas. In the Ising model, for example, the paramagnetic phase preserves symmetry, whereas ordered phases, such as the ferromagnetic and antiferromagnetic phases, correspond to regimes in which this symmetry is spontaneously broken. This model has the Hamiltonian, presented again for convenience, given by
\begin{equation}
    H = -J \sum_{\langle i,j\rangle} \sigma_i \sigma_j - h_0 \sum_i \sigma_i ,
    \label{ising2} 
\end{equation}

\noindent where $J$ is the coupling constant, $h_0$ is the external magnetic field of the system, which will be $0$ in this work, and $\sigma$ is the spin variable. The double sum $<i, j>$ runs over the nearest neighbors of the spins. The spins $\sigma$ can take values equal to $\pm 1$. The fluctuation of the spins around the mean is given by

\begin{equation}
    \psi(\vec{i})=\sigma_i-\langle \sigma_i\rangle,
\end{equation}

\noindent where $\langle\cdots \rangle$ refers to the ensemble average. The system correlation function $G(r)$ measures how fluctuations in the system interact with one another and is given by the equation

\begin{equation}
    G(r)=\langle\psi(\vec{r}+\vec{i})\psi(\vec{i})\rangle,
    \label{g(r)2}
\end{equation}

\noindent with inverse Fourier transform given by

\begin{equation}
    \tilde{G}(k) = \int d^{d}r e^{-ik \cdot r}G(r).
    \label{invg(r)2}
\end{equation}

For small values of $k$ and considering parity due to the symmetry of the system, equation \ref{invg(r)2} can be expanded simply in the form

\begin{equation}
    \tilde{G}(k) = \tilde{G}(0)(1-ak^2+O(k^4)) 
\end{equation}

\noindent and replacing this truncated series with a rational approximation, we can express $\tilde{G}(k)$ as

\begin{equation}
    \tilde{G}(r) \approx \tilde{G}(0)\frac{1}{1+ak^2} = \frac{\tilde{G}(0)}{a} \frac{1}{\frac{1}{a}+k^2}.
\end{equation}

\noindent Defining $\kappa^2 = 1/a$, we obtain the expression

\begin{equation}
    \tilde{G}(r) \approx \frac{C}{\kappa^2 + k^2},
    \label{g(r)final}
\end{equation}

\noindent where $C = \tilde{G(0)/a}$ and $\kappa= \rho^{-1}$ is the inverse correlation length. The inverse transform of \ref{g(r)final}, which brings $\tilde{G}(r)$ into real space, is given by function \ref{g_de_r_4}.\\

As a justification for the difference between the expected result $G(r) \sim r^{2-d}$ and the result presented by Fisher $G(r) \sim r^{2-d-\eta}$, it has been attributed to the failure of the fluctuation-dissipation theorem (FDT) at the phase transition of these systems. Our hypothesis is that the deviation from the expected result is associated with the geometry of the problem, which would be fractal at the critical temperature \cite{Lima24, Lima25}.

\begin{figure}[h!]
    \centering
    \includegraphics[width=0.85\linewidth]{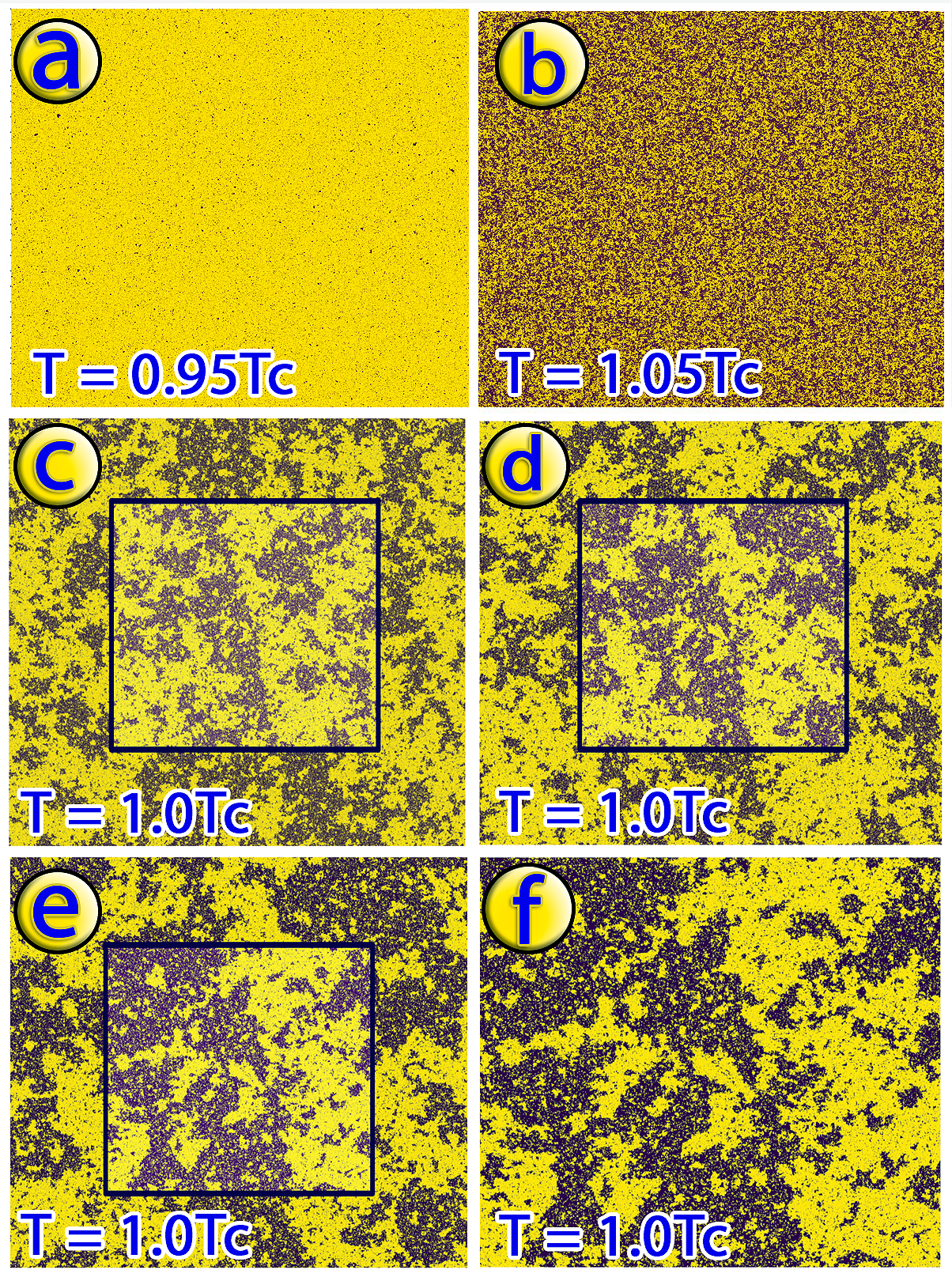}
    \caption[Typical configurations of the Ising model on a two-dimensional lattice for different temperatures around $T_c$.]{Typical configurations of the Ising model on a two-dimensional lattice for different temperatures around $T_c$.
(a) $T=0.95\,T_c$, ferromagnetic phase, subcritical regime, with predominance of one magnetized phase and only small islands of the opposite orientation.
(b) $T=1.05\,T_c$, paramagnetic phase, supercritical regime, in which long-range order disappears and the configuration fragments into smaller and more disordered domains in a regime that is symmetric under the change in spin value $1 \rightarrow-1$.
(c)-(f) $T=T_c$, critical point, characterized by large fluctuations and self-similar patterns at any scale, evidencing the absence of a characteristic length and the fractal organization of the domains.
The colors (yellow/purple) indicate the two possible spin orientations. Figure taken from \cite{Lima25}.}
    \label{spins_1}
\end{figure}

\subsection{Fractal Dynamics}

Attempting to describe the origins of this correction, in recent works, Lima and collaborators~\cite{Lima24,Lima25,Carrasco26} conducted a complete analysis of the correlation function (\ref{g_de_r_1}). The authors were inspired by three main observations:
\begin{enumerate}
\item Fluctuation-Dissipation Theorem (FDT): Correlation functions represent a simple form of the FDT (see the collection of articles~\cite{Nowak22} and the reviews \cite{Oliveira19,GomesFilho25}), which are directly related to susceptibility \cite{Goldenfeld18}. It is known that the FDT fails when ergodicity is violated \cite{Costa03,Costa06,Lapas08,Wen23}, a phenomenon widely studied in structural glasses \cite{Ricci-Tersenghi00,Crisanti03,Grigera99,Barrat98,Bellon02,Bellon06}, random-exchange Heisenberg chains \cite{Vainstein05}, proteins \cite{Hayashi07}, KPZ dynamics \cite{Kardar86,Cordeiro01,Rodrigues15,Henkel26,Gomes19,GomesFilho20,GomesFilho21b,Anjos21,GomesFilho24}, mesoscopic radiative heat transfer \cite{Perez-Madrid09,Averin10}, and ballistic diffusion \cite{Oliveira01,Morgado02,Morgado04,Bao03,Bao06,Lapas07,Vainstein06,Vainstein05a,Ferreira12,Ferreira22};
\item Response Functions in Growth Phenomena: Significant progress has been made in understanding response functions in growth phenomena by assuming an underlying fractal dynamics \cite{Barabasi95,Mello01,Mello01,Rodrigues14,Feder22, GomesFilho24,Anjos21,Luis22,Luis23};
\item Scaling and Fractal Geometries: Scaling relations are fundamental at the phase transition. We have, for example, the hyperscaling hypothesis that relates the specific-heat exponent $\alpha$ to $\nu$, given by
\begin{equation}
\alpha = 2 - d\nu,
\label{alfa_eq_1}
\end{equation}
the Fisher relation
\begin{equation}
  \gamma = (2-\eta)\nu,
  \label{gamma_eq_1}
\end{equation}
and the Rushbrooke equality
\begin{equation}
    \label{rushbroke_1}
    \alpha + 2\beta + \gamma = 2.
\end{equation}
Finally, the relation between the order-parameter exponent $\beta$ and
the fractal dimension of the ordered phase $d_f$ proposed by Suzuki \cite{Suzuki8}
\begin{equation}
\label{dl}
d_f=d-\frac{\beta}{\nu}\;
\end{equation}
\textcolor{black}{At the critical point of the percolation model, this fractal structure corresponds to the infinite percolating cluster\cite{Grimmett06,Cruz23}, whereas for general systems it is associated with the largest ordered cluster~\cite{Kroger00}.}
\end{enumerate}

These relations arise from scale invariance at the transition and relate thermodynamic exponents to geometric dimensions~\cite{Family85,Rodrigues24}. These concepts lead us to the need to understand the geometry at the transition as a fractal geometry.

 In Figure~\ref{spins_1} we provide a qualitative illustration of the change of regimes when crossing criticality; Fig.~\ref{spins_1} presents representative configurations of the 2D Ising model around $T_c$.
  The colors (yellow/purple) indicate the two possible spin orientations (+1/-1). For $T=0{.}995\,T_c$ (Fig.~\ref{spins_1}a), the predominance of one spin orientation is observed,
with the ordered phase occupying almost the entire system and only small islands of the opposite orientation,
consistent with the existence of spontaneous magnetization below $T_c$.
Above the transition, at $T=1{.}015\,T_c$ (Fig.~\ref{spins_1}b), long-range order is lost
and the spatial texture becomes more fragmented, with smaller and more irregular domains, reflecting the
disordered character of the paramagnetic regime.
At the critical point (Fig.~\ref{spins_1}c), the configuration exhibits fluctuations on multiple scales,
with large and small domains coexisting without a well-defined characteristic length, which is consistent
with the divergence of the correlation length and with the statistical self-similarity expected at criticality;
this critical morphology provides geometric intuition for the emergence of effectively fractal structures
associated with order-parameter fluctuations and correlations. The square inserted in (Fig.~\ref{spins_1}c) is (Fig.~\ref{spins_1}d), and so on until we reach (Fig.~\ref{spins_1}f). In all these transformations, we observe scale invariance. \\

Based on these concepts, we analyze the correlation function starting from the following ideas:
\begin{enumerate}
\item Mean-field theory proves inefficient for analyzing the order parameter at the critical point because it suppresses fluctuations when they are relevant. Notably, according to equation \ref{rhodivergence}, $\rho$ diverges when $T \rightarrow T_c$, indicating long-range correlations. This is associated with the emergence of a scale-invariant fractal structure, where fluctuations manifest themselves at all length scales. Consequently, this function should provide an accurate description if the appropriate geometry is used.
\item In physics, fractal geometry can emerge from predefined rules, such as anomalous infiltration in partially saturated porous media~\cite{Voller25} or fractal circuits~\cite{Chen17,Boyle07,Anjos24}. However, when derivatives are defined in Euclidean space, a fractional derivative may be necessary when extending the analysis to a fractal medium, as in the fractional heat equation~\cite{Angulo00} or the fractional Fokker--Planck equation~\cite{Metzler99,Barkai01,Sokolov01}. This is the case of Eq. (\ref{g_de_r_1}), as suggested by Fig. \ref{spins_1}, where we show some representations of a two-dimensional Ising lattice. At the critical temperature $T_{c}$, the system exhibits a fractal (scale-invariant) structure of spin clusters, together with the divergence of $\rho(T)$. The variation of $G(r)$, and therefore of $\nabla^2 G$, is more relevant at the fractal boundary of the clusters, which is where equation \ref{g_de_r_3} becomes nontrivial.
\end{enumerate}

As $T \rightarrow T_c$ and $\rho$ diverges according to Eq. (\ref{rhodivergence}), the correlation extends to long range owing to the emergence of a fractal structure, which can be better explained by using a fractal Laplacian. We then replace equation \ref{g_de_r_3} by
\begin{equation}
\label{G3}
(-\nabla^2)^\zeta G(r)=\delta^{(d_R)}(r),
\end{equation}
where $d_R$ is a fractal dimension associated with the Riesz fractional derivative of order $\zeta$, where $1/2 < \zeta < 1$, associated with a fractal dimension $d_{R}$, with $d-1 < d_{R} < d$. The preference for the Riesz fractional derivative is due to the existence of a well-defined Laplacian for it and to the identity~\cite{Muslih10,Muslih10b}
\begin{equation}
\label{SG3}
(-\nabla^2)^\zeta \left(\frac{1}{|\vec{r}-\vec{r'}|^{d_R-2\zeta}} \right) = C_{\zeta,d_R} \delta^{(d_R)}(\vec{r}-\vec{r'}),
\end{equation}
where $\delta^{(d_R)}(\vec{r}-\vec{r'})$ is a $\delta$ function in a space of fractal dimension $d_R$ and $C_{\zeta,d_R}=[4^{\zeta} \pi^{d_R/2} \Gamma(\zeta)]/[ \Gamma(d_R/2-\zeta)]$ is a constant. 
 The Dirac delta function in fractional dimension is given by
\begin{equation}
\label{dd1}
 \int \delta^{d_R}(\vec{r}-\vec{r'})f({ \vec{r'}})d^{d_R}\vec r'=f({\vec{r}}), 
\end{equation}
 in which the function $f(r)$ is any continuous function. Another important mathematical aspect is the nonlocality of the fractional derivative, which is consistent with the collective phenomenon that is a phase transition.

Comparing Eq. (\ref{G2}), in the limit $\rho \rightarrow \infty$, with (\ref{G3}) and (\ref{SG3}), we obtain $ \eta=2(1-\zeta)+d_R-d$. 
Finally, using a Gauss theorem for fractal distributions, the Fisher exponent is derived exactly as~\cite{Lima24}
\begin{equation}
\label{eta}
\eta=d-d_R=1-\zeta.
\end{equation}

\noindent that is, the Fisher exponent $\eta$ can be interpreted as the deviation between the integer dimension in which the structure of the system is contained and the effective dimension of the system dynamics at criticality, which is fractal, ensuring that the operator preserves its fractional nature and that the support of the correlation transitions between an almost hypersurface-like subset and the complete Euclidean space.\\

An intuitive way to understand Eq.~(\ref{G3}) is to recall that, at the critical point, the system becomes scale invariant; therefore, if we enlarge all distances by a factor $\lambda$, that is, $|\textbf{r}| \to \lambda |\textbf{r}|$, the equation governing $G(r)$ should not change in form; at most, its terms may rescale by power factors. Requiring the term on the left-hand side and the source on the right-hand side to exhibit the same behavior under this transformation imposes a homogeneity condition that, in turn, suggests replacing the usual Laplacian with an operator of fractional order. This choice preserves the required scaling structure and can be interpreted as an effective way of capturing nonlocality and/or dynamics constrained to geometries more general than standard Euclidean space. Operators of this type were presented by Laskin \cite{Laskin07} when deriving the fractional Schr\"odinger equation within what may be called fractional quantum mechanics.\\

\subsubsection{Relation between the fractal dimensions.}

An important relation between the fractal dimensions can be established by combining equations \ref{alfa_eq_1}, \ref{gamma_eq_1}, the equation proposed by Suzuki \ref{dl}, and the Rushbrooke equality, \ref{rushbroke_1}
to obtain
\begin{equation}
    \label{dl2}
    \boxed{d_{R} = 2(d_{f}-1)}.
\end{equation}

Before proceeding with our numerical analysis, we draw attention to the fractional dimensions obtained in percolation clusters, as described, for example, by Suzuki and Coniglio; in total, five fractal dimensions are presented for percolation clusters. The dimension $d_R$ introduced here represents a new fractal dimension, distinct from those reported previously. However, Eq. (\ref{dl2}) established a connection between them.

Table \ref{Table1} presents values of the critical exponents of the Ising universality class for $d = 2, 3$ and $4$. We thus determine the values for the quantity $\eta$, which measures the deviation between the correlation fractal dimension and the integer dimension of the system, and for the fractal dimension $d_{R}$. In addition, as $d = d_{c} = 4$, the upper critical dimension, is approached, the fractal dimension $d_{f}$ becomes equal to $d_{c}$.

%----------------------------------------------------
\begin{table}[h!]
        \centering
       	\begin{tabular}{c|ccc}
		\hline	\hline
\hspace{5mm}	$d$\hspace{5mm}  & \hspace{5mm} $2$ \hspace{5mm}   & \hspace{5mm} $3$ \hspace{5mm} & \hspace{5mm} $4$ \hspace{5mm}   \\ \hline
%		$\alpha$  & $0 (\log)$  &        $0.110(1)$        & $0$        \\ 
{	$\beta$} & $1/8 $   &  $ 0.3265(3) $ &  $ 1/2$  \\ 
{		$\nu$} & $1$   &     $ 0.6300(3) $    & $1/2$  \\ 
{  		$d_f$} & $15/8$   &     $ 2.4817(5) $    & $3$  \\ 
  %       $\gamma$ & $7/4$   &     $ 1.2372(5)$   &  $ 1$ \\ 
		$\eta$ & $1/4$   &     $ 0.0364(5) $    & $0$  \\ 
		$d_R$ & $7/4$   &     $ 2.9636(5)$  &  $4$ \\
        $\zeta$ & $3/4$   &     $ 0.9636(5)$  &  $1$ \\
		 \hline
		 	\hline
	\end{tabular}	
	\caption[Values of the critical exponents of the Ising universality class for dimensions $d = 2,3$ and $4$.]{  Values of the critical exponents of the Ising universality class for dimensions $d = 2,3$ and $4$. The exponent $d_{f}$ is obtained from \cite{Coniglio89}, $d_R$ from equation \ref{dl2}, and $\eta$ from equation~\ref{eta}. Note the symmetric deviation both for the exponent $d_{R} = d-\eta$ and for the fractional-derivative exponent $\zeta=1-\eta$. For $d=3$ we use the results of Pelissetto and Vicari \cite{Pelissetto02}.}
   \label{Table1}
\end{table}
%----------------------------------------------------

\subsection{Numerical Analysis}

\paragraph{} We then use a series of mathematical and computational resources to perform and analyze our experiments. Using the Monte Carlo method, the Metropolis and \textit{checkerboard} algorithms (computational application presented in Appendix \ref{ap:checkerboard}) to perform the simulations in parallel, and using \textit{CUDA} technology (\textit{Compute Unified Device Architecture}) \cite{CudaSite, sitecuda2, sitecuda3, sitecuda1}, we carried out a set of simulations with lattices of size $L \times L$, where $L$ took values equal to $1024, 2048, 3072, 4096$, and $5120$ in the simulations. The exact size used in each simulation will be specified throughout the text. In addition, we chose the notation $T_O$, which will be used throughout the work, to refer to the Onsager critical temperature, which has the value $T_O \sim 2.269J/k_B$. The critical temperature is then $T_c = 1$ in units of $T_O$ for the case of the classical system without disorder. \textbf{This work will later present systems with disorder, in which, for each associated disorder value $\sigma$, we will have a value $T_c(\sigma)$. }

\paragraph{} The magnetization per \textit{spin} of the system is given by the equation

\begin{equation}
    m = \frac{1}{N}\sum_{i}^{N} \sigma_{i},
\end{equation}

\noindent where $N$ is the total number of \textit{spins} and $\sigma$ is the value of the \textit{spin} with index $i$. Because our model is an equilibrium-physics model, each Monte Carlo interaction seeks the most probable \textit{ensemble}; thus, enough simulations must be run until a stable region of lattice samples is reached. In our work, we verify how the value of the order parameter $m$ varies as a function of the number of interactions performed, thereby making it possible to understand the number required to exclude the lower-probability \textit{ensembles} from our experiment.\\

\begin{figure}[h!]
    \centering
    \includegraphics[width=0.85\linewidth]{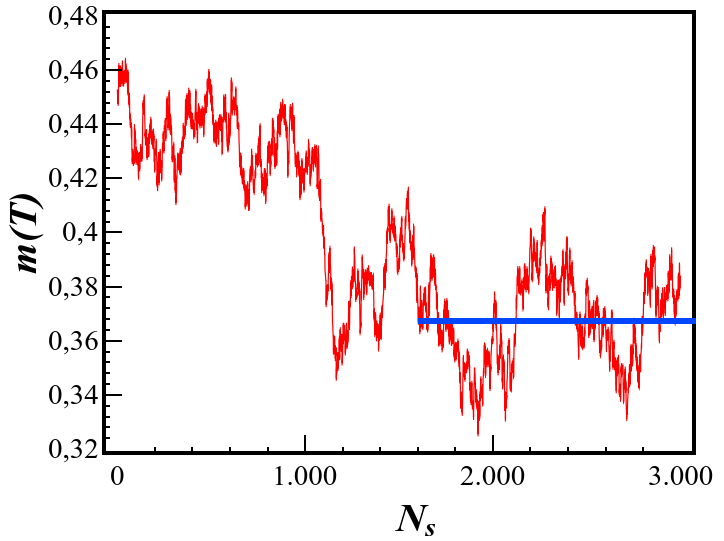}
    \caption[Order parameter $m(T)$ for a temperature $T = 1.000T_c$ as a function of the number of parallel Monte Carlo interactions using the $checkerboard$ method.]{Order parameter $m(T)$ for a temperature $T = 1.000T_O$ as a function of the number of parallel Monte Carlo interactions using the $checkerboard$ method. The horizontal blue line was drawn as a visual guide to represent the approximate mean value of \(m\) in the region \(N_s \geq 1600\), where the system is considered sufficiently stable for us to define the value of $m(T)$ and for which \(\bar m \approx 0{,}37 \pm 0{,}02\) was estimated.}
    \label{m_nsim}
\end{figure}

Figure \ref{m_nsim} shows values of the order parameter $m(T)$ as the number of parallel Monte Carlo interactions using the $checkerboard$ method is increased. The method performs $L^{2}$ simultaneous interactions, one interaction per \textit{spin} site; thus, 3000 interactions are equivalent to $3000 \times L^{2}$ individual interactions. For each temperature we perform an average of 281200 parallel Monte Carlo interactions, each interaction involving $L^{2}$ \textit{spins}, and, for each simulation, 200 temperatures. Having defined the working method, we performed computational experiments in order to analyze the parameters of interest.

\subsection{Extraction of the exponents $\beta$ and $\nu$}

\paragraph{} To obtain the critical exponents $\beta$ and $\nu$, we explore both the asymptotic behavior of the order parameter $m(T)$ and the correlation-length curve $\rho(T)$ in the paramagnetic phase, that is, for temperatures above $T_c$. The central hypothesis is that, sufficiently close to the transition, the order parameter obeys a universal power law of the type
\begin{equation}
m(T)\propto |T_c-T|^{\beta},
\label{m(T)_fit}
\end{equation}
while the function $\rho(T)$ follows \ref{rhodivergence}. First, we determine $T_c$ using the order-parameter curve $m(T)$, analyzing the dependence of the fit of function \ref{m(T)_fit} on the chosen data window; we keep $T_{\min}$ fixed and vary $T_{\max}$, identifying a stability interval in which $T_c(T_{\max})$ remains approximately constant. With this final value of $T_c$ selected, we then construct a log--log plot of $m$ as a function of $(T_c-T)/T_O$, where $T_O$ corresponds to the Onsager critical temperature, approximately 2.2692 for $\sigma = 0$, using only points sufficiently close to criticality, but still outside the region in which finite-size effects, statistical noise, or corrections to the dominant scaling may distort the power-law regime.\\

In this representation, the power law becomes a linear relation,
\begin{equation}
\log m = \beta\,\log\!\left(T_c-T\right) + \mathrm{const},
\end{equation}
so that the exponent $\beta$ is obtained directly as the slope of the linear fit in the range where the data exhibit scaling behavior. The uncertainty associated with $\beta$ is extracted from the fit error and, when appropriate, supplemented by the variation of the result when the interval of points included in the fit is slightly modified, ensuring that the final value reflects the critical regime and not a particular choice of window.\\

\begin{figure}[H]
    \centering
    \includegraphics[width=0.75\linewidth]{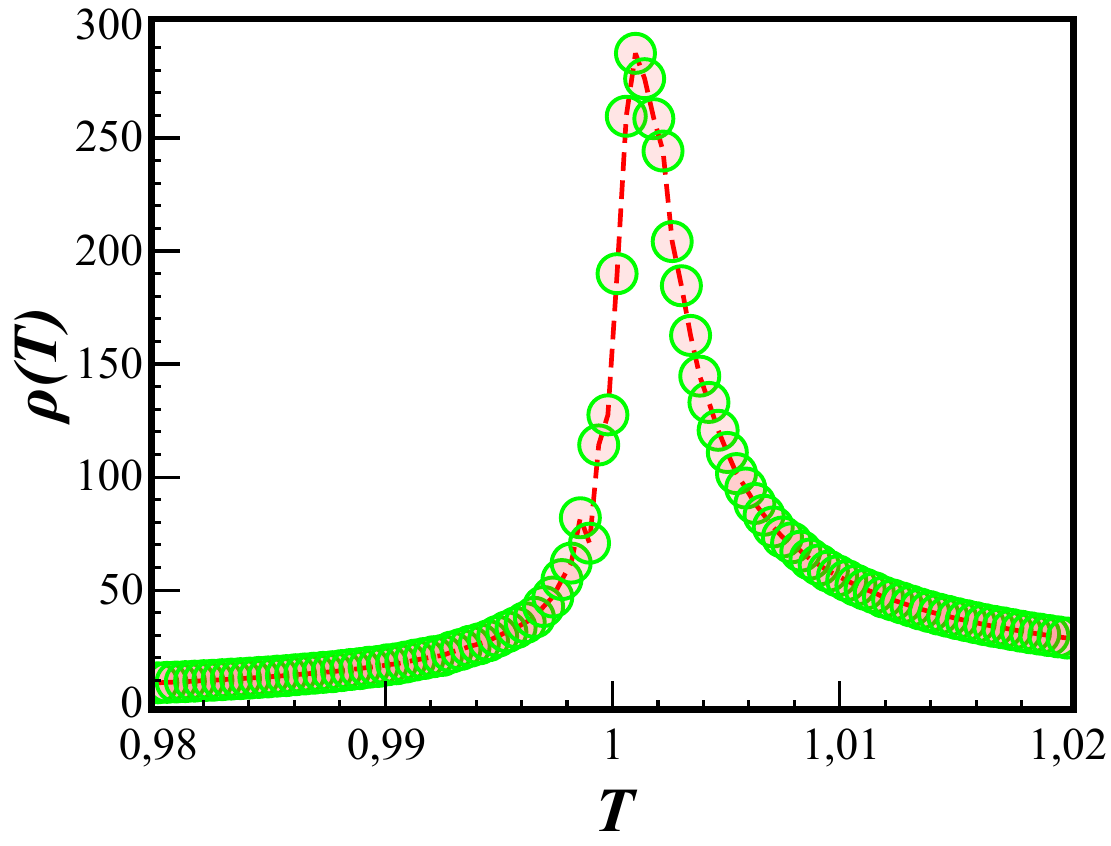}
    \caption[Typical behavior of the correlation length $\rho(T)$ as a function of temperature $T$.]{Typical behavior of the correlation length $\rho(T)$ as a function of the reduced temperature $T$, here defined in units of $T/T_O$. The sharp growth of $\rho(T)$ near criticality is observed, with a maximum around $T=1$, evidencing the characteristic divergence of the correlation length in the vicinity of the phase transition.}
    \label{rho_curve_0}
\end{figure}

\begin{figure}[H]
    \centering
    \includegraphics[width=0.85\linewidth]{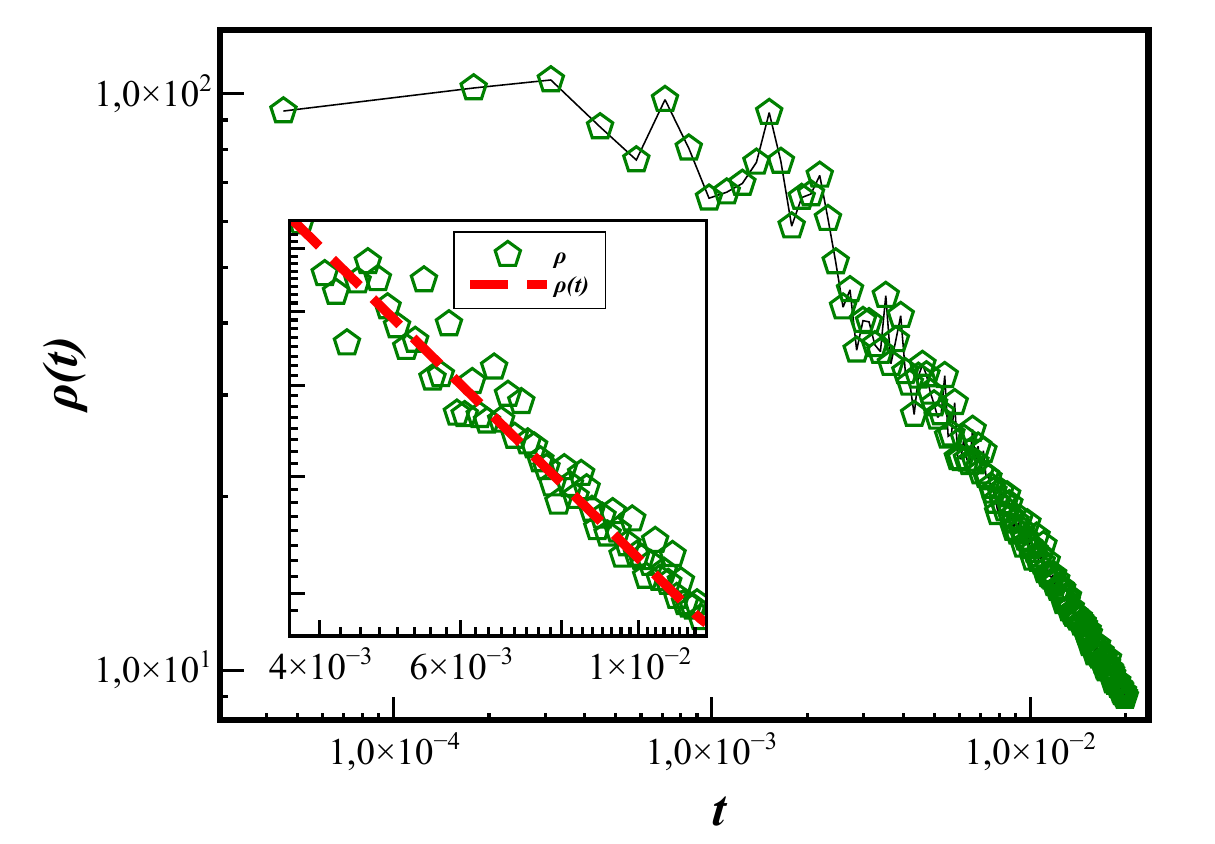}
    \caption{Log--log plot of the correlation length $\rho\left(t\right)$, with $t\equiv (T-T_c)$, in the vicinity of criticality, with the power-law fit used to determine $\nu$.}
    \label{rho_adjusted}
\end{figure}

To extract the exponent $\nu$, we need to measure the correlation-length curve $\rho(T)$, a quantity obtained from fitting a general correlation function

\begin{equation}
    G(r) \sim A\frac{e^{-r/\rho}}{r^{d-2+\eta}},
\label{g(r)_fit}
\end{equation}
for several values of the temperature $T$, where $A$ is a fitting parameter. Figure \ref{rho_curve_0} shows the behavior of the correlation length \(\rho(T)\) as a function of the reduced temperature \(T/T_O\). It is observed that, as the system approaches the critical temperature, \(\rho(T)\) grows sharply, reaching a maximum near \(T=1\). Once the $\rho(T)$ curve is obtained, we find the value of the critical exponent in a manner similar to that used previously for the exponent $\beta$, fitting the curve $\rho(T)$ with a function of the form

\begin{equation}
\log \rho = -\nu\,\log\!\left(T_c-T\right) + \mathrm{const}.
\end{equation}
The fitted curve is shown in Figure \ref{rho_adjusted}.\\

\begin{figure}
    \centering
    \includegraphics[width=0.95\linewidth]{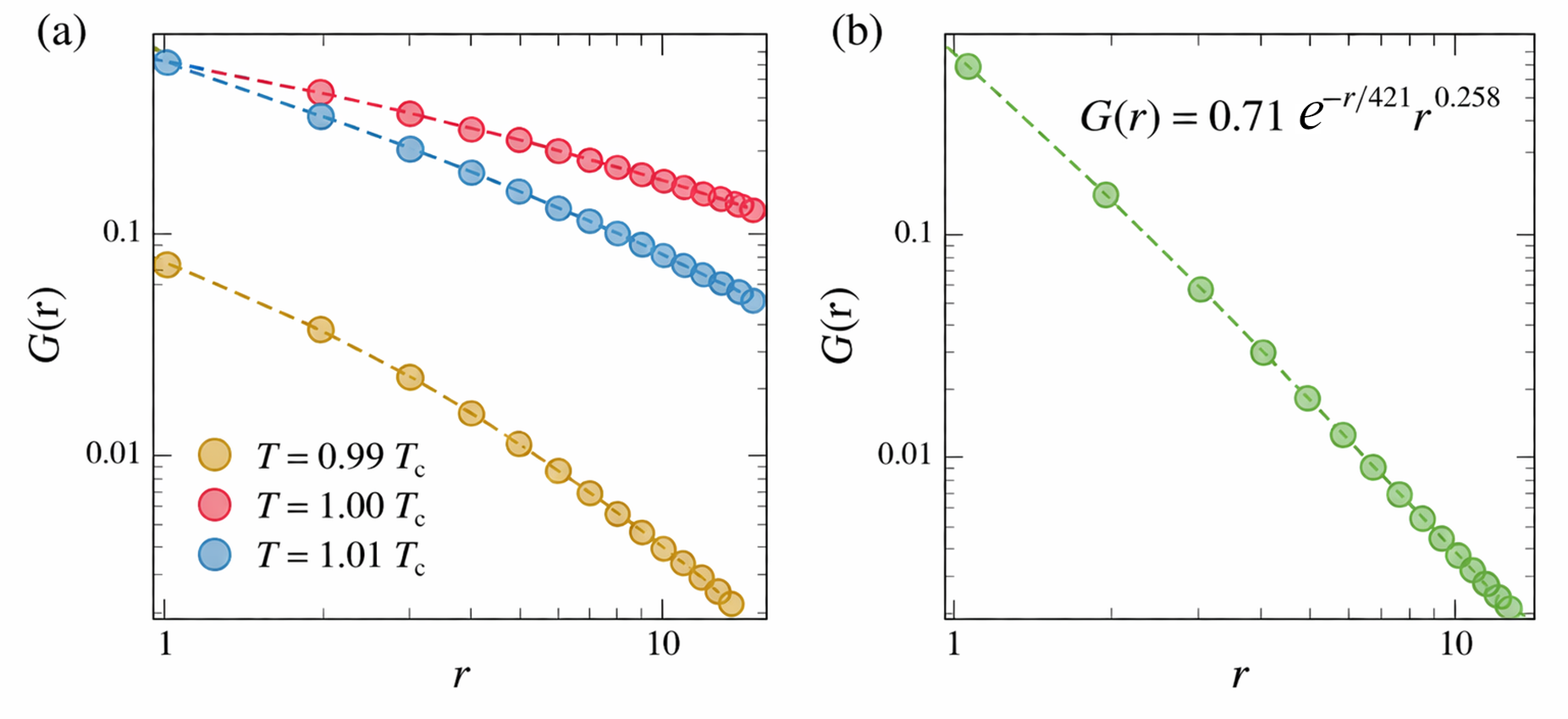}
    \caption[Correlation function $G(r)$ as a function of distance $r$.]{(a) Correlation function $G(r)$ as a function of distance $r$ for the temperatures considered. The dashed lines correspond to the fits obtained from equation \ref{g(r)_fit}. (b) Enlargement of the result for $T=T_c$, showing the agreement between the numerical data and the fit, from which the parameters $\rho=421(3)$ and $\eta=0.258(1)$ are obtained. Figure adapted from the supplementary material of \cite{Lima24}.}
    \label{g(r)_fig}
\end{figure}

To determine the spatial correlation function \(G(r)\), we apply the spectral-amplitude method directly to the two-dimensional spin lattice of the Ising model. Considering the final configuration of the system as a discrete function \(S(x,y)\), defined on the lattice sites, its two-dimensional discrete Fourier transform is first calculated; next, the power spectrum associated with the configuration is constructed, given by the squared modulus of this transform, \(|\hat{S}(k_x,k_y)|^2\). Based on the Wiener--Khinchin theorem, the inverse transform of this spectrum yields the spatial autocorrelation of the configuration as a function of the displacements \((dx,dy)\), that is, the correlation between spins separated by different distances on the lattice. By the Wiener--Khinchin theorem, the spatial autocorrelation can be obtained by the inverse Fourier transform of this spectrum, that is,
\[
C(dx,dy)=\mathcal{F}^{-1}\left(|\hat{S}(k_x,k_y)|^2\right).
\]
Thus, the correlation between spins separated by different displacements \((dx,dy)\) on the lattice is obtained. Since the physical interest lies in dependence only on the scalar distance \(r\), and not on direction, the two-dimensional function \(C(dx,dy)\) was subsequently subjected to a radial average, resulting in \(G(r)\). This procedure makes it possible to calculate the correlation in a numerically efficient manner, avoiding the direct sum over all pairs of \textit{spins} on the lattice. Figure \ref{g(r)_fig} presents the behavior of the correlation function $G(r)$ for three distinct temperature values. In particular, for $T=T_c$, the curve becomes approximately linear on a log--log scale, evidencing asymptotic power-law behavior. This result is compatible with the critical regime, in which the correlations are no longer described by exponential decay.\\

We therefore have a way to measure the correlation between \textit{spins} at positions $\textbf{x}$ and $\textbf{x}+r$ for a significant number of sizes $r$; in this work we use a maximum value of $r=200$. Technical details about the method, as well as its explanation, are presented in the appendix of this document. For each temperature $T$, we then have a curve of the function $G(r)$ from which the parameters $\rho$ and $\eta$ can be extracted.

\begin{figure}[h!]
    \centering
    
    % Panel (a)
    \begin{subfigure}[h!]{0.56 \linewidth}
        \centering
        \includegraphics[width= \linewidth]{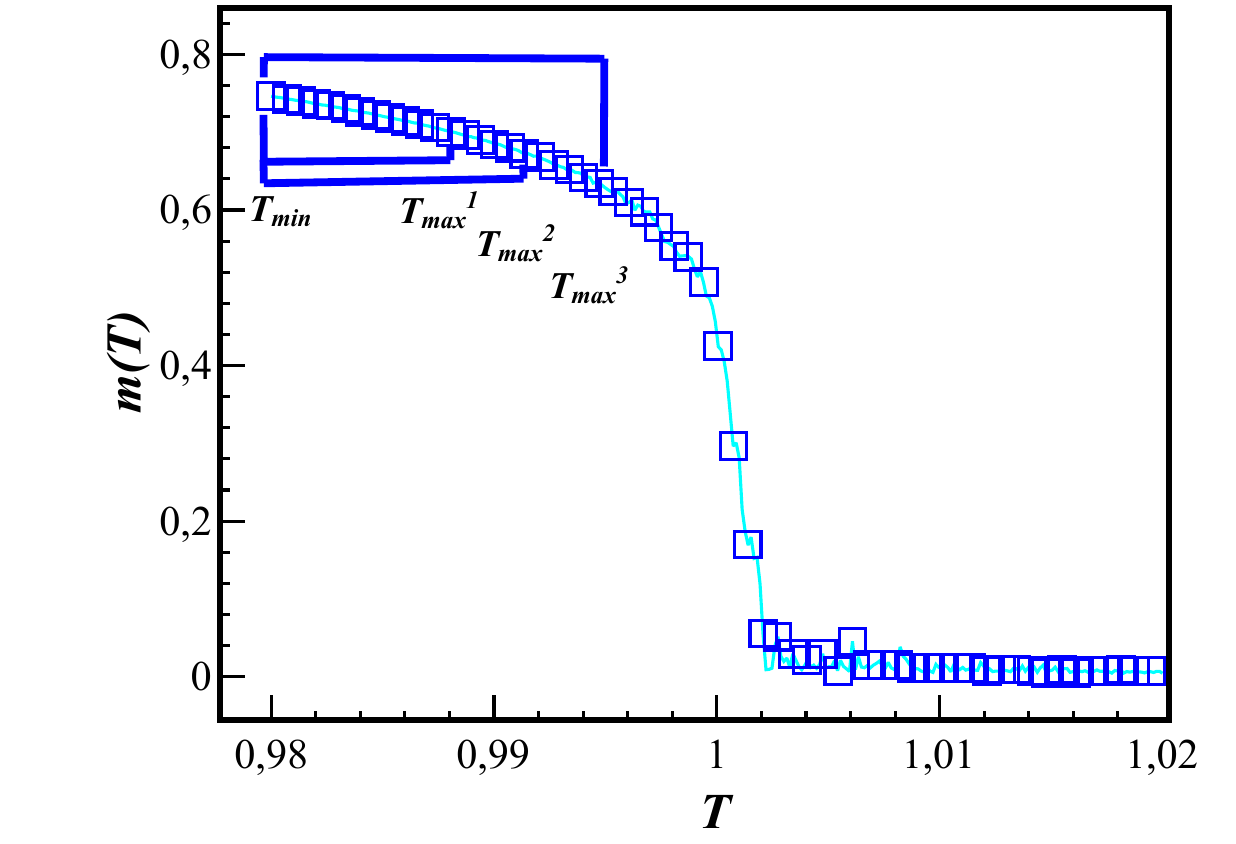}
        \caption{}
        \label{mag_tc_tmax}
    \end{subfigure}
    % Panel (b)
    \begin{subfigure}[h!]{0.56 \linewidth}
        \centering
        \includegraphics[width=\linewidth]{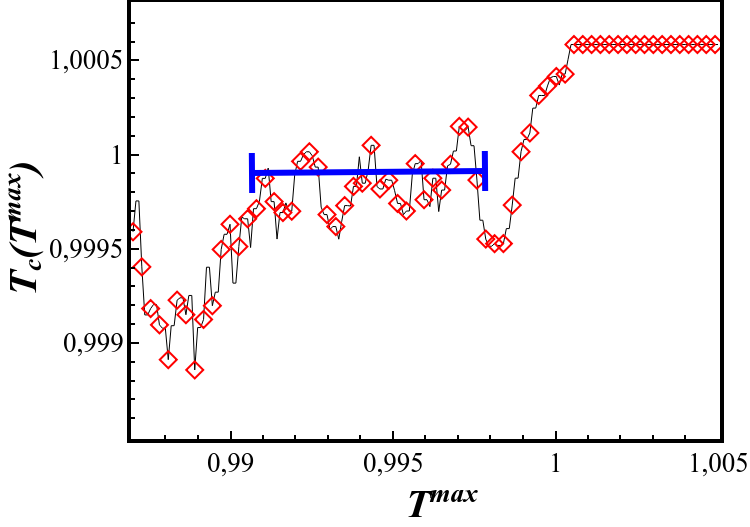}
        \caption{}
        \label{tc_tmax}
    \end{subfigure}
    % Panel (c)
    \begin{subfigure}[h!]{0.59 \linewidth}
        \centering
        \includegraphics[width=\linewidth]{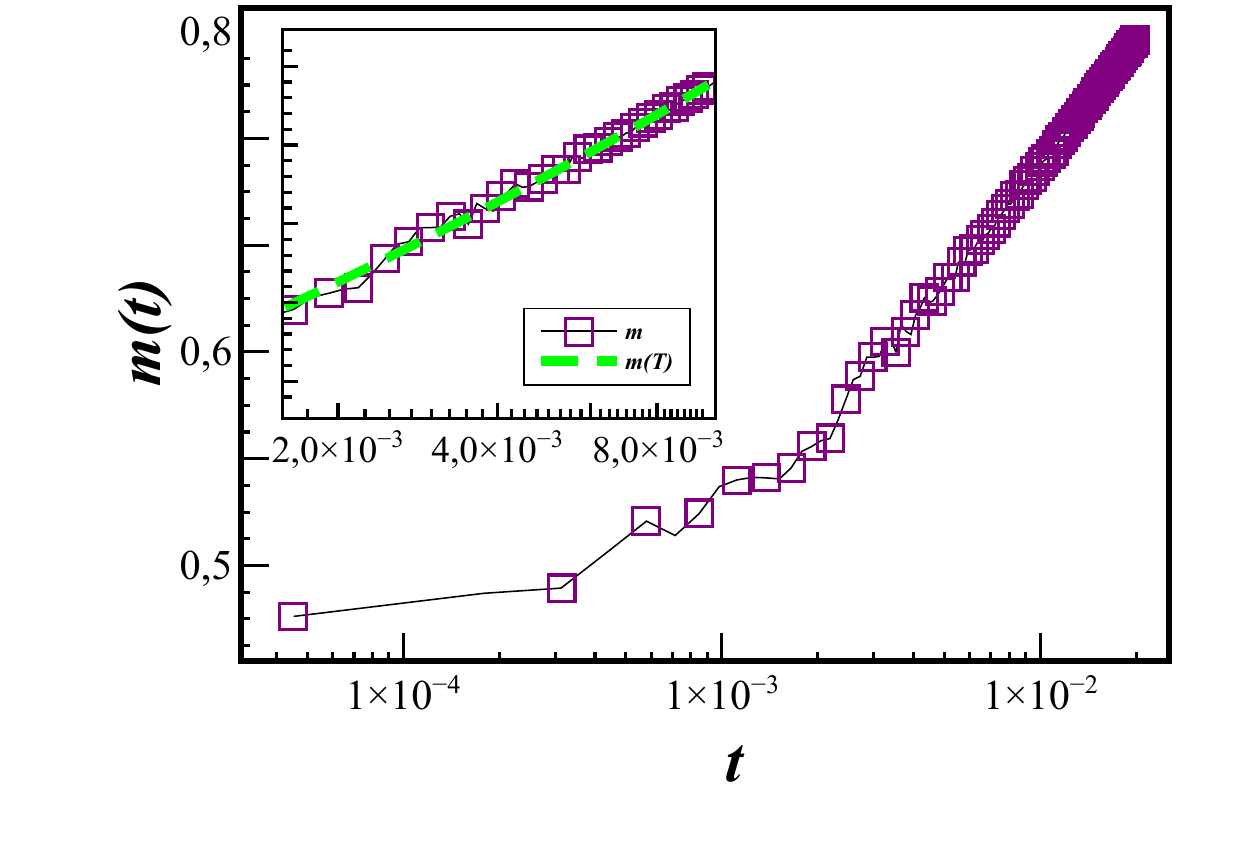}
        \caption{}
        \label{mag_fit_1}
    \end{subfigure}

    \caption[Sequence of steps for determining the critical temperature and the critical exponent $\beta$.]{Sequence of steps for determining the critical temperature and the critical exponent $\beta$.
(a) Order-parameter curve $m(T)$ illustrating the choice of fitting windows, in which $T_{\min}$ is kept fixed and $T_{\max}$ is varied to evaluate the sensitivity of the fit to the window selection.
(b) Estimated critical temperature $T_c(T_{\max})$ as a function of the upper fitting limit.
The highlighted region indicates the stability interval from which the final value of $T_c$ is selected.
(c) Log--log plot of the order parameter $m(t)$, with $t\equiv (T_c - T)$, in the vicinity of criticality, with the power-law fit used to determine $\beta$.
The exponent extracted for the data in the $inset$ of the figure is $\beta = 0.1251(9)$. Figures produced by the author. Results associated with a manuscript in preparation.
    }
    \label{tc_steps}
\end{figure}

\subsection{Numerical analysis of the fractal dimension $d_{R}$}

\paragraph{} Seeking to obtain numerically the value of the fractal dimension $d_R$, we resort to the computational \textit{box counting} method, developing a parallel-programming version of the method. \textbf{We describe the method in detail, as well as its application to the \textit{spin} lattices, in the appendix of this document}. Figure \ref{dr_sigma0} shows the general behavior of the $d_R (T)$ curve in the temperature interval $0.98T_{c}$ to $1.02T_{c}$.

\begin{figure}[H]
    \centering
    \includegraphics[width=0.85\linewidth]{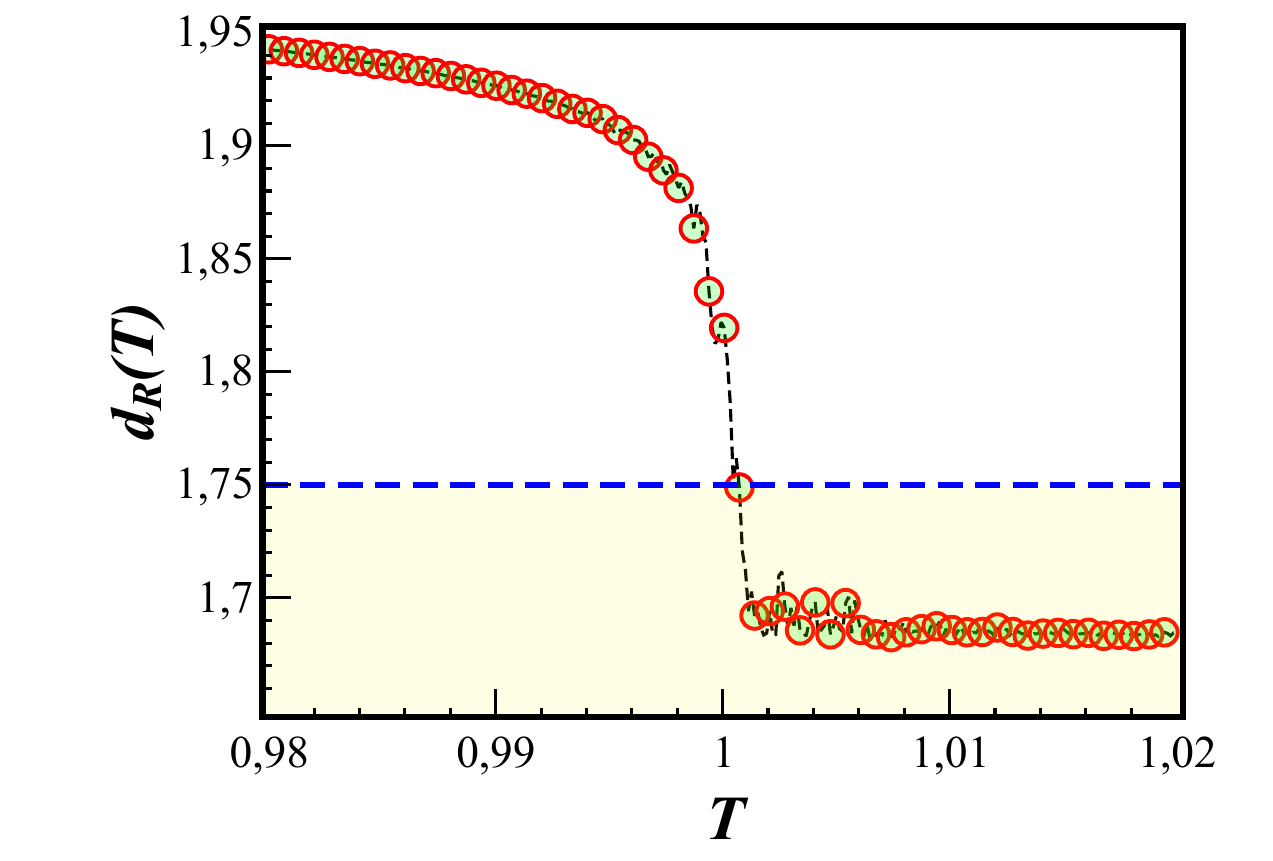}
    \caption[Behavior of the fractal dimension $d_R(T)$ as a function of temperature $T$.]{Behavior of the fractal dimension $d_R(T)$ as a function of the reduced temperature $T$. The horizontal dashed line at $d_R=1.75$ highlights the characteristic value associated with criticality, while the curve evidences the transition between two distinct regimes near $T=T_c$.}
    \label{dr_sigma0}
\end{figure}

The $d_{R}(T)$ curve exhibited, near criticality, a divergence similar to what we expect for the order parameter; therefore, to ensure sufficient resolution around the critical temperature, we performed simulations with 20 temperature points in the interval $0.995T_c$ to $1.005T_c$. Simulations were carried out in this interval with lattice sizes $L = 2048$ and $L = 5120$, using for the fit of the points the function

\begin{equation}
    d_R(T) = 1.75 + a\,(T-T_c)+b\,(T-T_c)^{2}.
    \label{dr_fit_curve}
\end{equation}
The use of \ref{dr_fit_curve} to fit the points is justified because we want to analyze only points around the value $d_{R} = 1.75$ in order to determine the critical temperature. As a result of the fit, we obtained the values $T_c=0.99996 \pm 0.00025$ for $L=2048$ and $T_c=1.00007 \pm 0.0003$ for $L=5120$, values consistent with the expected theoretical result $T_{c} = 1$. 

\begin{figure}[H]
    \centering
    \includegraphics[width=0.9\linewidth]{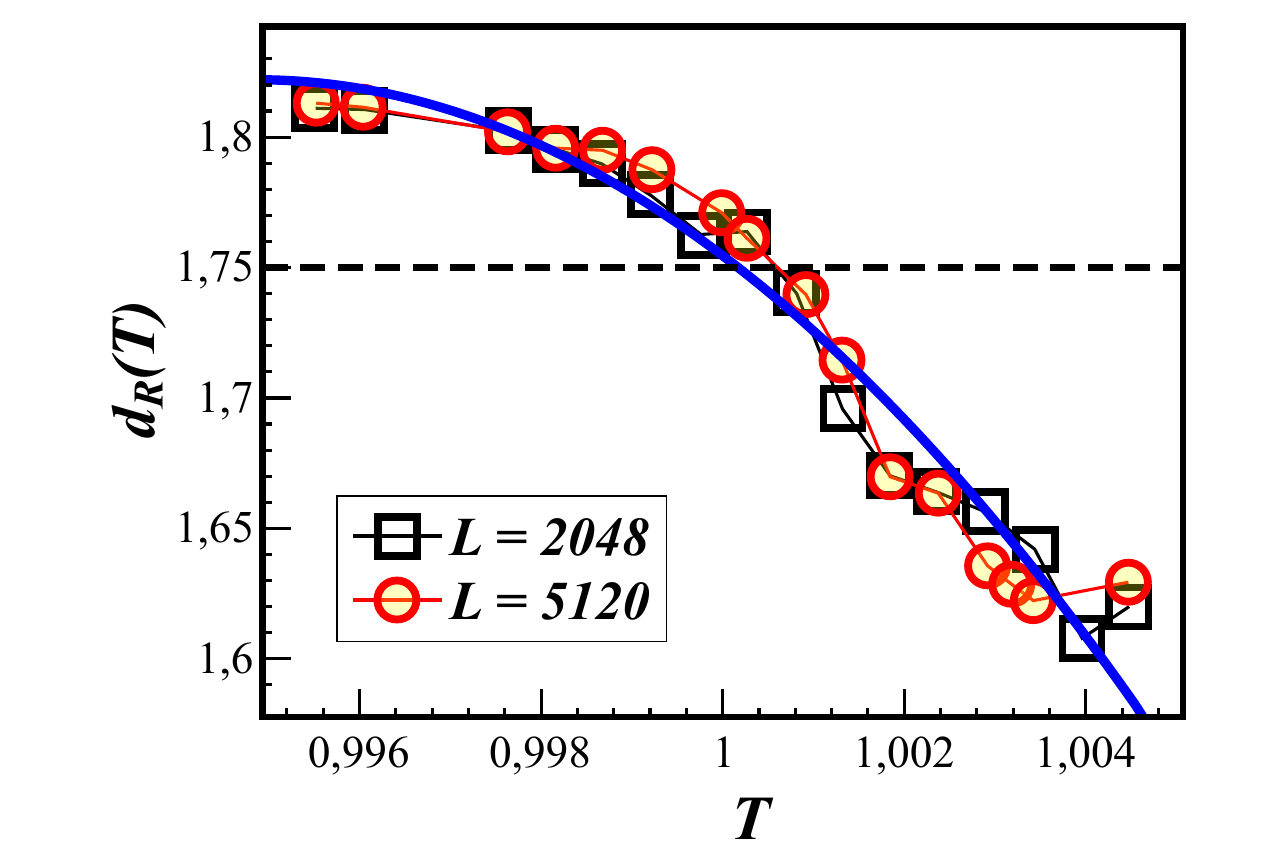}
    \caption[Dynamic fractal properties of the two-dimensional Ising model.]{Dynamic fractal properties of the two-dimensional Ising model. The fractal dimension $d_R(T)$ is presented as a function of the reduced temperature $T/T_O$ for a square lattice with sizes $L=2048$ (black squares) and $L=5120$ (red circles). The horizontal black dashed line indicates the theoretical value $d_R=7/4=1.75$ near the critical point $T=T_c$. The continuous blue curve represents the least-squares fit to the data, described by $f(T)=1.75+a\,(T-T_c)+b\,(T-T_c)^2$, where $a$ and $b$ are fitting parameters.}
    \label{dr_curve1}
\end{figure}

Figure \ref{dr_curve1} presents the simulation results for sizes $L = 2048$ and $L = 5120$, the fit made by function \ref{dr_fit_curve} to the $L = 5120$ points, and the black dashed line referring to the region $d_R=7/4=1.75$. The computational results are consistent with what is expected from our analytical values. The value found for $T_{c} = 1.00007(30)$ reveals fractal dynamics at criticality that explains the geometric deviation characterized by the exponent $\eta$.\\ 

This work therefore allows us to conclude that criticality can be interpreted more deeply from a geometric perspective, in which the dynamics associated with the correlations of the system begins to occur in a fractal subspace. The main contribution of the article is to show that the Fisher exponent can be understood through the relation $ \eta = d - d_R$, which directly connects critical behavior to the effective fractal dimension of the correlations. From this viewpoint, the exponent $\eta$ ceases to be merely a device required to correct classical theory and instead admits a clear geometric interpretation associated with the reduction of the space effectively accessible to the dynamics of the system at criticality. In the case of the two-dimensional Ising model, this proposal leads to the result $d_f=7/4$, in agreement with the exact value $\eta=1/4$, which reinforces the consistency of the central hypothesis of the work. In addition, the apparent failure of the fluctuation-dissipation theorem at criticality should not be understood merely as a formal limitation of the usual theory, but as an indication that the relevant geometry of the system ceases to be entirely Euclidean. In this way, our work not only offers a new interpretation for the origin of critical exponents, but also points to a broader conceptual structure capable of connecting phase transitions, fractal dimensions, and dynamical phenomena. Finally, the results obtained open perspectives for applying this approach in other contexts, such as disordered systems, dynamical transitions, and growth models, suggesting that fractal geometry may play an even more general role in the description of critical systems.\\

\chapter{Fractal dynamics in structures of non-integer dimension}

\paragraph{} In this chapter, we address the phase transition when the spin lattice itself is non-integer. That is, we keep the interactions among spins as in the Ising model, but the lattice dimension $d$ may take non-integer values in the interval $1 \leq d \leq 4$. In this case, we keep the same definition for the Riesz Laplacian, Eq. (\ref{G3}), and obtain the main analytical results~\cite{Lima25} of the previous chapter, such as Eqs. (\ref{eta}) and (\ref{dl2}).
Therefore, we will compare our results with values obtained in the literature using other methods.\\

 Le Guillou and Zinn-Justin obtain the critical exponents for the Ising model in non-integer dimension \(d\) using \(\phi^{4}\) field theory via the renormalization group. Starting from perturbative expansions in \(\varepsilon = 4-d\) for the exponents of interest, for example, \(\nu\), \(\eta\), and \(\gamma\), standard resummation techniques for asymptotic series are employed, in particular the Borel--Leroy scheme combined with conformal mapping, with a scan over auxiliary resummation parameters to ensure numerical stability~\cite{Guillou85,Guillou87}. In addition, a transformation in the parameter \(\varepsilon\) is introduced to improve the effective analytical behavior of the series, and the procedure is anchored by imposing the exact values of the Ising model at \(d=2\), which increases the accuracy of the estimates at intermediate dimensions. In this way, reliable values of the critical exponents are obtained as continuous functions of \(d\), allowing interpolation between integer and fractional dimensions.\\

\begin{table}[htbp]
\centering
\label{tab:ising_noninteger_exponents}
\begin{tabular}{c|cccc}
\hline
\hline
\(d\) & \(\gamma\) & \(\nu\) & \(\beta\) & \(\eta\) \\
\hline

4      & 1                    & 1/2                 & 1/2                 & 0 \\
3.75   & \(1.046060 \pm 30\)  & \(0.523405 \pm 15\) & \(0.458355 \pm 20\) & \(0.001435 \pm 10\) \\
3.5    & \(1.10055  \pm 35\)  & \(0.55215  \pm 20\) & \(0.41600  \pm 20\) & \(0.00685  \pm 20\) \\
3.25   & \(1.1642   \pm 15\)  & \(0.5873   \pm 10\) & \(0.3723   \pm 10\) & \(0.0180   \pm 10\) \\
3      & \(1.2390   \pm 25\)  & \(0.6310   \pm 15\) & \(0.3270   \pm 15\) & \(0.0375   \pm 25\) \\
2.75   & \(1.328    \pm 5\)   & \(0.686    \pm 3\)  & \(0.2800   \pm 30\) & \(0.067    \pm 6\)  \\
2.5    & \(1.436    \pm 8\)   & \(0.758    \pm 5\)  & \(0.2305   \pm 30\) & \(0.11     \pm 1\)  \\
2.25   & \(1.571    \pm 10\)  & \(0.857    \pm 6\)  & \(0.1790   \pm 25\) & \(0.17     \pm 1\)  \\
2      & 7/4                 & 1                   & 1/8               & 1/4 \\
1.875  & \(1.862    \pm 15\)  & \(1.10     \pm 1\)  & \(0.097    \pm 3\)  & \(0.30     \pm 3\)  \\
1.75   & \(1.99     \pm 4\)   & \(1.23     \pm 3\)  & \(0.068    \pm 6\)  & \(0.35     \pm 5\)  \\
1.65   & \(2.11     \pm 8\)   & \(1.37     \pm 7\)  & \(0.045    \pm 10\) & \(0.40     \pm 10\) \\
1.5    & \(2.35     \pm 20\)  & \(1.65     \pm 20\) & \(0.010    \pm 15\) & \(0.50     \pm 15\) \\
1.375  & \(2.6      \pm 4\)   & \(2.1      \pm 5\)  & \(-0.02    \pm 3\)  & \(0.55     \pm 25\) \\
1.25   & \(3.0      \pm 10\)  & \(3.0      \pm 15\) & \(-0.05    \pm 5\)  & \(0.65     \pm 35\) \\
1      & \(\infty\) & \(\infty\) & 0 & 1 \\
\hline
\hline
\end{tabular}
\caption[Critical exponents of the Ising universality class in non-integer dimension $1\leq d \leq 4$.]{Critical exponents of the Ising universality class in non-integer dimension $1\leq d \leq 4$. Values taken from \cite{isingnoninteger}.}
\label{tablenoninteger}
\end{table}

Table \ref{tablenoninteger} presents values for the critical exponents of the Ising universality class obtained by the methods described above. We wish to verify the validity of the results obtained from equations \ref{dl} and \ref{dl2} for the values presented in the previous table. Immediately, we can use the values of the exponents $\beta$ and $\nu$ in table \ref{tablenoninteger} and apply them to equation \ref{dl} to find values of $d_f$ as a function of the dimension $d$. The values of $d_f$, with the associated errors, are described in table \ref{tabeladlxd} as a function of $d$.\\

\begin{table}[ht]
\centering
\begin{tabular}{c|ccc}
\hline\hline
$d$ & $\nu$ & $\beta$ & $d_f = d-\beta/\nu$ \\
\hline
4     & 1/2           & 1/2            & 3 \\
3.75  & 0.523405(15)  & 0.458355(20)   & 2.87428(5) \\
3.5   & 0.55215(20)   & 0.41600(20)    & 2.7466(5) \\
3.25  & 0.5873(10)    & 0.3723(10)     & 2.6161(20) \\
3     & 0.6310(15)    & 0.3270(15)     & 2.4818(27) \\
2.75  & 0.686(3)      & 0.2800(30)     & 2.342(5) \\
2.5   & 0.758(5)      & 0.2305(30)     & 2.196(4) \\
2.25  & 0.857(6)      & 0.1790(25)     & 2.041(3) \\
2     & 1             & 1/8          & 15/8 \\
1.875 & 1.10(1)       & 0.097(3)       & 1.7868(28) \\
1.75  & 1.23(3)       & 0.068(6)       & 1.695(5) \\
1.65  & 1.37(7)       & 0.045(10)      & 1.617(7) \\
1.5   & 1.65(20)      & 0.010(15)      & 1.494(9) \\
1.375 & 2.1(5)        & -0.02(3)       & 1.385(14) \\
1.25  & 3.0(15)       & -0.05(5)       & 1.267(19) \\
1     & $\infty$      & 0              & 1 \\
\hline\hline
\end{tabular}
\caption{
Values of the critical exponents $\nu$ and $\beta$ as a function of $d$ and the corresponding fractal dimension $d_f$, calculated from the relation $d_f = d - \beta/\nu$.
}
\label{tabeladlxd}
\end{table}

For $d = 2, 4$, the values presented in table \ref{tabeladlxd} coincide with the exact values from the work of \cite{Coniglio89}. Using equation \ref{eta} together with the values of $d_f$ defined in table \ref{tabeladlxd}, we calculate values for the exponent $\eta$ and can compare them with those presented in \cite{isingnoninteger} for the same exponent. Table \ref{tabelaetas} contains the values of $\eta$ calculated from equation \ref{eta} for a series of values of the dimensions $d$ and $d_R$.

\newpage

\begin{table}[H]
\centering
\begin{tabular}{cccc}
\hline\hline
$d$ & $d_R = 2(d_f-1)$ & $\eta = d-d_R$ & $\eta^{*}$ \\
\hline
4     & 4          & 0          & 0 \\
3.75  & 3.7486(1)  & 0.0014(1)  & 0.001435(1) \\
3.5   & 3.493(1)   & 0.007(1)   & 0.00685(2) \\
3.25  & 3.232(4)   & 0.018(4)   & 0.0180(1) \\
3     & 2.964(5)   & 0.036(5)   & 0.0375(3) \\
2.75  & 2.68(1)    & 0.07(1)    & 0.067(6) \\
2.5   & 2.392(8)   & 0.108(8)   & 0.11(1) \\
2.25  & 2.082(6)   & 0.168(6)   & 0.17(1) \\
2     & 7/4       & 1/4       & $1/4$ \\
1.875 & 1.574(6)   & 0.301(6)   & 0.30(3) \\
1.75  & 1.39(1)    & 0.36(1)    & 0.35(5) \\
1.65  & 1.23(1)    & 0.42(1)    & 0.40(10) \\
1.5   & 0.99(2)    & 0.51(2)    & 0.50(15) \\
1.375 & 0.77(3)    & 0.60(3)    & 0.55(25) \\
1.25  & 0.53(4)    & 0.72(4)    & 0.65(35) \\
1     & 0          & 1          & 1 \\
\hline\hline
\end{tabular}
\caption[Values of the anomalous exponent $\eta$, obtained from equation \ref{eta}, as a function of the spatial dimension $d$ and the parameter $d_R$, obtained from the relation $d_R = 2(d_f-1)$.]{
Values of the anomalous exponent $\eta$, obtained from equation \ref{eta}, as a function of the spatial dimension $d$ and the parameter $d_R$, obtained from the relation $d_R = 2(d_f-1)$.
The $\eta^{*}$ column presents the reference values taken from \cite{isingnoninteger}.
}
\label{tabelaetas}
\end{table}

\begin{figure}[H]
    \centering
    \includegraphics[width=0.85\linewidth]{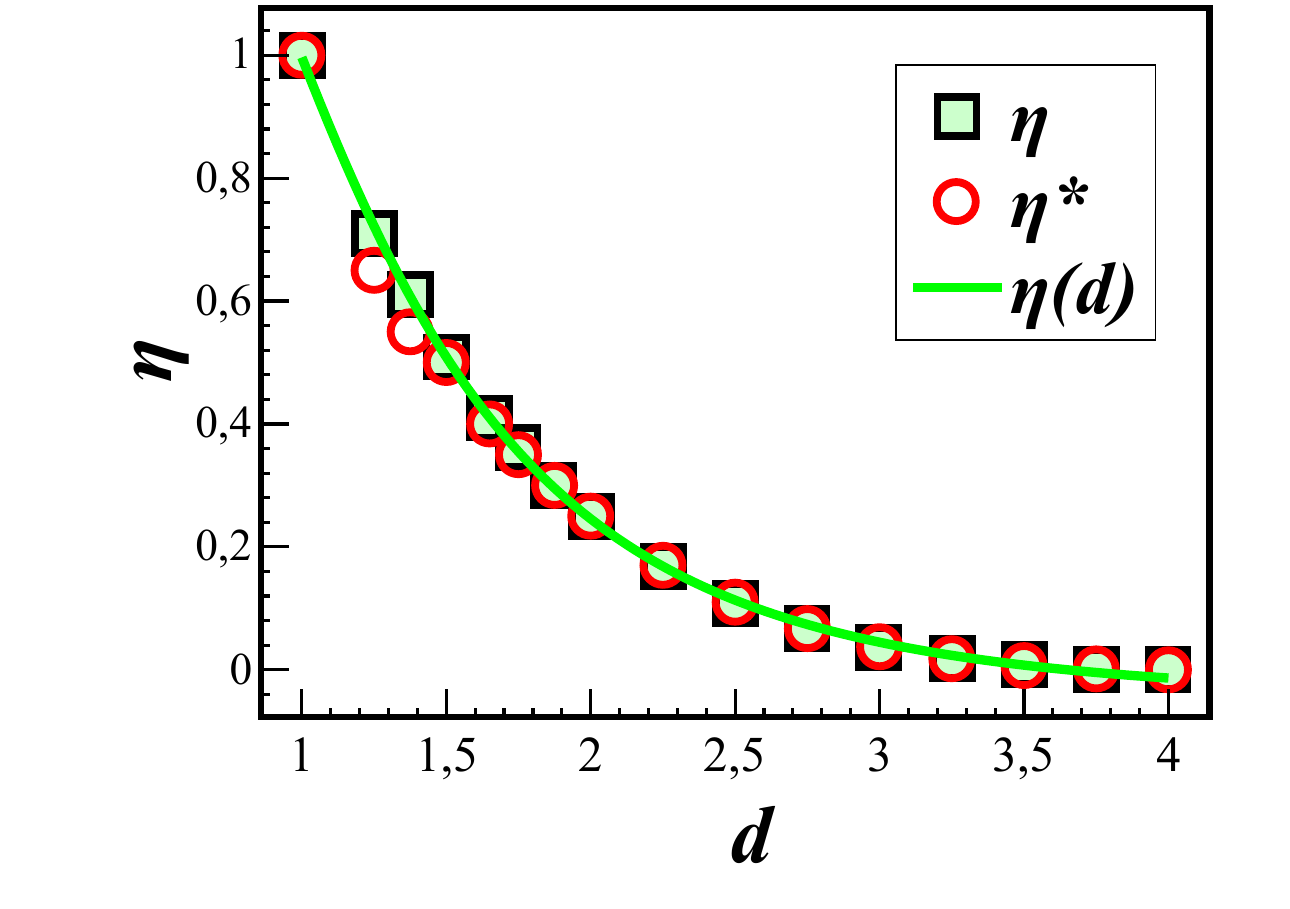}
    \caption[Anomalous exponent $\eta$ as a function of the dimension $d$.]{$\eta$ as a function of $d$. $\eta^{*}$ represents the values extracted from the literature and $\eta$ are values calculated from equation \ref{eta} for dimensions in the interval $1 \leq d \leq 4$. All values are described in table \ref{tabelaetas}, while the fit $\eta(d)$, represented by the green curve, represents the function $\eta(d)$ proposed by the equation. Figure adapted from \cite{Lima25}.}
    \label{eta_fig}
\end{figure}

\textcolor{black}{We now propose expressions for the critical exponents as functions of dimension, based on three guiding principles: first, they should be as simple as possible; second, they should reproduce the known exponents exactly for $d=1,2$ and $4$, which are captured by the term outside the brackets below, for example $\beta(d)=f(d)[...]$, we suggest $f(d)=\frac{d-1}{a+bd}$, and fitting $\beta(2)=1/8$ and $\beta(4)=1/2$ we obtain $a=10$ and $b=-1$; and, finally, they should fit the data in table~\ref{tabeladlxd}, for which we introduce a small correction inside the brackets. Thus}
\begin{equation}
	\label{curve_beta_fit}
	\beta(d) = \frac{d - 1}{10 - d}\left[1 - C_1 (4 - d)(2 - d)\right],
\end{equation}
and
\begin{equation}
	\label{curve_nu_fit}
	\nu(d) = \frac{6 - d}{4(d - 1)}\left[1 - C_2 (4 - d)(2 - d)\right],
\end{equation}
the constants $C_1 = 149/1000$ and $C_2 = 37/1000$ are determined by a least-squares fit in the interval $1.5 \leq d \leq 4$. The region $d < 1.5$ is disregarded in this procedure because of the significantly lower accuracy of the data, as already discussed above. The results found for the exponent $\eta$ in table \ref{tabelaetas} have higher accuracy than the reference values $\eta^{*}$ extracted from \cite{isingnoninteger}, revealing the validity and usefulness of equation \ref{eta} for calculating the anomalous exponent also for non-integer dimensions. From table \ref{tabelaetas} we fit the function

\begin{equation}
\label{curve_eta_d}
    \eta(d)=\frac{4-d}{5d-2} \left[1-\frac{1772(d-1)(d-2)}{625(5d-2)} \right],
\end{equation}
using exact values of $\eta$ for integer dimensions $1,2,4$ and approximate values for $d=3$. The fit represented by equation~\ref{curve_eta_d} and the comparison between the values of $\eta$ and $\eta^{*}$ are presented in figure \ref{eta_fig}.

\begin{figure}[H]
    \centering
    \includegraphics[width=0.75\linewidth]{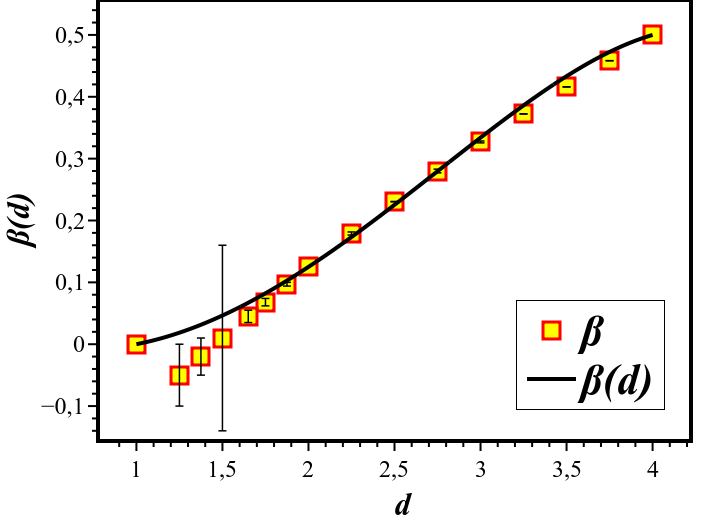}
    \caption[Critical exponent $\beta$ as a function of the dimension $d$. Curve for non-integer values of $d$.]{Critical exponent $\beta$ as a function of the dimension $d$. The fitting curve was obtained from equation \ref{curve_beta_fit}, using theCritical exponent $\beta$ as a function of the dimension $d$.s data points listed in table~\ref{tabeladlxd} in the interval $1.5<d<4$. It should be noted that the fitted curve does not incorporate the uncertainties associated with negative values of $\beta$. Figure taken from \cite{Carrasco26}.}
    \label{non_beta(d)}
\end{figure}

\begin{figure}[H]
    \centering
    \includegraphics[width=0.75\linewidth]{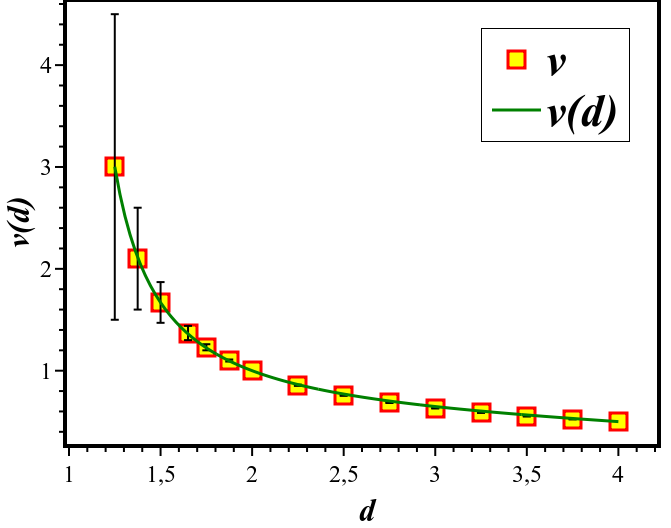}
    \caption[Critical exponent $\nu$ as a function of the dimension $d$. Curve for non-integer values of $d$.]{Critical exponent $\nu$ as a function of the dimension $d$. The fitting curve was obtained from equation \ref{curve_nu_fit}, using the data points listed in table~\ref{tabeladlxd} in the interval $1.5 \leq d \leq 4$. Figure taken from \cite{Carrasco26}.}
    \label{non_nu(d)}
\end{figure}

Analogously, and using tables \ref{tabeladlxd} and \ref{tabelaetas}, the effective dimensions $d_R$ and $d_f$ can be fitted by
\begin{equation}
	\label{non_curve_dr}
	d_R(d) = \frac{56(d - 1)}{5d + 22}\left[1 - \frac{2152(4 - d)(d - 2)}{100000}\right],
\end{equation}
and from Eq.~(\ref{dl2}),
\begin{equation}
	\label{non_curve_df}
	d_f(d) = 1 + \frac{28(d - 1)}{5d + 22}\left[1 - \frac{2152(4 - d)(d - 2)}{100000}\right].
\end{equation}

\begin{figure}[H]
    \centering
    \includegraphics[width=0.75\linewidth]{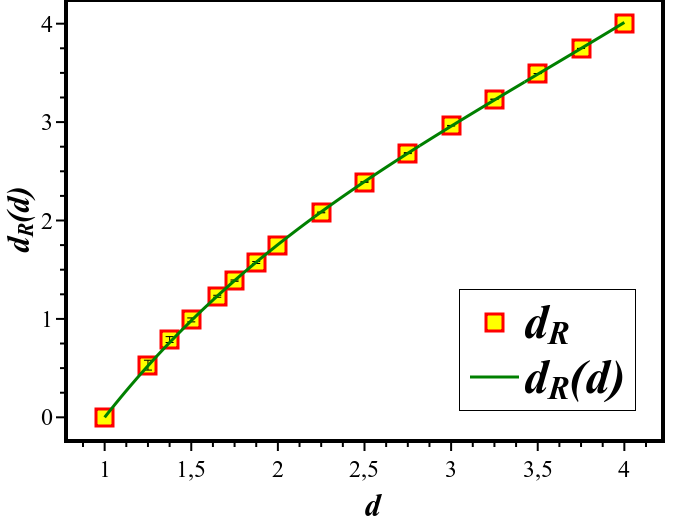}
    \caption[Fractal dimension $d_R$ as a function of the dimension $d$. Curve for non-integer values of $d$.]{Fractal dimension $d_R$ as a function of the dimension $d$. The fitting curve was obtained from equation \ref{non_curve_dr}, using the data points listed in table~\ref{tabelaetas} in the interval $1.5 \leq d \leq 4$. Figure taken from \cite{Carrasco26}.}
    \label{non_dr(d)_fig}
\end{figure}

\begin{figure}[H]
    \centering
    \includegraphics[width=0.75\linewidth]{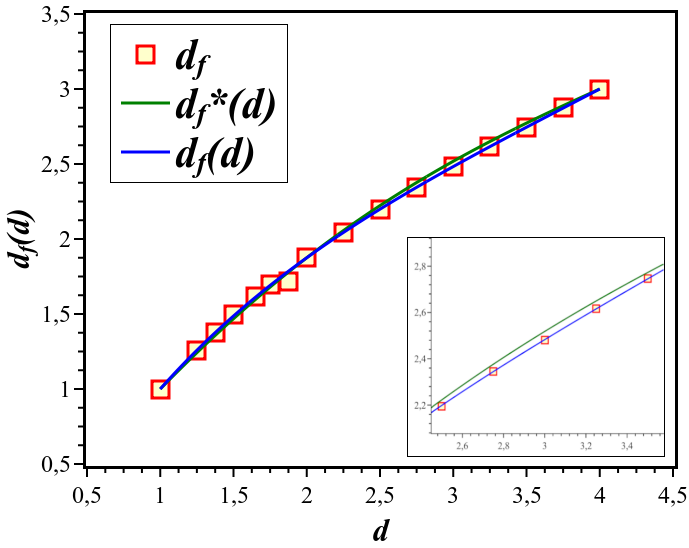}
    \caption[Fractal dimension $d_f$ as a function of the dimension $d$. Curve for non-integer values of $d$.]{Fractal dimension $d_f$ as a function of the dimension $d$. The fitting curve was obtained from equation \ref{non_curve_df}, using the data points listed in table~\ref{tabeladlxd} in the interval $1.5 \leq d \leq 4$. Figure taken from \cite{Carrasco26}.}
    \label{non_df(d)_fig}
\end{figure}

Analyzing scaling relations, represented by equations \ref{alfa_eq_1} and \ref{gamma_eq_1} for the exponents $\alpha$ and $\gamma$, together with the Rushbrooke equation \ref{rushbroke_1}, we analyze how these exponents and relations behave for non-integer dimensions.

\begin{table}[h!]
\centering
\begin{tabular}{c|cccc}
\hline\hline
$d$     & $\alpha$       & $\gamma$    & $\Sigma^*$  & $\Sigma$ \\ \hline
2     & 0          & 1.75     & 2    &  2  \\
1.875 & -0.0625(1) & 1.862(1) & 1.993(7) & 2.00(3)\\
1.75  & -0.15(5)  & 1.99(4)  & 1.97(5) & 2.00(2) \\
1.65  & -0.3(1)      & 2.11(8)  & 1.93(6) & 2.00(3) \\
1.5   & -0.5(3)  & 2.35(2)  & 1.89(4) & 2.0(3) \\
1.375 & -0.9(7)    & 2.6(4)   & 1.67(25) & 2.0(1)    \\
1.25  & -1.75(1.85)   & 3.0(1)   & 1.15(1) & 2.0(3) \\
1     & $-\infty$   & $\infty$ & $?$ & 2\\ \hline\hline
\end{tabular}

\caption[Values of the critical exponents for the Ising universality class with non-integer dimensions $d$.]{Values of the critical exponents for the Ising universality class with non-integer dimensions $d$. The values for $\beta$, $\gamma$, and $\nu$ taken from \cite{isingnoninteger} were used in the calculations of $\alpha$ and $\Sigma^{*}$, while $\Sigma$ is calculated using \ref{new_rushbrooke2}.}
\label{tabelasigmas}
\end{table}

\noindent Rewriting equation \ref{dl} as

\begin{equation}
\label{beta_dl}
\beta = (d-d_f)\nu,
\end{equation}
and then substituting equations~\eqref{alfa_eq_1}, \eqref{beta_dl}, and \eqref{gamma_eq_1} into the Rushbrooke relation \(\alpha+2\beta+\gamma\), we obtain
\begin{align}
\alpha + 2\beta + \gamma
&= (2-d\nu) + 2(d-d_f)\nu + (2-\eta)\nu \nonumber\\
&= 2 + \bigl[-d + 2(d-d_f) + (2-\eta)\bigr]\nu \nonumber\\
&= 2 + (d + 2 - \eta - 2d_f)\nu, 
\end{align}
thus, the Rushbrooke identity is recovered when the term proportional to $\nu$ vanishes, that is, when we have
\begin{equation}
    (d + 2 - \eta - 2d_f)\nu = 0,
    \label{new_rushbrooke}
\end{equation}
\textcolor{black}{obtaining}

\begin{equation}
\label{new_rushbrooke2}
    \Sigma = \alpha + 2\beta + \gamma = 2.
\end{equation}
From equation~\ref{new_rushbrooke2}, the combination \(\Sigma=\alpha+2\beta+\gamma\) can be estimated in terms of quantities directly associated with the geometry of the problem, $d$ and $d_f$, in addition to the anomalous exponent $\eta$. The comparison between $\Sigma^*=\alpha+2\beta+\gamma$ and the values obtained via equation~\ref{new_rushbrooke2} is presented in table~\ref{tabelasigmas}. Thus, we observe that the previous relations directly imply the condition
\begin{equation}
d + 2 - \eta - 2d_f = 0,
\end{equation}
which, when substituted into equation \ref{new_rushbrooke2}, guarantees the immediate recovery of the Rushbrooke identity. In physical terms, this equivalence shows that the set of additional relations obtained from differential fractal geometry is not merely redundant; it acts as a consistency criterion between the critical exponents and the geometric quantities of the system, connecting the fractal dimensions $d_R$, $d_f$, and the anomalous exponent \(\eta\) in a manner compatible with the usual scaling relations. Consequently, even in situations in which some standard scaling relation may be violated or show deviations (for example, due to effects of effective dimensionality, disorder, finite-size scaling corrections, or limitations of the numerically accessible asymptotic regime), this geometric formalism provides a robust framework for constraining the exponents and producing mutually consistent estimates, preserving the overall consistency structure expected for the set \(\{\alpha,\beta,\gamma,\nu,\eta,d_f\}\).\\

In summary, the results discussed in this chapter reinforce that the introduction of fractal geometry concepts provides a solid and natural extension of the formalism of critical phenomena beyond integer-dimensional lattices, making it possible to reinterpret scaling relations in terms of known quantities. In particular, the relation connecting the anomalous exponent \(\eta\) to the difference between the Euclidean dimension \(d\) and a fractal dimension associated with the critical structure (such as \(d_f\) or, equivalently in the adopted formalism, \(d_R\)) establishes a direct link between long-range correlations and the morphology of the critical set at the critical temperature. The internal consistency of the method can be tested by simultaneously verifying the derived identities (for example, those imposing \(d + 2 - \eta - 2d_f = 0\)), which provides an additional robustness criterion even when usual relations suffer practical limitations due to finite-size corrections, restricted asymptotic regimes, or system-specific features. Thus, by expressing critical observables through geometric parameters, a promising route is opened for investigating phase transitions in non-integer effective dimensions and, more broadly, for extending these ideas to scenarios in which heterogeneities, disorder, or nonequilibrium dynamics play a central role, topics that naturally motivate the next stages of this work.

\chapter{Fractal dynamics in disordered systems}

\paragraph{} Disordered systems constitute a natural and, at the same time, challenging class of problems in critical phenomena, because they incorporate microscopic heterogeneities that are unavoidable in real materials, such as impurities, vacancies, local stresses, and composition variations. Magnetic materials, such as ferromagnetic alloys (for example, Fe--Co or Ni--Cu), doped semiconductors, and even complex oxides inevitably exhibit impurities, crystal defects, and local variations in the interactions among their constituents. In systems such as spin glasses, this disorder is not merely a detail, but the central element governing the collective behavior of the system. In the context of the Ising model, this disorder can be represented by randomness in the couplings \(J_{ij}\) (bond disorder) or by local fields \(h_i\) (random fields), making the system spatially nonhomogeneous and introducing a competition between thermal fluctuations and ``frozen'' disorder fluctuations (\emph{quenched}). As a consequence, properties near criticality can be profoundly modified: the critical temperature may shift, amplitudes and scaling corrections may intensify and, in certain cases, even the universality class may change. Thus, the study of disorder offers not only a more realistic approximation to experimental materials, but also a theoretical laboratory for understanding how robust scaling relations and universality emerge (or fail) in the presence of intrinsic heterogeneities. In our work, we want to verify the validity of our new relations for systems that exhibit structural impurities; in the case of the Ising model, we will work with controlled levels of disorder in the couplings of the \textit{spins}.\\

A very general way to describe this type of disorder in the magnetic systems considered in this work is through the Hamiltonian

\begin{equation}
\label{ising3}
    H(s) = -\sum_{<i,j>}J_{i,j}s_{i}s_{j}-\sum_{i}h_i s_{i},
\end{equation}

\noindent where $s_i=\pm1$ represents the \textit{spins}, $i$ runs over all elements of the system, and $j$ runs over the nearest neighbors of $i$. Unlike Eq.~\eqref{ising2}, in which the coupling $J$ is assumed constant and equal to $1$, Eq.~\eqref{ising3} introduces a random coupling $J_{i,j}$ between \textit{spins} $i$ and $j$, with a value drawn from a probability distribution $P(J_{i,j})$. In this work, we consider the absence of an external field, that is, $h_i=0$, and adopt for $P(J_{i,j})$ the form

\begin{equation}
\label{dis}
    P(J_{i,j})=\frac{1}{\sqrt{2 \pi\sigma^2 }}\exp\left(-\frac{(J_{i,j}-J)^{2}}{2 \sigma^{2}}\right).
\end{equation}

\noindent The distribution $P(J_{i,j})$ gives the probability of occurrence of a given coupling value $J_{i,j}$ between neighboring spins. Because this distribution is centered at $J$, this parameter represents the mean value of the system couplings. In turn, $\sigma$ determines the width of the distribution and, therefore, quantifies the intensity of disorder: the larger $\sigma$, the greater the dispersion of the values of $J_{i,j}$ around $J$. This type of formulation is particularly suitable for describing real magnetic systems in which microscopic imperfections, compositional fluctuations, or structural disorder locally modify the intensity of exchange interactions. Situations of this nature can be found, for example, in dilute magnetic alloys, in which random substitution of lattice atoms changes the local environment of the magnetic moments, in spin glasses, characterized by competition among interactions and by the intrinsic presence of randomness, and also in amorphous materials, where the absence of long-range crystalline order produces a spatially heterogeneous distribution of the effective couplings. Thus, introducing a distribution for $J_{i,j}$ constitutes a natural way of modeling \textit{quenched} disorder, allowing us to investigate how such fluctuations influence the collective behavior and critical properties of the system. \\

\begin{figure}[h!]
    \centering
    \includegraphics[width=0.85\linewidth]{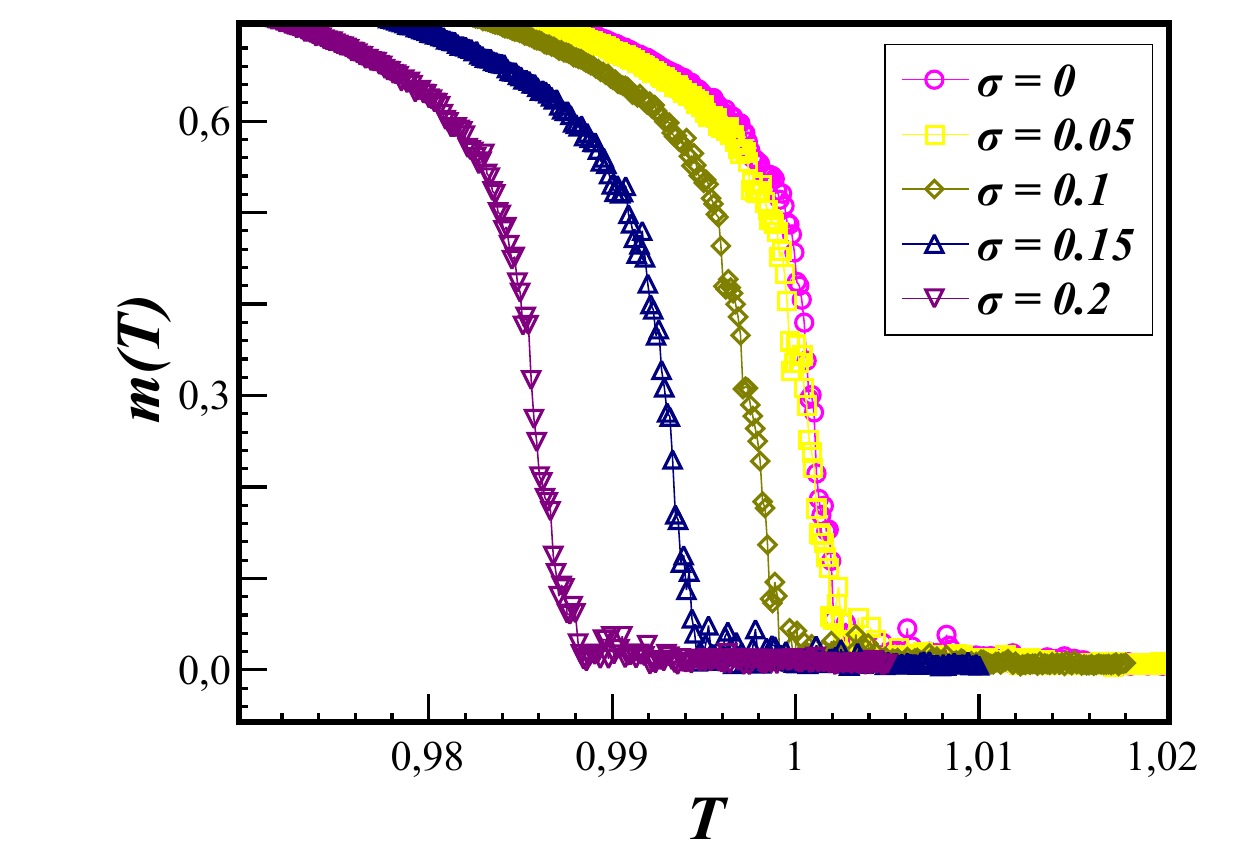}
    \caption[Order parameter $m$ as a function of temperature $T$ for different values of the standard deviation $\sigma$ of the probability distribution $P(J_{i,j})$.]{Order parameter $m$ as a function of temperature $T$ for different values of the standard deviation $\sigma$ of the probability distribution $P(J_{i,j})$. From this point onward, $T$ represents the system temperature in units of the Onsager critical temperature $T_{O} \sim 2.269J/k_{B}$. }
    \label{m_sigmas}
\end{figure}

\begin{figure}[h!]
    \centering
    \includegraphics[width=0.85\linewidth]{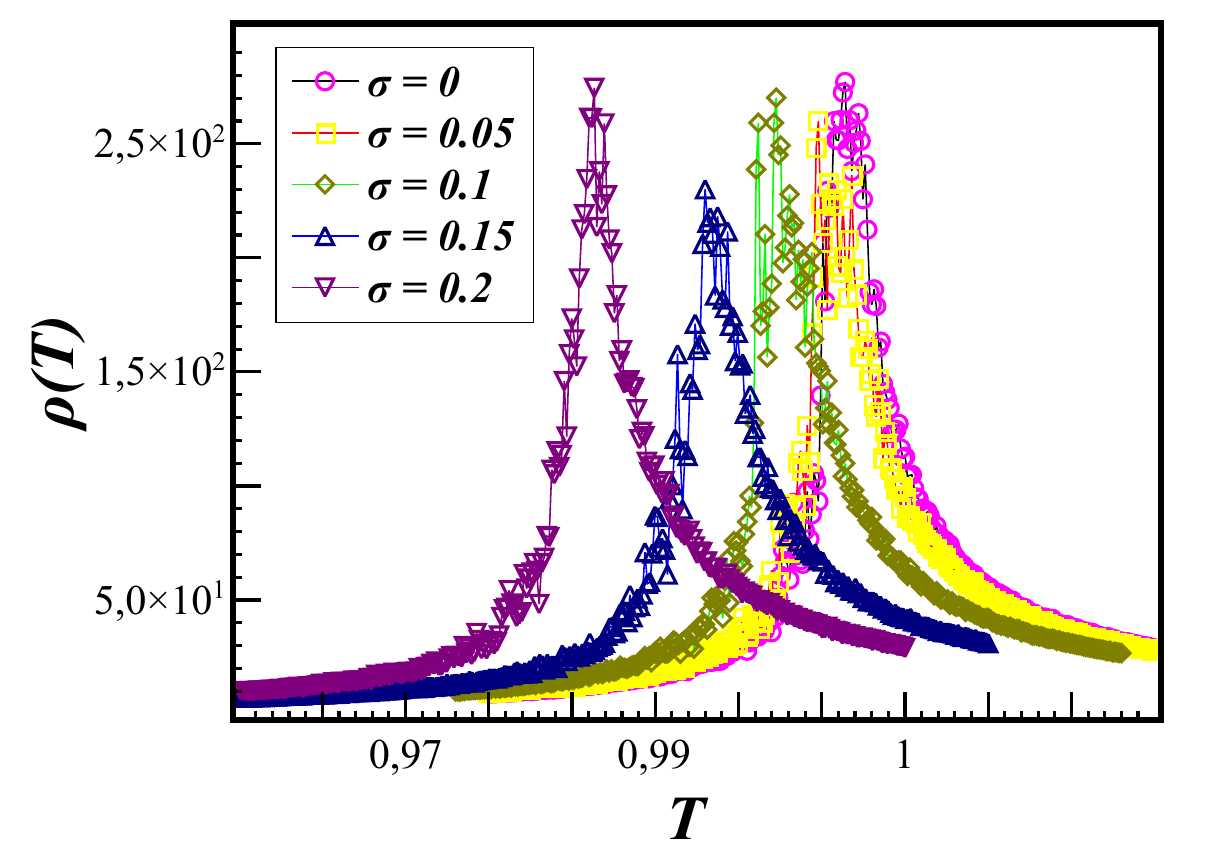}
    \caption{Correlation length $\rho$ as a function of temperature $T$ for different values of the standard deviation $\sigma$ of the probability distribution $P(J_{i,j})$. }
    \label{rho_sigmas}
\end{figure}

\begin{figure}[h!]
    \centering
    \includegraphics[width=0.9\linewidth]{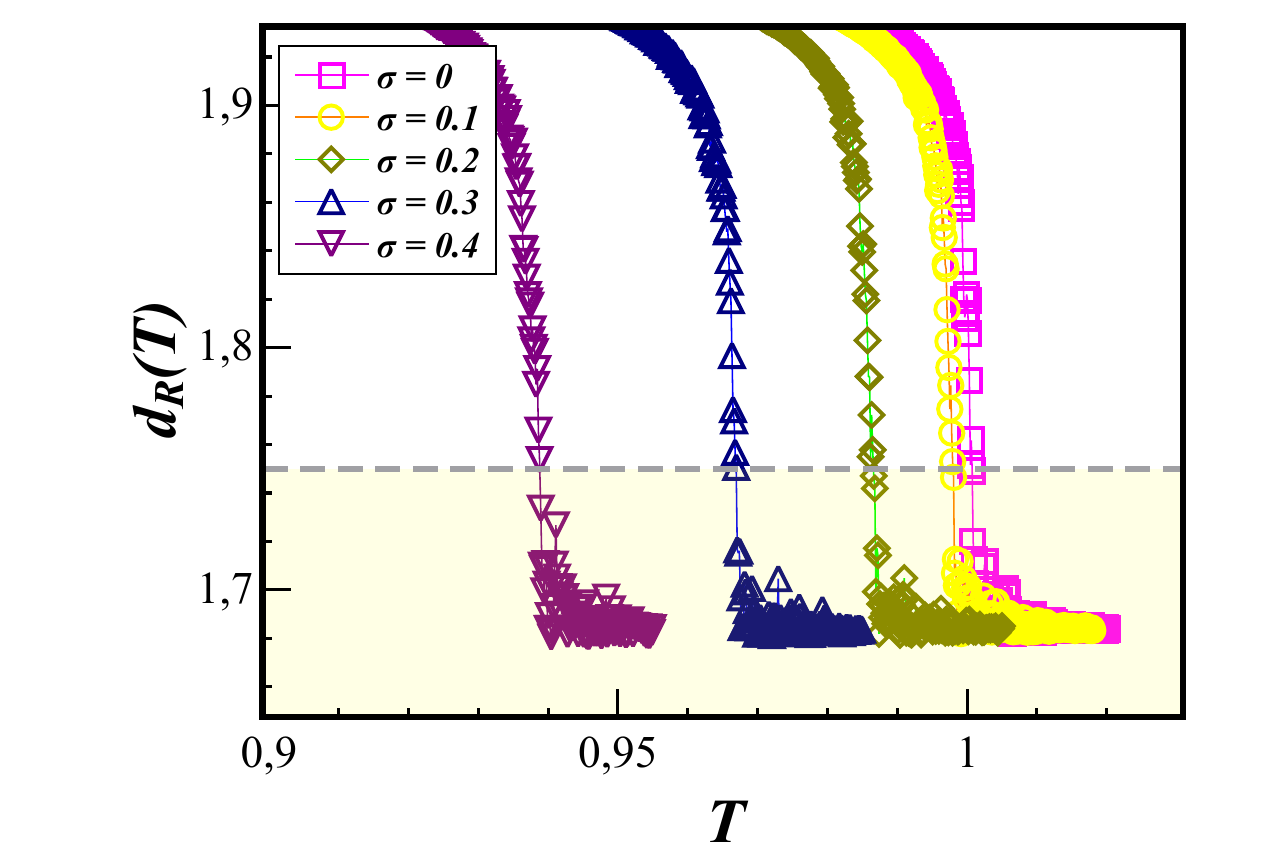}
    \caption[Fractal dimension $d_{R}$ as a function of temperature $T$ for different values of the standard deviation $\sigma$ of the probability distribution $P(J_{i,j})$.]{Fractal dimension $d_{R}$ as a function of temperature $T$ for different values of the standard deviation $\sigma$ of the probability distribution $P(J_{i,j})$. The gray dashed line marks the region for which $d_{R} = 1.75$.}
    \label{dr_sigmas}
\end{figure}

We begin our analyses by examining the behavior of the order parameter $m(T)$ under the influence of disorder for some values of $\sigma$. Figure \ref{m_sigmas} shows the tendency of the $m(T)$ curves to shift toward $T \rightarrow0$ due to the contribution of disorder applied to the \textit{spin} couplings. The first point to highlight is the direct influence of disorder in anticipating the critical point. Figure \ref{m_sigmas} shows that, by increasing the value of the dispersion $\sigma$, we anticipate the phase transition of our system, which leads to the understanding that the impurity added to the system through disorder in the couplings acts as an extra temperature, a disorder temperature. Similar behavior can be observed in figure \ref{rho_sigmas} for the correlation-length curves $\rho(T)$. This same trend is also observed in figure \ref{dr_sigmas}, which presents the fractal dimension $d_R(T)$ for different values of $\sigma$; as occurs for the order parameter and the correlation length, the curves undergo a systematic shift toward lower temperatures as disorder increases. In addition, the gray dashed line corresponding to $d_{R}=1.75$ provides a visual criterion for comparison among the different values of $\sigma$, highlighting the temperature range in which each curve reaches this same value.

\subsubsection{Simulations in disordered systems}

\paragraph{} In the present work, we consider the Hamiltonian given by equation \ref{ising3} with $h_i=0$, corresponding to the Edwards--Anderson model, defined on a square lattice of side $L$. In this configuration, the system is composed of $N=L^2$ \textit{spins}, and disorder is introduced exclusively through the distribution of the couplings $J_{i,j}$. Seeking to systematically investigate the effects of disorder, we keep the mean value of the couplings fixed at $J=1$, taking the standard deviation $\sigma$ as the only control parameter; in this way, it becomes possible to analyze how increasing disorder affects the properties of the system. In particular, the objective is to follow the behavior of the critical exponents and fractal dimensions as $\sigma$ increases. As a reference, recall that in the pure case $\sigma=0$, the two-dimensional Ising model has the exact values $\beta=1/8$ for the order-parameter exponent, $\nu=1$ for the correlation length, and $\eta=1/4$ for the Fisher exponent \cite{Cardy96, kaufman_1949}. From a geometric point of view, this same limit leads to $d_R=7/4$ for the Riesz fractal dimension \cite{Lima24} and to $d_f=15/8$ for the fractal dimension of the largest ordered \textit{cluster} \cite{Suzuki8,Kroger00}.\\

The numerical simulations were performed on lattices of lateral size $L=4096$. Due to the high computational cost associated with systems of this size, we employed \textit{CUDA} technology to execute the simulations in parallel on graphics processing units. The dynamics of the Ising model was implemented using a Monte Carlo algorithm with \textit{checkerboard} (white/black) updating, in which the lattice is decomposed into two sublattices. This procedure makes it possible to simultaneously update all \textit{spins} of one sublattice while the other remains temporarily fixed, preserving the consistency of the dynamics. Each Monte Carlo step is thus composed of the sequential update of the two sublattices. A rigorous explanation of the \textit{checkerboard} update algorithm is provided in the appendix of this document. For each temperature $T$, $300000$ parallel \textit{Monte Carlo Sweeps} were performed, which means that for each \textit{sweep}, we interact with $L^{2}$ \textit{spins}. This procedure allows us to work with large lattices and obtain a sufficiently large number of samples to calculate statistically stable averages, ensuring consistent values for the quantities analyzed.\\

\subsection{Data analysis for disordered systems}

\begin{figure}[H]
    \centering
    \includegraphics[width=0.85\linewidth]{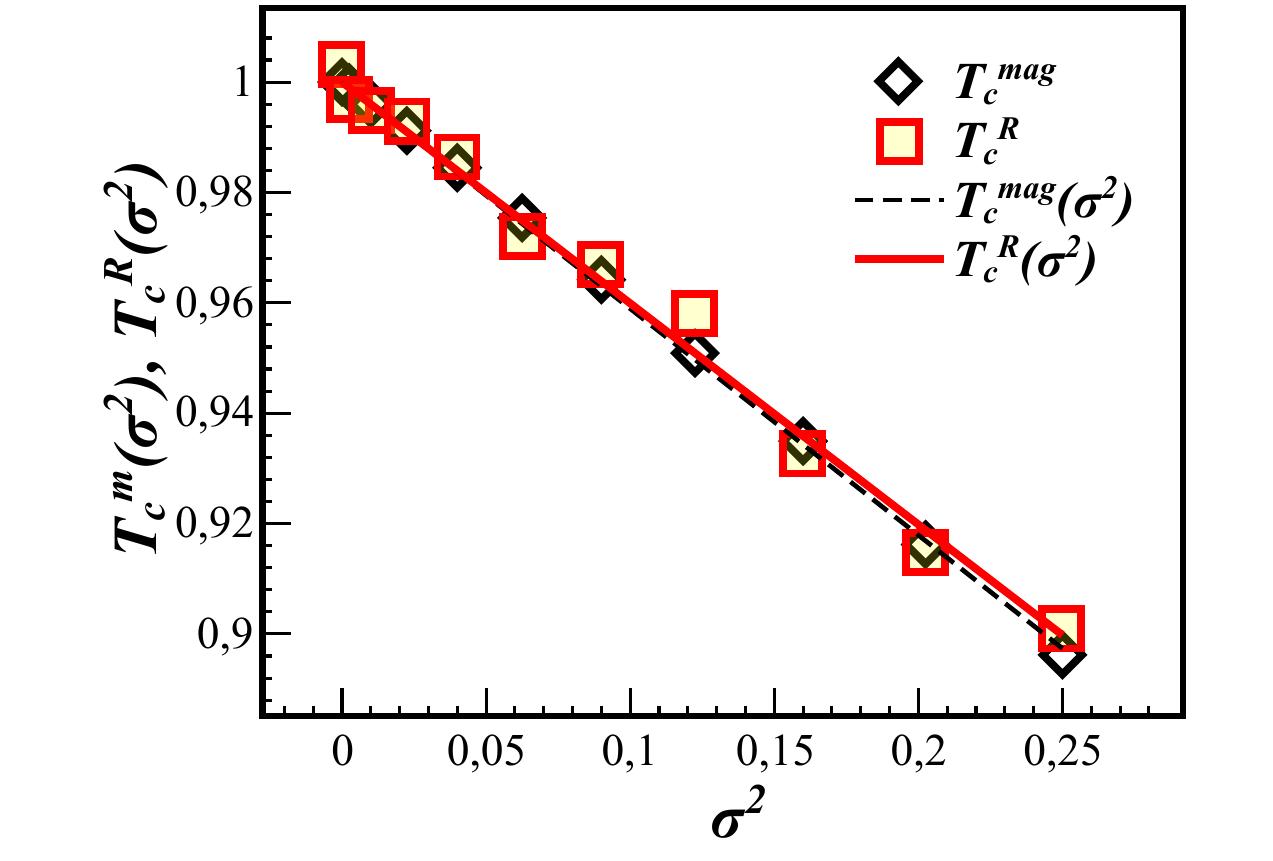}
    \caption[Normalized critical temperatures $T_c^{\mathrm{mag}}(\sigma^2)$ and $T_c^{R}(\sigma^2)$ as a function of $\sigma^2$, obtained from the magnetization and the Riesz fractal dimension.]{Normalized critical temperatures $T_c^{\mathrm{mag}}(\sigma^2)$ and $T_c^{R}(\sigma^2)$ as a function of $\sigma^2$, obtained from the magnetization and the Riesz fractal dimension. The lines represent the linear fits for each data set, highlighting the decrease of $T_c$ with increasing disorder.}
    \label{tcs_comp}
\end{figure}

\paragraph{} Throughout this chapter, we will devote ourselves to analyses of disordered systems using values of $\sigma = [0, 0.5]$ varying by $0.05$. The first, and perhaps most significant, observation to be made is how the critical temperature $T_{c}$ behaves with changes in the parameter $\sigma$. To do this, we will analyze critical temperatures extracted by two different methods and compare them; First, values of $T_{c}$ obtained from fitting the order-parameter curve $m(T)$, which behaves as equation \ref{m(T)_fit}; Second, values of $T_{c}$ obtained through a fit of the form

\begin{equation}
    d_R(T)=d_R + C_1(T-T_c^{d_{R}})+C_2(T-T_c^{d_{R}})^{2}.
\label{dr_fit}
\end{equation}
in the $d_R(T)$ curves over a range of points where $T\sim T_{c}$. We will call $T_{c}^{mag}$ and $T_{c}^{R}$ the critical temperatures obtained by the first and second methods, respectively. The comparison is useful because it allows us to verify whether there are substantial differences between the critical temperatures found by these two methods, which could indicate some inconsistency in the analyses or in the methodology employed. Table \ref{tabelatccomp} presents the values found for the critical temperatures $T_{c}^{mag}$ and $T_{c}^{R}$ for different values of the probability-dispersion parameter $\sigma$. Figure \ref{tcs_comp} presents the values of $T_{c}^{mag}$ and $T_{c}^{R}$ as a function of $\sigma^{2}$, which provides a linear behavior for fitting the points, performed using a function of the form $T_{c}(\sigma^2)=1+a  \cdot\sigma^2$. The values found for the variable $a$ from the fits of $T_{c}^{mag}(\sigma^{2})$ and $T_{c}^{R}(\sigma^{2})$ are $-0.410(5)$ and $-0.400(9)$, respectively.\\

We chose $T_{c}^{mag}$ as the critical temperature to be used in the subsequent analyses, for example, for extracting the critical exponents $\beta$, $\nu$, and $\gamma$. Using the values found for the critical temperature as a function of $\sigma$, we will reproduce the same steps shown in figure \ref{tc_steps} to extract the values of the critical exponents.

\begin{table}[H]
\centering
\begin{tabular}{c|c c}
\hline\hline
$\sigma$ & $T_c^{\mathrm{mag}}$ & $T_c^{R}$ \\
\hline
0.00 & 1.0000(2)   & 1.0031(2)   \\
0.05 & 0.99901(8)  & 0.997(1)    \\
0.10 & 0.99607(3)  & 0.995(1)    \\
0.15 & 0.9912(2)   & 0.9928(1)   \\
0.20 & 0.98448(3)  & 0.98652(3)  \\
0.25 & 0.9754(7)   & 0.972(2)    \\
0.30 & 0.96419(3)  & 0.9668(1)   \\
0.35 & 0.95088(4)  & 0.958(2)    \\
0.40 & 0.93494(11) & 0.9327(7)   \\
0.45 & 0.9162(3)   & 0.9148(4)   \\
0.50 & 0.8962(2)   & 0.90093(5)  \\
\hline\hline
\end{tabular}
\caption{Normalized critical temperatures obtained from analyses of the magnetization and the fractal dimension $d_R$ for different values of the standard deviation $\sigma$.}
\label{tabelatccomp}
\end{table}

The critical exponent $\gamma$ was determined from the magnetic susceptibility $\chi(T)$, which in the numerical code employed in this work is calculated from magnetization fluctuations according to the relation
\begin{equation}
    \chi(T) = \frac{\langle m^2 \rangle - \langle m \rangle^2}{N\,T},
\end{equation}
where $N$ represents the total number of lattice sites, $T$ the system temperature, and $\langle m \rangle$ and $\langle m^2 \rangle$ correspond, respectively, to the averages of the magnetization and of the square of the magnetization obtained over the numerical realizations. This definition is directly associated with the fluctuational character of the susceptibility, since this quantity measures the intensity of fluctuations of the order parameter and, therefore, constitutes an observable particularly sensitive to the approach to the critical point. In the vicinity of criticality, the function $\chi(T)$ behaves as

\begin{equation}
    \chi(T) \sim |T-T_c|^{-\gamma},
\end{equation}
a function that we use to perform the corresponding fits and extract the exponent $\gamma$. With the curve determined, we perform the same procedure described in figures \ref{mag_fit_1} and \ref{rho_adjusted} to calculate the exponent $\gamma$. The values of the exponents $\beta$, $\nu$, $\gamma$, and of the ratio $\beta/\nu$ are presented as a function of $\sigma$ in table \ref{tabelaexpoentes}.\\

\begin{table}[H]
\centering
\begin{tabular}{c|c c c c}
\hline\hline
$\sigma$ & $\beta$ & $\nu$ & $\beta/\nu$ & $\gamma$ \\
\hline
0.00 & 0.1253(6)  & 1.000(5)   & 0.1252(8)  & 1.753(65) \\
0.05 & 0.1234(5)  & 0.985(13)  & 0.1253(17) & 1.699(34) \\
0.10 & 0.1213(7)  & 0.965(15)  & 0.1257(21) & 1.71(12)  \\
0.15 & 0.1197(8)  & 0.956(20)  & 0.1252(28) & 1.673(76) \\
0.20 & 0.125(1)   & 0.995(85)  & 0.1255(15) & 1.75(11)  \\
0.25 & 0.1197(10) & 0.952(14)  & 0.1257(22) & 1.66(17)  \\
0.30 & 0.119(1)   & 0.955(13)  & 0.1248(20) & 1.82(21)  \\
0.35 & 0.1197(14) & 0.9591(96) & 0.1248(19) & 1.71(11)  \\
0.40 & 0.114(1)   & 0.898(24)  & 0.1272(38) & 1.66(24)  \\
0.45 & 0.1176(4)  & 0.9601(59) & 0.1226(9)  & 1.83(15)  \\
0.50 & 0.1246(4)  & 0.9002(69) & 0.1385(12) & 1.66(15)  \\
\hline\hline
\end{tabular}
\caption{Values of the critical exponents $\beta$, $\nu$, $\gamma$, and of the ratio $\beta/\nu$ for different values of the standard deviation $\sigma$.}
\label{tabelaexpoentes}
\end{table}

\begin{figure}[H]
    \centering
    \includegraphics[width=0.85\linewidth]{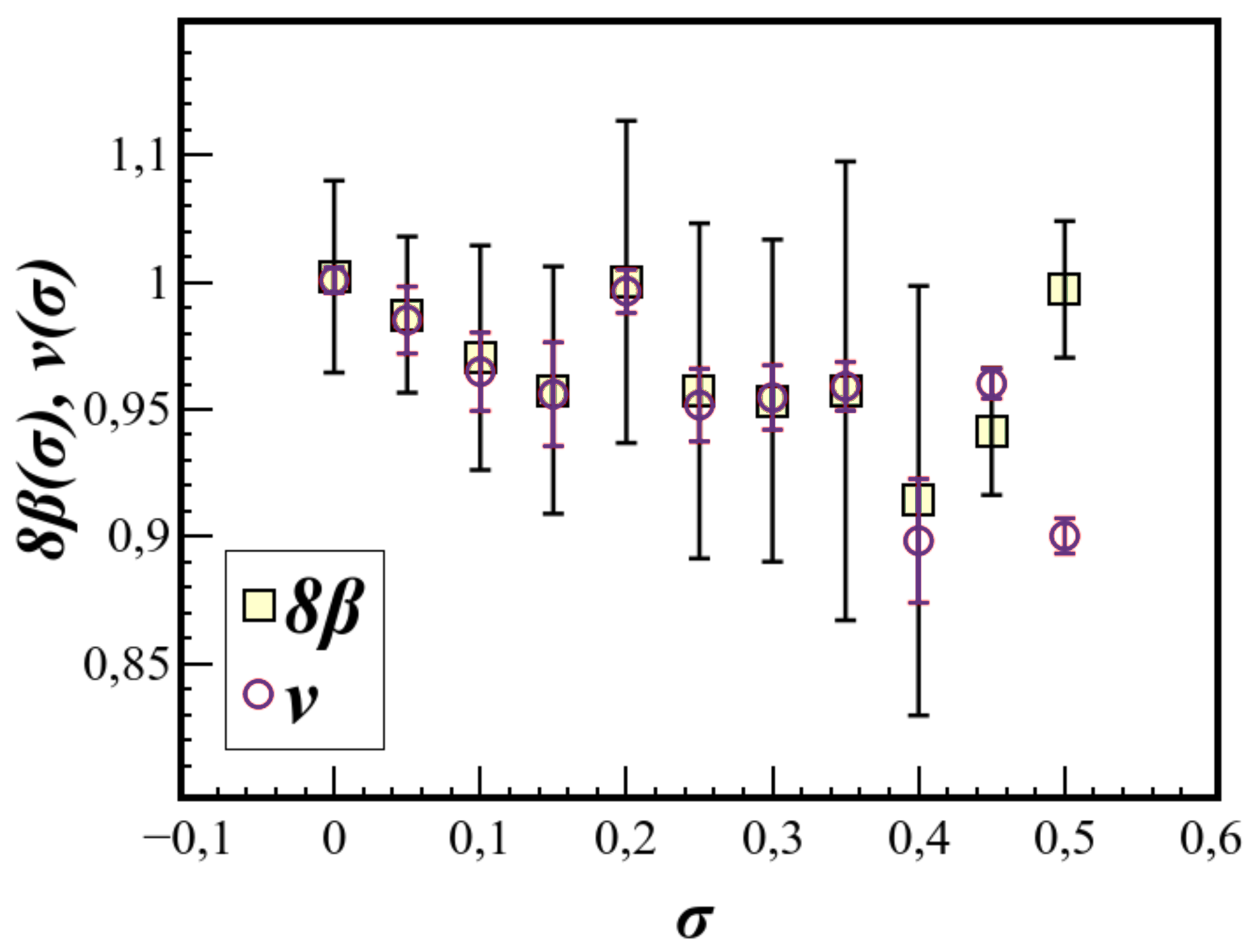}
    \caption[Evolution of the critical exponents $\beta$ and $\nu$ as a function of disorder.]{Evolution of the critical exponents $\beta$ and $\nu$ as a function of disorder. The exponents $\nu$ and $8\beta$ are presented as a function of the standard deviation $\sigma$. The values of $\beta$, as well as their respective errors, were multiplied by $8$ in order to allow both exponents to be represented on the same scale.}
    \label{beta_e_nu}
\end{figure}

\begin{figure}[H]
    \centering
    \includegraphics[width=0.85\linewidth]{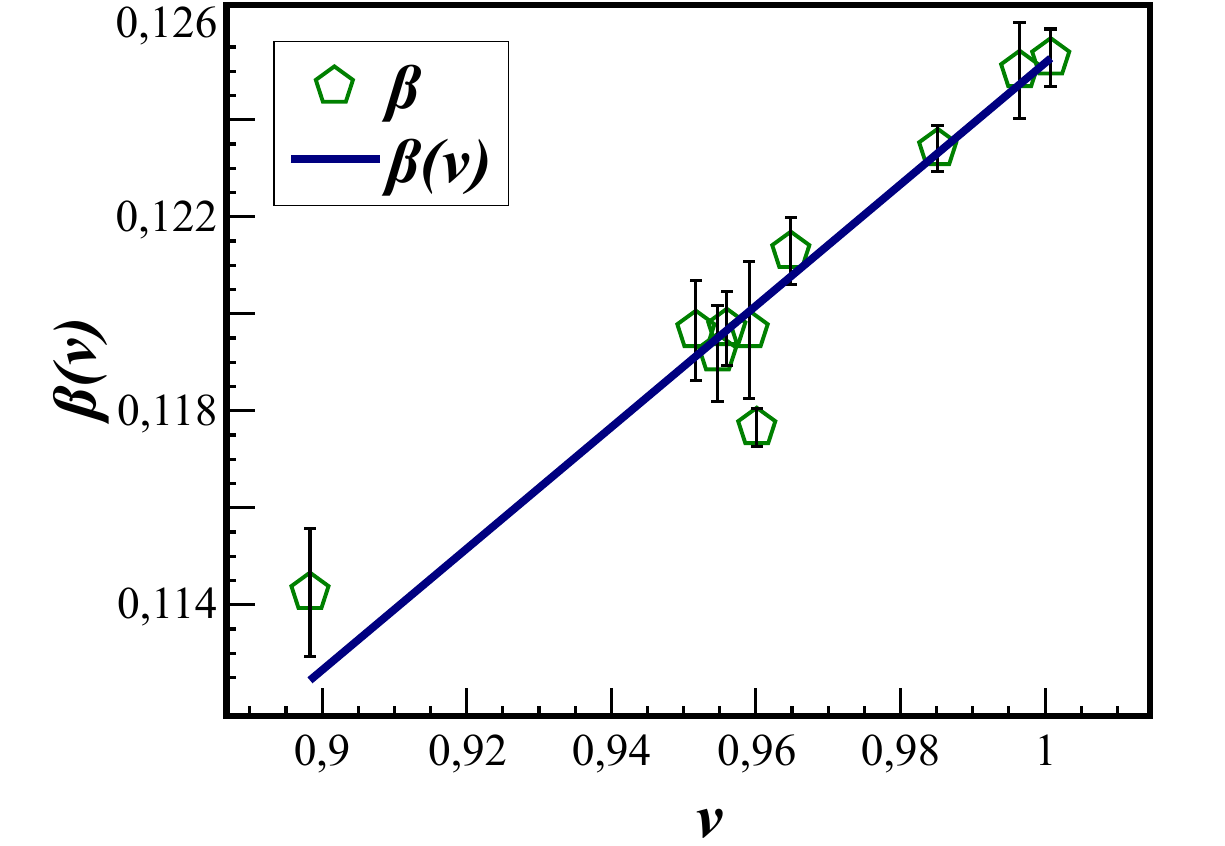}
    \caption[Critical exponent $\beta$ as a function of $\nu$.]{Critical exponent $\beta$ as a function of $\nu$. The blue line indicates the linear relation $\beta = a\cdot\nu$ between these parameters, which remains valid also for disordered systems, with $a = 0.12517(36)$.}
    \label{beta(nu)}
\end{figure}

The column referring to the ratio $\beta/\nu$ in table \ref{tabelaexpoentes} shows that the values of this quantity fluctuate around the value $\beta/\nu = 0.125$, which is expected for the case $\sigma = 0$. This ratio is of particular interest because it is fundamental for calculating quantities such as $d_{f}$, $d_{R}$, and $\eta$. The results show the interesting result of a relation between the exponents $\beta$ and $\nu$: \textbf{Despite the individual change of the exponents as a function of disorder $\sigma$, the ratio $\beta/\nu$ remains constant, fluctuating within an error margin, for small and intermediate values of impurity in the system} \cite{reis96}. Figure \ref{beta_e_nu} presents the values of the exponents $\beta$ and $\nu$ on the same scale; for this purpose, we multiply $\beta$ and its errors by $8$, obtaining a curve of the form $8\beta(\sigma)$. A linear relation of the form $\beta = a\cdot \nu$ was used as a fit and presented in figure \ref{beta(nu)}, with the result $a = 0.12517(36)$, which shows that there is a strong tendency for the exponents to maintain the ratio $\beta/\nu$ close to the value $0.125$.\\

Equations \ref{dl} and \ref{dl2} are immediate relations of the ratio $\beta/\nu$, which allows us to easily find the values for the fractal dimensions $d_f$ and $d_R$. The values calculated from these relations and the associated errors are shown in table \ref{tabeladfs}. 

\begin{table}[h!]
    \centering
\begin{tabular}{c|cc}
\hline\hline
$\sigma$ & $d_R$ & $d_f$ \\
\hline
0    & 1.750(7)   & 1.875(4) \\
0.05 & 1.750(2)   & 1.875(1) \\
0.1  & 1.750(2)   & 1.875(1) \\
0.15 & 1.7486(18) & 1.8743(9) \\
0.2  & 1.750(4)   & 1.875(2) \\
0.25 & 1.752(2)   & 1.876(1) \\
0.3  & 1.7496(18) & 1.8748(9) \\
0.35 & 1.748(2)   & 1.874(1) \\
0.4  & 1.748(4)   & 1.874(2) \\
0.45 & 1.752(2)   & 1.876(1) \\
0.5  & 1.744(4)   & 1.872(2) \\
\hline\hline
\end{tabular}
\caption{  Fractal dimensions $d_{R}$ and $d_{f}$ as a function of $\sigma$. The values found fluctuate around $d_R=7/4$ and $d_f=15/8$.}
\label{tabeladfs}
\end{table}

\begin{figure}[H]
    \centering
    \includegraphics[width=0.85\linewidth]{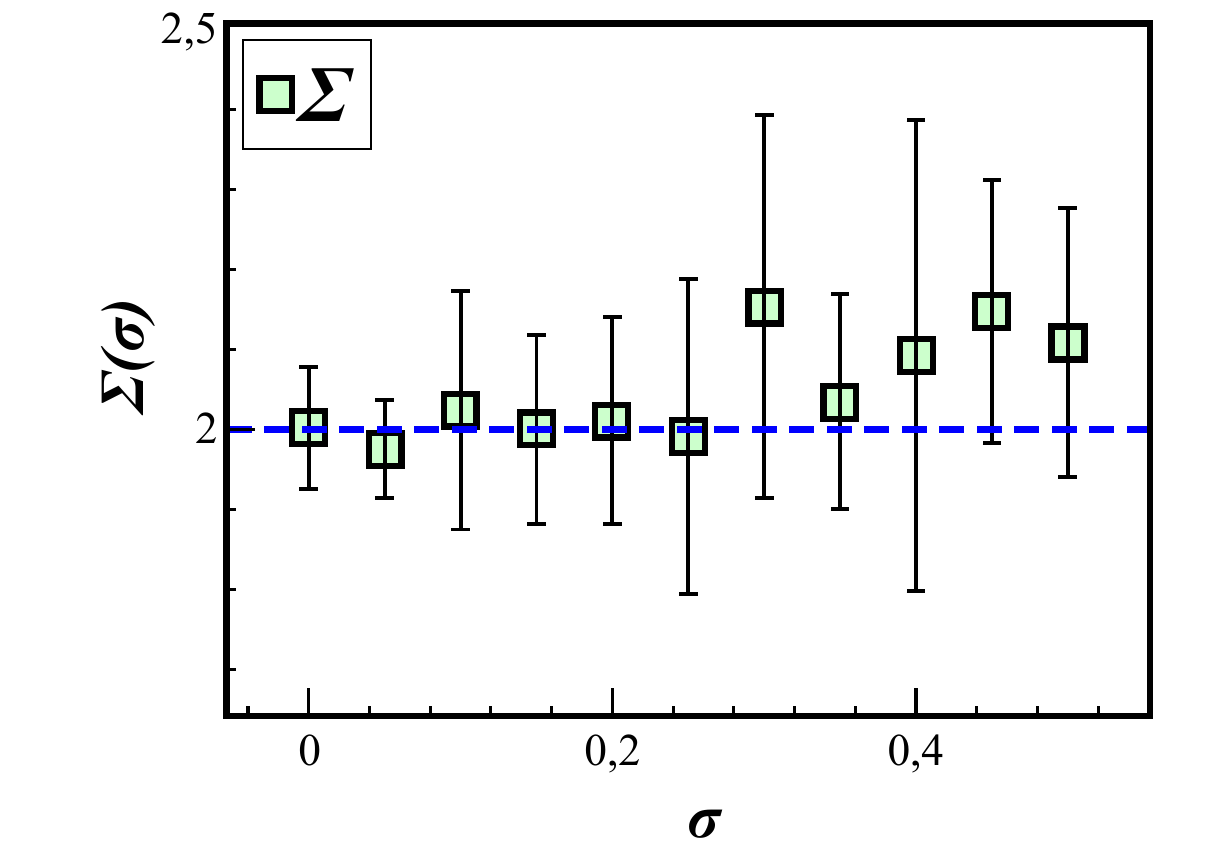}
    \caption[Behavior of the quantity $\Sigma$ as a function of the standard deviation $\sigma$.]{Behavior of the quantity $\Sigma$ as a function of the standard deviation $\sigma$. The horizontal dashed line at $\Sigma=2$ indicates the value predicted by the Rushbrooke relation for $\sigma = 0$, while the symbols represent the values obtained numerically for the disordered system. It can be observed that, within the error bars, the results remain compatible with this value throughout the entire range of disorder analyzed.}
    \label{figurarushbrooke}
\end{figure}

Once the main critical exponents of the disordered system have been determined, we now turn to the analysis of the Rushbrooke relation, given by equation \ref{rushbroke_1}, with the objective of verifying to what extent this thermodynamic identity remains valid in the presence of disorder. The Rushbrooke relation is particularly interesting in this context, because it connects critical exponents associated with different thermodynamic observables, serving as an important consistency criterion for the description of the critical behavior of systems. The exponent $\alpha$, calculated from the specific-heat curve $C_v(T)$, can also be obtained from the relation given by equation \ref{alfa_eq_1} as a direct function of the exponent $\nu$. As in \ref{new_rushbrooke2}, we will define the quantity $\Sigma$, which will be expressed by

\begin{equation}
    \Sigma = \alpha+2\beta+\gamma,
\label{sigma_eq}
\end{equation}
and then we will find the values of $\Sigma$ as a function of disorder $\sigma$. Table \ref{tabelarushbrooke} presents the values of the exponents $\alpha$, $2\beta$, $\gamma$, and $\Sigma$ as a function of disorder $\sigma$. The behavior of the points $\Sigma(\sigma)$ is presented in figure \ref{figurarushbrooke}, where it can be seen that, considering the values of the associated errors, all points touch the value $\Sigma = 2$. 

\begin{table}[H]
\centering
\begin{tabular}{c|c c c c}
\hline\hline
$\sigma$ & $\alpha$ & $2\beta$ & $\gamma$ & $\Sigma$ \\
\hline
0.00 & -0.0016(99) & 0.2506(12) & 1.753(65) & 2.002(76) \\
0.05 & 0.030(26)   & 0.2468(10) & 1.699(34) & 1.975(62) \\
0.10 & 0.070(31)   & 0.2426(14) & 1.71(12)  & 2.02(15)  \\
0.15 & 0.088(41)   & 0.2394(16) & 1.673(76) & 2.00(12)  \\
0.20 & 0.007(17)   & 0.250(2)   & 1.75(11)  & 2.01(13)  \\
0.25 & 0.097(29)   & 0.2394(20) & 1.66(17)  & 1.99(20)  \\
0.30 & 0.091(25)   & 0.238(2)   & 1.82(21)  & 2.15(24)  \\
0.35 & 0.082(19)   & 0.2394(28) & 1.71(11)  & 2.03(13)  \\
0.40 & 0.203(49)   & 0.228(2)   & 1.66(24)  & 2.09(29)  \\
0.45 & 0.080(12)   & 0.2352(8)  & 1.83(15)  & 2.15(16)  \\
0.50 & 0.200(14)   & 0.2492(8)  & 1.66(15)  & 2.11(17)  \\
\hline\hline
\end{tabular}
\caption{Values of $\alpha$, $2\beta$, $\gamma$, and $\Sigma$ for different values of the standard deviation $\sigma$.}
\label{tabelarushbrooke}
\end{table}

In general, the results presented in this chapter indicate that introducing disorder into the couplings $J_{ij}$ of the Ising model does not destroy the fundamental structure of critical behavior, but promotes systematic modifications in the parameters that characterize it. We observe that increasing the parameter $\sigma$ leads to a shift of the critical temperature toward lower values (a kind of disorder temperature), both when estimated from the order parameter and through the fractal dimension $d_R$, highlighting the role of disorder as an agent that effectively anticipates the phase transition. In addition, the analysis of the critical exponents revealed that, although they vary as a function of $\sigma$, their scaling relations remain, to a large extent, preserved within the uncertainties. As observed in table \ref{tabelaexpoentes}, for example, the change in the values of the exponents $\beta$ and $\nu$ did not produce substantial changes in the relation $\beta/\nu$, which translates into a particularly interesting result for the geometric quantities $d_f$ and $d_R$. From a geometric point of view, the results obtained for the dimension $d_R$ show that this quantity continues to be an informative tool for characterizing the critical system, even when the interactions are perturbed by disorder. The persistence of well-defined behaviors for $d_R(T)$, as well as its consistency with the expected values at criticality, reinforces the idea that describing the critical regime in terms of effective dimensions remains relevant even for disordered systems. In particular, we find that combinations such as those involved in the Rushbrooke relation remain compatible with the expected values, suggesting that the global structure of criticality theory retains its validity even in the presence of disorder that is not too pronounced.\\

Thus, by introducing disorder into the couplings of the Ising model, we not only move closer to more realistic conditions, but also put to the test the central hypothesis of this work, namely that critical behavior can be understood as an effective dynamics restricted to a space of fractal nature, even when the microscopic structure of the system contains imperfections.

\chapter{Beyond the Ising model}

In the previous chapters, we presented a geometric interpretation for Fisher's anomalous exponent $\eta$, associating it with the difference between the Euclidean dimension of the system and a fractal correlation dimension, denoted here by $d_R$. The objective of this chapter is to show that this formulation is not restricted to the two-dimensional Ising model, but can be verified in other fundamental models of statistical mechanics, such as the Potts, XY, and Heisenberg models.

\section{The Potts model}

The Potts model is a natural generalization of the Ising model to more than two states \cite{Potts52, f.y.wu}. A simple Hamiltonian for this model can be written as
\begin{equation}
    H=-J\sum_{\langle i,j\rangle}\delta_{s_i,s_j}
    -h_0\sum_i\delta_{s_i,0},
\end{equation}
where the sum $\langle i,j\rangle$ is performed over pairs of nearest neighbors, $\delta$ is the Kronecker delta, $J$ is the spin--spin coupling constant, and $h_0$ is an external field. The variables $s_i$ can take the values
\begin{equation}
    s_i=0,1,2,\ldots,q-1,
\end{equation}
thus defining the $q$-state Potts model.\\

This model is of great importance in statistical mechanics, because it is related to different problems in network physics and critical phenomena \cite{Xu25}. For $q=1$, the model is associated with percolation and complex networks; for $q=2$, the Ising model is recovered; and for $q=4$, there is a connection with the Baxter--Wu model. In subsequent studies devoted to exploring its properties, the Potts model became established as an important test system for different methods and approaches in the study of critical-point theory~\cite{f.y.wu}. Thus, the Potts model constitutes a privileged system for testing the validity of relations between critical exponents and fractal geometry.\\

A useful representation for the local order parameter of the $q$-state Potts model is given by a vector with $(q-1)$ components. In this simplex representation, each state $\sigma_i$ is mapped onto a unit vector $\mathbf{e}_{\sigma_i}$ pointing toward one of the vertices of a simplex. Thus, the local order parameter at site $i$ can be written as
\begin{equation}
    \mathbf{m}_i=\mathbf{e}_{\sigma_i}.
\end{equation}
The global order parameter is then
\begin{equation}
    \mathbf{M}=\frac{1}{N}\sum_i \mathbf{m}_i,
\end{equation}
and the corresponding fluctuation field can be defined as
\begin{equation}
    \phi_i=\mathbf{m}_i-\langle \mathbf{M}\rangle.
\end{equation}

In Table \ref{tab:potts_fractal}, we present the values of the critical exponents and fractal dimensions for the universality class of the two-dimensional Potts model, for $q=0,1,2,3$, and $4$. The values of $d_f$ are obtained from relation \ref{dl}, while $d_R$ is determined by \ref{dl2} and $\eta$ is calculated from \ref{eta}. These values agree with the results obtained by Coniglio~\cite{Coniglio89}.

\begin{table}[h!]
\centering
\begin{tabular}{c|ccccc}
\hline\hline
$q$ & $0$ & $1$ & $2$ & $3$ & $4$ \\
\hline
$\beta$ & $1/6$ & $5/36$ & $1/8$ & $1/9$ & $1/12$ \\
$\nu$ & $\infty$ & $4/3$ & $1$ & $5/6$ & $2/3$ \\
$d_f$ & $2$ & $91/48$ & $15/8$ & $28/15$ & $15/8$ \\
$d_R$ & $2$ & $43/24$ & $7/4$ & $26/15$ & $7/4$ \\
$\eta$ & $0$ & $5/24$ & $1/4$ & $4/15$ & $1/4$ \\
$\zeta$ & $1$ & $19/24$ & $3/4$ & $11/15$ & $3/4$ \\
\hline\hline
\end{tabular}
\caption{Values of the critical exponents and fractal dimensions for the universality class of the two-dimensional Potts model. The dimension $d_f$ is obtained from the relation $d_f=d-\beta/\nu$, the dimension $d_R$ from $d_R=2(d_f-1)$, and the exponent $\eta$ from $\eta=d-d_R$.}
\label{tab:potts_fractal}
\end{table}

The results in Table \ref{tab:potts_fractal} show that the formulation is satisfied for all values of $q$ considered. From the exponents $\beta$ and $\nu$, we determine $d_f$, comparing it with the results presented in \cite{Coniglio89, f.y.wu}, then $d_R$, $\zeta$, and finally $\eta$. The values of $\eta$ obtained in this way coincide with the exact values known for the universality class of the two-dimensional Potts model.\\

A particularly important case is $q=2$, which corresponds to the two-dimensional Ising model. In this case,
\begin{equation}
    d_f=\frac{15}{8},
\end{equation}
and, therefore,
\begin{equation}
    d_R=2\left(\frac{15}{8}-1\right)=\frac{7}{4}.
\end{equation}
Consequently,
\begin{equation}
    \eta=2-\frac{7}{4}=\frac{1}{4},
\end{equation}
in agreement with the exact values known for the two-dimensional Ising model, already presented in this work.\\

Another interesting point occurs for $q=4$. Although the values of $\beta$ and $\nu$ are different from those found for $q=2$, the ratio $\beta/\nu$ is the same. Thus, the Suzuki relation provides the same value of $d_f$; consequently, the same value of $d_R$ and the same value of $\eta$ are obtained. Therefore, the equality between the values of $\eta$ for $q=2$ and $q=4$ should not be interpreted as a coincidence, but as an imposition of fractal geometry, since both cases have the same values of $d_f$ and $d_R$. This result reinforces the validity of relation \ref{eta} both for determining and for physically interpreting the Fisher exponent.

\section{The XY and Heisenberg models}

\paragraph{} Following a natural order of complexity, after the Potts model, we can investigate models with continuous symmetry, such as the XY and Heisenberg models. Unlike the Potts model, the classical XY and Heisenberg models have dynamical degrees of freedom given by unit vectors with two or three components. A general Hamiltonian for these systems can be written as
\begin{equation}
    H=-J\sum_{\langle i,j\rangle}
    \left(
    \Delta S_i^zS_j^z+S_i^xS_j^x+S_i^yS_j^y
    \right),
\end{equation}
where $J$ defines the energy scale and $\Delta$ is an anisotropy parameter. The XY model is recovered by taking $\Delta=0$, while the isotropic Heisenberg model corresponds to $\Delta=1$ \cite{Kosterlitz74, PhysRevB.43.6087}.\\

In Table \ref{tab:ising_xy}, we present the values of $d_f$ and $d_R$ for the three-dimensional Ising model and for the XY model in two and three dimensions, together with their respective critical exponents. The notation $\eta^\ast$ represents the value of $\eta$ obtained from the literature, distinguishing it from the value $\eta$ calculated by our geometric formulation.\\

\begin{table}[h!]
\centering
\begin{tabular}{c|ccc}
\hline\hline
Model & Ising 3D & XY 2D & XY 3D \\
\hline
$\beta$ & $0.326418(2)$ & $0$ & $0.3485(2)$ \\
$\nu$ & $0.629971(4)$ & $\infty$ & $0.67155(27)$ \\
$d_f$ & $2.481856(2)$ & $2$ & $2.481(2)$ \\
$d_R$ & $2.963713(5)$ & $2$ & $2.962(5)$ \\
$\zeta$ & $0.963713(5)$ & $1$ & $0.9619(4)$ \\
$\eta^\ast$ & $0.0362978(20)$ & $1/4$ & $0.0380(4)$ \\
$\eta$ & $0.036288(5)$ & $0$ & $0.0380(5)$ \\
\hline\hline
\end{tabular}
\caption{Values of the critical exponents and fractal dimensions for the 3D Ising, 2D XY, and 3D XY models. The values of $\eta^\ast$ represent the values known in the literature, while $\eta$ is obtained from the relation $\eta=d-d_R$.}
\label{tab:ising_xy}
\end{table}

For the three-dimensional Ising model and the three-dimensional XY model, the values of $\eta$ obtained from relation \ref{eta} agree, within the error margins, with the values known from the literature. This indicates that the geometric interpretation of the Fisher exponent remains valid even in three-dimensional systems and with different order-parameter symmetries.\\

However, a particularly interesting result arises in the case of the two-dimensional XY model. This system does not exhibit spontaneous symmetry breaking, in agreement with the Mermin--Wagner theorem, but it has a topological transition of the Berezinskii--Kosterlitz--Thouless type. In this case, the parameter $\eta$ exhibits the behavior
\begin{equation}
\eta(T)=
\begin{cases}
0, & T<T_{BKT},\\
1/4, & T=T_{BKT},\\
\infty, & T>T_{BKT},
\end{cases}
\end{equation}
where $T_{BKT}$ is the topological transition temperature.\\

The main characteristic of this case is the universal value $\eta=1/4$ at $T=T_{BKT}$. However, because this transition is not a conventional order--disorder transition, this value does not agree with the result predicted by our formulation. Under this extreme condition, our approach predicts flat surfaces with
\begin{equation}
    d_R=d_f=d=2,
\end{equation}
and, therefore, a return to the Euclidean mean-field result,
\begin{equation}
    \eta=0,
\end{equation}
with fractional order
\begin{equation}
    \zeta=1.
\end{equation}
Consequently, we conclude that the proposed formulation does not apply to topological phase transitions.

\subsection{The three-dimensional Heisenberg model}

The two-dimensional Heisenberg model represents an even more restrictive case than the two-dimensional XY model, because it does not even exhibit a topological transition. For this reason, we consider only the three-dimensional Heisenberg model.\\

In Table \ref{Table3}, we present the values of the fractal dimensions $d_f$ and $d_R$ for the three-dimensional Heisenberg model, together with their respective critical exponents.\\

Again, we use the notation $\eta^\ast$ to represent the values of $\eta$ obtained in the literature, distinguishing them from the value $\eta$ calculated from our geometric relation.\\

\begin{table}[H]
\centering
\begin{tabular}{ c | cccc }
\hline \hline
\hspace{4mm} Ref.      \hspace{4mm} &\hspace{4mm} \cite{Guillou80} \hspace{4mm} &\hspace{4mm} \cite{Brezin85}       \hspace{4mm} &\hspace{4mm}  \cite{Holm93}      \hspace{4mm} &\hspace{4mm} \cite{Campostrini02}   \hspace{4mm}\\ \hline
$\beta$ &   $0.3645(25) $   &  $0.368(4) $   & 0.362(4)      &  $0.3689(3)$ \\
$\nu$   &   $0.705(3)$    &  $0.710(7) $   & $0.706(9)$     &  $0.71120(5)$ \\
$d_{f}$ &    $2.483(5)$   &  $2.482(11) $   &  $2.49(1)$  & $2.4813(5)$ \\
$d_R$ &    $2.97(1)$    &  $2.963(22) $   &  $2.97(2)$  &  $2.9626(9)$\\
  $\zeta$ & $0.965(10)$ &  $0.963(22) $   & $0.975(2) $   &  $0.9626(9)$  \\
  $\eta*$  &  $0.033(4)$   &   $0.040(3) $   & $0.027(2)$   & $0.0375(3)$ \\ 
$\eta$  &   $0.035(10)$ &   $0.037(40) $   & $0.025(2) $  &  $0.0374(9)$ \\ \hline \hline
\end{tabular}
\caption{Values of the critical exponents and fractal dimensions for the three-dimensional Heisenberg model. The values of $d_f$ are obtained from $d_f=d-\beta/\nu$, the values of $d_R$ from $d_R=2(d_f-1)$, while $\eta$ and $\zeta$ are obtained from $\eta=d-d_R=1-\zeta$.}
   \label{Table3}
\end{table}

The results in Table \ref{Table3} show that the values of $\eta$ obtained by our formulation agree, within the uncertainties, with the values of $\eta^\ast$ reported in the literature. Therefore, even for the three-dimensional Heisenberg model, whose continuous symmetry differs from that of the Ising model, the proposed geometric relation remains consistent.

\subsection{Discussion}

\paragraph{} The results presented in this chapter indicate that the fractal structure associated with critical correlations is not a peculiarity of the two-dimensional Ising model. Relation \ref{eta} is satisfied exactly for the universality class of the two-dimensional Potts model, and it also reproduces, within the numerical uncertainties, the known values for the Ising, XY, and Heisenberg models in three dimensions.\\

This agreement reinforces the interpretation that Fisher's exponent $\eta$ characterizes how much the fractal correlation dimension deviates from the integer spatial dimension of the system. The traditional formulation introduces $\eta$ as a correction required for the behavior of the correlation function, while the fractal formulation shows that this exponent can be understood as a direct consequence of the effective geometry of the correlations at the critical point.\\

The case of the two-dimensional XY model shows, on the other hand, the limit of the approach. Because the BKT transition is a topological transition, and not a conventional order--disorder transition associated with a local order parameter, the fractal correlation geometry considered here does not capture the universal value $\eta=1/4$ at $T_{BKT}$. Thus, the proposed formulation should be understood as valid for symmetry-breaking thermodynamic phase transitions, but not necessarily for topological transitions.\\

Thus, we show that the fractal formulation of the correlation function provides a consistent interpretation for the Fisher exponent in different phase-transition models. Replacing the usual Euclidean equation with an equation based on the Riesz fractional derivative makes it possible to associate the critical decay of correlations with a fractal correlation dimension $d_R$.\\

From this structure, the exponent $\eta$ can be written as \ref{eta}, showing that it represents the difference between the integer spatial dimension and the effective dimension in which critical correlations develop \cite{Lima24}. The results for the two-dimensional Potts model, as well as for the three-dimensional Ising, XY, and Heisenberg models, provide strong evidence that this interpretation is effective for equilibrium phase transitions associated with symmetry breaking.\\

Thus, this chapter extends the central hypothesis of the thesis, showing that the fractal dynamics of critical correlations is not only a property of the Ising model, but a more general geometric structure present in different universality classes.

\chapter{Conclusion}

\paragraph{} Throughout this thesis, we investigated the hypothesis that criticality should not be understood only as a singular regime from the thermodynamic point of view, but also as a geometric reorganization of the space effectively accessible to the system dynamics. Starting from the classical problem of the failure of the fluctuation-dissipation theorem and the ad hoc introduction of the Fisher exponent, we sought to develop an interpretation according to which critical correlations become established in a subspace of fractal nature. In this context, the central objective of the work was to articulate results from phase-transition theory, fractal geometry, and high-performance numerical simulation in a framework capable of assigning a deeper physical meaning to the anomalous exponent $\eta$. By defining the expression $\eta = d-d_R$, we present a general solution, for a variety of systems, that serves as an answer to why theories of critical behavior fail. Throughout this thesis, the ferromagnetic Ising model played the role of a reference system for validating the central hypotheses of the work, proving fully compatible with the numerical analyses developed.\\

The analysis developed throughout the chapters showed that this perspective is consistent at different levels. Initially, we revisited the foundations of phase transitions and the way classical theory fails to adequately describe the critical regime, highlighting the generalization proposed by Fisher for the correlation function. We again highlighted the importance of the advances in fractal calculus developed by Muslih and Agrawal \cite{Muslih10, Muslih10b}, essential to the construction of this thesis, which enabled us to present the anomalous exponent as a correction to the effective geometry of the system dynamics. We then found, through analysis of the fractal dimension using the \textit{box counting} method, that the fractal dimension of the \textit{clusters} of \textit{spins} in the Ising model at the phase transition has a value compatible with the theoretical result $d_R = 1.75$. In particular, this result is important for consistently defining which geometric structure was evaluated to calculate this result, as in the case of critical percolation, Barabasi \textit{et al.} present a set of values for different fractal dimensions associated with distinct parts of the percolating structure \cite{Barabasi95}. Thus, determining the fractal dimension should be understood not only as a numerical procedure, but as a central part of identifying the effective geometry associated with the critical regime. Once the relevance of interpreting this fractal geometry for describing the critical regime had been established, the next step chosen was to examine the validity of this proposal in systems defined in non-integer dimensions.\\

Using the equation proposed by Suzuki $d_f = d - \beta/\nu$, the relation $d_{R} = 2(d_f - 1)$, and the fundamental result of this thesis $\eta = d - d_R$, we compared the values presented in \cite{isingnoninteger} with others obtained from quantities directly associated with the geometry of the problem. This analysis showed that, even when extending the system to non-integer dimensions, the relations between critical exponents and fractal dimensions remain consistent, suggesting that the scaling structure of the system is preserved in this more general context. In particular, we observed that the geometric quantities introduced maintain well-defined relations among themselves for all non-integer dimensions evaluated, indicating that the interpretation of critical behavior does not strictly depend on the Euclidean dimensionality of the system, but rather on the way correlations are spatially organized. In this way, the results reinforce the hypothesis that criticality can be understood as a manifestation of dynamics restricted to a fractal subspace, whose effective dimensionality directly governs the system dynamics. This consistency across different dimensions suggests that the approach adopted in this thesis does not constitute only a particular description of the two-dimensional Ising model, but points to a more general principle in which the geometry of the system plays a central role in determining its critical properties.\\

The idealization of perfectly homogeneous systems, although extremely useful from a theoretical point of view, rarely has a direct counterpart in real physical systems; this was therefore the motivation for developing the analyses performed for disordered systems. Figure \ref{tcs_comp} demonstrates the proximity of the compared critical temperatures $T_c^{mag}$ and $T_c^R$, a fundamental result for understanding that the fractal dimension $d_{R} = 1.75$ lies within the critical region of the system. Up to the limit at which disorder was studied in this thesis, $\sigma = 0.5$, the systems respond almost as pure shifts of the model as a function of the critical temperature, which naturally leaves room for questioning what disorder limit would still make sense for the model. In a complementary way, the analysis of the critical exponents as a function of disorder showed that, although quantitative variations are observed, the scaling relations remain, to a large extent, preserved within the numerical uncertainties. In particular, we verified that the Rushbrooke relation remains compatible with the expected values, indicating that the presence of disorder, at least at the levels studied, does not compromise the structural coherence of criticality theory, at least within the interval considered. Among the critical exponents analyzed, the ratio $\beta/\nu$ assumes a central role because, even in the presence of disorder, it establishes a direct connection between the classical scaling relations and the geometric interpretation of criticality developed in this thesis.\\

The validity of the proposed formulation was also examined in other fundamental models of statistical mechanics, such as the two-dimensional Potts model and the Ising, XY, and Heisenberg models in three dimensions. This extension showed that the relation $\eta=d-d_R$ is not limited to the two-dimensional Ising model, but correctly reproduces known values of the Fisher exponent in different universality classes associated with conventional thermodynamic transitions. In particular, the Potts model made it possible to verify the consistency of the relation for different values of $q$, while the three-dimensional models showed that the geometric interpretation remains compatible even in systems with different order-parameter symmetries. The case of the two-dimensional XY model, in turn, revealed a natural boundary of the approach, since the Berezinskii--Kosterlitz--Thouless transition has a topological nature and does not fit directly into the fractal formulation developed in this thesis.\\

Throughout the execution of this work, we investigated the critical phenomenon from the perspective of fractal geometry, seeking to understand to what extent the spatial structure of correlations could provide a deeper physical interpretation for the critical exponents. Throughout the observations, we showed that criticality can be understood not only as a singular regime of thermodynamic quantities, but also as a manifestation of geometric reorganization of the system, in which the effective dynamics begins to develop in a subspace of non-integer nature. The results obtained throughout the thesis indicate that this approach is consistent with the models and scenarios investigated, reinforcing the idea that fractal geometry is not merely a descriptive tool, but a central element for understanding critical behavior. At the end of this thesis, the analyses developed and the results obtained consistently support the conclusion that critical behavior admits a geometric interpretation. Several possibilities remain for deepening this line of investigation, among which are the advancement of analyses in disordered systems, disorder parameters with higher values, and the study of models with temporal dynamics, such as synchronization~\cite{Pecora91,Longa96,Ciesla01,Morgado07}, pattern formation~\cite{Turing52,Cross93,DaCunha09,DaCunha11,Barbosa17,Fuentes03,Aranda20-1,Aranda21,Aranda20-2,Fuentes04}, modeling of drug release\cite{Carwood26,Barbosa17,GomesFilho16,GomesFilho22}, and dynamic phase transition \cite{Ziff86,Fernandes18,Santos24,Pinto16,Pinto17}. Although such topics were not incorporated into the scope of this work, they constitute natural developments of the research carried out here and will be the subject of future investigations.

\nocite{ Kardar86, Kadar85, Alves16, Krug92, Krug97, feder2013fractals, Derrida98, Meakin86, Daryaei20, Edwards82, Hansen00, Wolf, Rodrigues24, Merikoski03, Odor10, Takeuchi13, Gwa92, Vega85, Plischke87, Corwin18, nahum17, ljubotina19, DeNardis19, Moca23, Rodriguez19, Grigera99, Ricci-Tersenghi00, Crisanti03, Barrat98, Bellon02, Bellon06, Vainstein06, Hayashi07, Perez-Madrid09, Averin10, Costa03, Costa06, Lapas07, Lapas08, Luis22, Luis23, Amorim23, bhattacharyya06, Zhang12, Wen23, Barkai01, Dumouchel05, Nowak22, Kuramoto84, Edwards82, Carrasco18, Cruz23, Sokolov01, Pham23, reis96, Reis04, Chen17, Boyle07, Angulo00, Wallace79, Ziff86, Fernandes18, Kawabata19, salman24, Grimmett06, Sherrington75, Almeida78, Kalosakas22, Metzler99,Lin89}
\apptocmd{\thebibliography}{\small\setlength{\itemsep}{4pt}\setlength{\parskip}{0pt}}{}{}
% Bibliography precompiled for arXiv compatibility

% ======================================================================
% APPENDIX X — BOX COUNTING (COMPUTATIONAL FOCUS)
% ======================================================================

\appendix
\chapter{\textit{Box Counting} Method}
\label{ap:box_counting}

\textbf{The images in this appendix were produced with the aid of artificial intelligence through the chatGPT® platform by providing the corresponding \textit{scripts} related to the content presented here}.\\

The geometric characterization of clusters and interfaces in statistical systems can be carried out by different scale-analysis methods. Among them, the \textit{box counting} method occupies a prominent position because of its conceptual simplicity and numerical efficiency in estimating fractal dimensions. In the present work, this method was employed to estimate the fractal dimension associated with the spin configurations obtained in the Ising model.\\

The fundamental idea of the procedure consists of covering the system with square boxes of linear side \(s\) and, for each scale considered, counting the number of boxes that intersect the set of interest. In the context of this work, the analyzed set is formed by the sites whose spin variable takes the value
\begin{equation}
    s_i = +1,
\end{equation}
that is, the so-called \textit{spin up} sites. In other words, with each final lattice configuration we associate a subset of occupied sites, defined precisely by the lattice points at which the spin is oriented upward. The objective of the method is to quantify how this set fills space as the observation scale is changed.\\

Let \(N(s)\) be the number of boxes of side \(s\) required to cover the set formed by the \(+1\) spins. When this set exhibits fractal behavior over a range of scales, the quantity \(N(s)\) is expected to satisfy a power law of the form
\begin{equation}
    N(s) \sim s^{-d_f},
\end{equation}
where \(d_f\) represents the fractal dimension of the set. This relation expresses the fact that the occupation of space does not, in general, occur as in a regular Euclidean object, but rather in an intermediate way between different effective dimensionalities.

Taking the logarithm on both sides of the previous expression, we obtain
\begin{equation}
    \ln N(s) = -d_f \ln s + C,
\end{equation}
where \(C\) is a constant. In this way, the fractal dimension can be estimated from the slope of the straight line obtained in the plot of \(\ln N(s)\) as a function of \(\ln s\). Thus, the method reduces the problem of determining \(d_f\) to a linear fit on a logarithmic scale. In the present case, the method was applied directly to the two-dimensional matrix representing the spin configuration of the system. Because the interest is focused on sites with value \(+1\), the counting is performed only over these elements. Initially, the total magnetization of the configuration is calculated. When the global sum of the spins is negative, all spins in the lattice are inverted. This procedure does not alter the geometric structure of the system, but guarantees that the majority phase is represented by the \(+1\) spins, thereby standardizing the numerical analysis and the geometric interpretation of the results. Next, a discrete set of scales \(s\) is chosen, corresponding to the sizes of the boxes used to cover the lattice. For each value of \(s\), the lattice is partitioned into square blocks, and it is checked which of these blocks contain at least one spin up. Each block satisfying this condition contributes one unit to the value of \(N(s)\). Repeating this procedure for all scales considered yields the sequence of pairs \((s, N(s))\), from which the linear fit on a logarithmic scale can be performed.\\

From a computational point of view, the routine was implemented with CUDA parallelization, allowing the massively parallel structure of the problem to be exploited. The strategy consists of associating each lattice site with a GPU \textit{thread}. For a given scale \(s\), each \textit{thread} identifies the box to which the site belongs; if the spin at that site is \(+1\), the counter corresponding to that box is incremented through an atomic operation. At the end of this stage, the number of \(+1\) spins present in each block is obtained for each scale. Since the \textit{box counting} method is concerned only with whether the box is occupied or not, the values are subsequently binarized: boxes with a count greater than or equal to one receive unit value, while empty boxes remain zero. The sum of these binary values gives precisely the number \(N(s)\) of occupied boxes at that scale.\\

After obtaining the values of \(N(s)\), a linear fit of the function \(\ln N(s)\) as a function of \(\ln s\) is performed. If the slope of the fitted line is denoted by \(a\), then the fractal dimension is given by
\begin{equation}
    d_f = -a.
\end{equation}

This estimate corresponds to the quantity returned by the numerical routine used in this work. Below we present a simplified version of the code employed in terms of a function in the \textit{Python} language, preserving only the essential sections for implementing the method:

\begin{verbatim}
import numpy as np
import cupy as cp
from numba import cuda

def cuda_df_calc(x):
    if np.sum(x) < 0:
        x *= -1

    @cuda.jit
    def bc_cuda(x, bs, offset, result):
        n, m = x.shape
        i, j = cuda.grid(2)

        if i < n and j < m:
            nob = n / bs
            ib = i / bs
            jb = j / bs
            idx = int(jb * nob + ib + offset)

            if x[i, j] == 1:
                cuda.atomic.add(result, idx, 1)

    @cuda.jit
    def binarize_boxes(x):
        tid = cuda.blockIdx.x * cuda.blockDim.x + cuda.threadIdx.x
        if tid < x.size and x[tid] >= 1:
            x[tid] = 1

    threadsperblock = (32, 32)
    blockspergrid = (
        int(np.ceil(x.shape[0] / threadsperblock[0])),
        int(np.ceil(x.shape[1] / threadsperblock[1]))
    )

    scales = [1, 2, 3, 4, 5, 6, 8]
    offsets = [(len(x) / s) ** 2 for s in scales]
    result = cp.zeros(int(np.sum(offsets)))

    offset = 0
    for s in scales:
        bc_cuda[blockspergrid, threadsperblock](x, s, offset, result)
        offset += (len(x) / s) ** 2

    threadsperblock_1d = 32
    blockspergrid_1d = int(np.ceil(result.shape[0] / threadsperblock_1d))
    binarize_boxes[blockspergrid_1d, threadsperblock_1d](result)

    counts = []
    start = 0
    for size in offsets:
        end = start + int(size)
        counts.append(int(np.sum(result[start:end])))
        start = end

    coeffs = np.polyfit(np.log(scales), np.log(counts), 1)
    return -coeffs[0]
\end{verbatim}

In practical terms, the routine above implements exactly the logic discussed previously. First, the set of interest is defined as the set of \(+1\) spins. Next, for each scale \(s\), the lattice is partitioned into square boxes and it is checked whether each box contains at least one of these spins. Counting the number of occupied boxes produces the function \(N(s)\), whose dependence on scale makes it possible to estimate the fractal dimension of the set.\\

It should be noted that the quality of this estimate depends on the appropriate choice of the set of scales used in the fit, as well as on the existence of an approximately linear regime in the log--log plot. Even so, the \textit{box counting} method constitutes a particularly useful tool because it transforms a complex geometric property into a simple, robust numerical procedure with direct interpretation. For this reason, it is especially convenient in the analysis of spatial configurations obtained in lattice-defined statistical models, as in the case studied in this thesis.

% in the preamble:
% \usepackage{tikz}

\begin{figure}[ht]
\centering
\begin{tikzpicture}[x=1cm,y=1cm]

% ==========================================================
% Fixed list of "fine" positions (grid with step 0.5)
% These are the same elements in both panels
% ==========================================================

\def\occupiedcells{
  1/1,
  2/2,
  3/3,
  4/4,
  5/5,
  6/6,
  1/5,
  2/6,
  3/6,
  5/2,
  6/3
}

% ==========================================================
% Panel (a): larger epsilon
% Groups the fine grid into 1x1 boxes
% ==========================================================
\begin{scope}
    \node[anchor=south] at (2,4.55) {\small (a) larger $\varepsilon$};

    % fills occupied boxes of the coarse grid
    \foreach \i/\j in \occupiedcells {
        \pgfmathtruncatemacro{\I}{\i/2}
        \pgfmathtruncatemacro{\J}{\j/2}
        \fill[black!15] (\I,\J) rectangle ++(1,1);
    }

    % draws a single element at the center of each occupied box of the coarse grid
    % (list of occupied boxes without repetition)
    \foreach \I/\J in {
        0/0,
        1/1,
        2/2,
        3/3,
        0/2,
        1/3,
        2/1,
        3/1
    }{
        \fill (\I+0.5,\J+0.5) circle (1.8pt);
    }

    % coarse grid
    \draw[thick] (0,0) rectangle (4,4);
    \foreach \x in {1,2,3} \draw (\x,0) -- (\x,4);
    \foreach \y in {1,2,3} \draw (0,\y) -- (4,\y);
\end{scope}

% ==========================================================
% Panel (b): smaller epsilon
% Uses the fine grid directly
% ==========================================================
\begin{scope}[xshift=6cm]
    \node[anchor=south] at (2,4.55) {\small (b) smaller $\varepsilon$};

    \foreach \i/\j in \occupiedcells {
        \fill[black!15] (\i*0.5,\j*0.5) rectangle ++(0.5,0.5);
        \fill (\i*0.5+0.25,\j*0.5+0.25) circle (1.4pt);
    }

    % fine grid
    % fine grid
    \draw[thick] (0,0) rectangle (4,4);
    \foreach \x in {0.5,1.0,...,3.5} \draw (\x,0) -- (\x,4);
    \foreach \y in {0.5,1.0,...,3.5} \draw (0,\y) -- (4,\y);
\end{scope}

\end{tikzpicture}
\caption[Schematic representation of the \textit{box counting} method]{Schematic representation of the \textit{box counting} method. The same set of elements is maintained at both scales, with only the box size being varied. Each box may be empty or contain a single element at its center. For the larger scale, several fine cells belong to the same box, so the fraction of occupied boxes is relatively larger. When the scale is reduced, the total number of boxes increases, and the proportion of filled boxes decreases, although the number of occupied boxes \(N(\varepsilon)\) increases. Figure produced with the aid of artificial intelligence.}
\label{fig:boxcounting}
\end{figure}
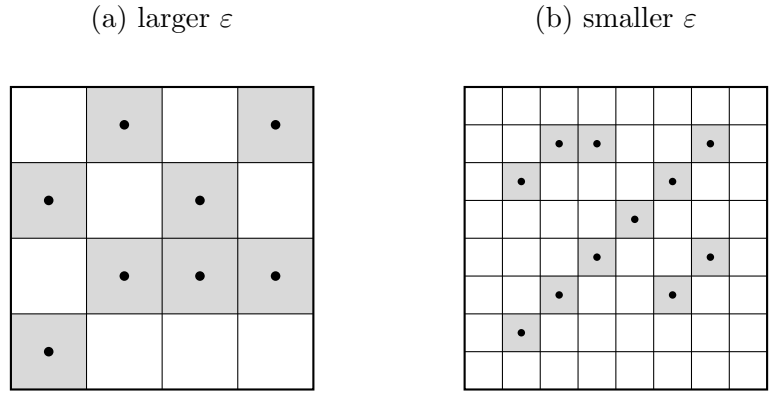

\subsection{Numerical example of the \textit{box counting} method}
\label{ap:exemplo_box_counting}

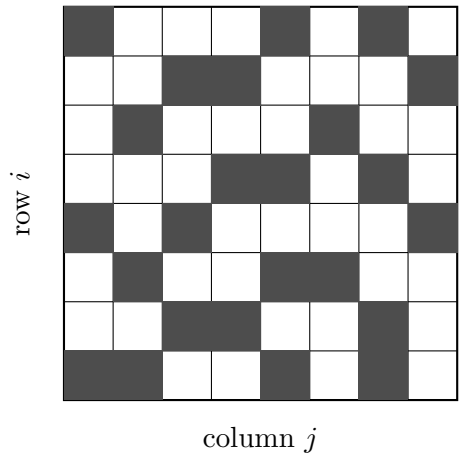
\begin{figure}[ht]
\centering
\begin{tikzpicture}[scale=0.65]

% 8x8 grid
\draw[thick] (0,0) rectangle (8,8);
\foreach \x in {1,...,7} \draw (\x,0)--(\x,8);
\foreach \y in {1,...,7} \draw (0,\y)--(8,\y);

% cells with spin +1
\foreach \x/\y in {
0/0,1/0,4/0,6/0,
2/1,3/1,6/1,
1/2,4/2,5/2,
0/3,2/3,7/3,
3/4,4/4,6/4,
1/5,5/5,
2/6,3/6,7/6,
0/7,4/7,6/7
}{
    \fill[black!70] (\x,\y) rectangle ++(1,1);
}

\node at (4,-0.8) {\small column \(j\)};
\node[rotate=90] at (-0.9,4) {\small row \(i\)};

\end{tikzpicture}
\caption[Schematic spin configuration on an \(8\times 8\) lattice.]{Schematic spin configuration on an \(8\times 8\) lattice. The dark squares represent the sites with \(s_i=+1\), which constitute the set of interest for applying the \textit{box counting} method. Figure produced with the aid of artificial intelligence.}
\label{fig:rede_8x8_boxcounting}
\end{figure}

In order to concretely illustrate the application of the \textit{box counting} method, let us consider a hypothetical spin configuration defined on an \(8\times 8\) square lattice. At each lattice site, the spin variable can take the values \(s_i=\pm1\). As discussed previously, the set of interest is formed exclusively by the sites with \textit{spin up}, that is, the sites for which \(s_i=+1\). Sites with \(s_i=-1\) do not participate in the counting. Figure~\ref{fig:rede_8x8_boxcounting} presents the configuration considered in this example. The dark squares represent sites occupied by \(+1\) spins, while the light squares correspond to \(-1\) spins. The objective of the method is to determine, for different scales \(s\), the number \(N(s)\) of boxes required to cover the set formed by the \(+1\) spins.\\

\begin{figure}[ht]
\centering
\begin{tikzpicture}[scale=0.48]

% =========================================================
% Panel (a): s = 1
% =========================================================
\begin{scope}
\node at (4,9.2) {\small (a) \(s=1\)};

\draw[thick] (0,0) rectangle (8,8);
\foreach \x in {1,...,7} \draw (\x,0)--(\x,8);
\foreach \y in {1,...,7} \draw (0,\y)--(8,\y);

\foreach \x/\y in {
0/0,1/0,4/0,6/0,
2/1,3/1,6/1,
1/2,4/2,5/2,
0/3,2/3,7/3,
3/4,4/4,6/4,
1/5,5/5,
2/6,3/6,7/6,
0/7,4/7,6/7
}{
    \fill[black!25] (\x,\y) rectangle ++(1,1);
    \fill (\x+0.5,\y+0.5) circle (1.7pt);
}
\end{scope}

% =========================================================
% Panel (b): s = 2
% =========================================================
\begin{scope}[xshift=11cm]
\node at (4,9.2) {\small (b) \(s=2\)};

% All 16 2x2 boxes are occupied
\foreach \x/\y in {
0/0,2/0,4/0,6/0,
0/2,2/2,4/2,6/2,
0/4,2/4,4/4,6/4,
0/6,2/6,4/6,6/6
}{
    \fill[black!25] (\x,\y) rectangle ++(2,2);
}

\draw[gray!40] (0,0) rectangle (8,8);
\foreach \x in {1,...,7} \draw[gray!40] (\x,0)--(\x,8);
\foreach \y in {1,...,7} \draw[gray!40] (0,\y)--(8,\y);

\draw[thick] (0,0) rectangle (8,8);
\foreach \x in {2,4,6} \draw[thick] (\x,0)--(\x,8);
\foreach \y in {2,4,6} \draw[thick] (0,\y)--(8,\y);

\foreach \x/\y in {
0/0,1/0,4/0,6/0,
2/1,3/1,6/1,
1/2,4/2,5/2,
0/3,2/3,7/3,
3/4,4/4,6/4,
1/5,5/5,
2/6,3/6,7/6,
0/7,4/7,6/7
}{
    \fill (\x+0.5,\y+0.5) circle (1.5pt);
}
\end{scope}

% =========================================================
% Panel (c): s = 4
% =========================================================
\begin{scope}[yshift=-11cm]
\node at (4,9.2) {\small (c) \(s=4\)};

\foreach \x/\y in {
0/0,4/0,
0/4,4/4
}{
    \fill[black!25] (\x,\y) rectangle ++(4,4);
}

\draw[gray!40] (0,0) rectangle (8,8);
\foreach \x in {1,...,7} \draw[gray!40] (\x,0)--(\x,8);
\foreach \y in {1,...,7} \draw[gray!40] (0,\y)--(8,\y);

\draw[thick] (0,0) rectangle (8,8);
\draw[thick] (4,0)--(4,8);
\draw[thick] (0,4)--(8,4);

\foreach \x/\y in {
0/0,1/0,4/0,6/0,
2/1,3/1,6/1,
1/2,4/2,5/2,
0/3,2/3,7/3,
3/4,4/4,6/4,
1/5,5/5,
2/6,3/6,7/6,
0/7,4/7,6/7
}{
    \fill (\x+0.5,\y+0.5) circle (1.5pt);
}
\end{scope}

% =========================================================
% Panel (d): s = 8
% =========================================================
\begin{scope}[xshift=11cm,yshift=-11cm]
\node at (4,9.2) {\small (d) \(s=8\)};

\fill[black!25] (0,0) rectangle (8,8);

\draw[gray!40] (0,0) rectangle (8,8);
\foreach \x in {1,...,7} \draw[gray!40] (\x,0)--(\x,8);
\foreach \y in {1,...,7} \draw[gray!40] (0,\y)--(8,\y);

\draw[thick] (0,0) rectangle (8,8);

\foreach \x/\y in {
0/0,1/0,4/0,6/0,
2/1,3/1,6/1,
1/2,4/2,5/2,
0/3,2/3,7/3,
3/4,4/4,6/4,
1/5,5/5,
2/6,3/6,7/6,
0/7,4/7,6/7
}{
    \fill (\x+0.5,\y+0.5) circle (1.5pt);
}
\end{scope}

\end{tikzpicture}
\caption[Schematic application of the \textit{box counting} method to the configuration considered in Figure~\ref{fig:rede_8x8_boxcounting}.]{Schematic application of the \textit{box counting} method to the configuration considered in Figure~\ref{fig:rede_8x8_boxcounting}. The shaded regions indicate the occupied boxes at each scale \(s\). Figure produced with the aid of artificial intelligence.}
\label{fig:boxcounting_escalas_8x8}
\end{figure}
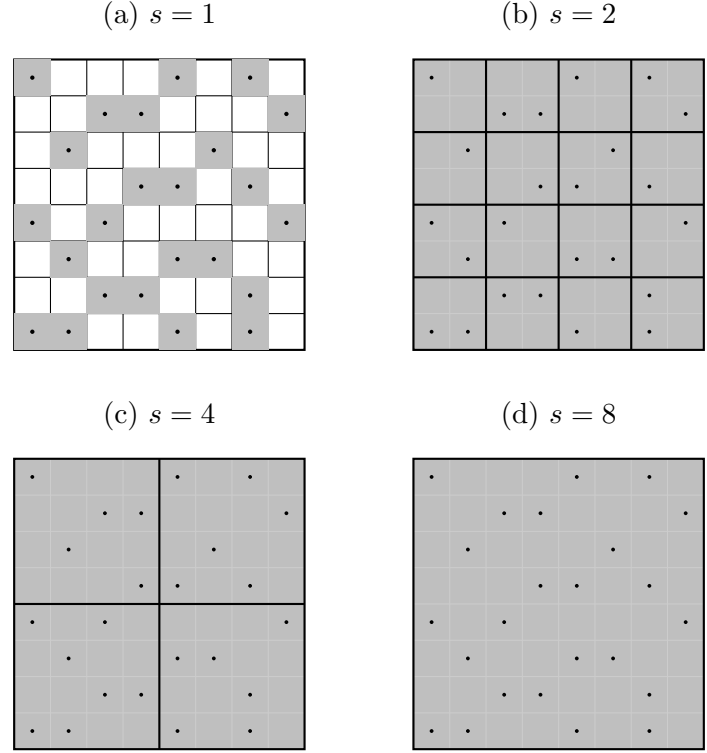

Once the configuration has been defined, the next step consists of covering the lattice with square boxes of side \(s\) and counting how many of these boxes contain at least one \(+1\) spin. Figure~\ref{fig:boxcounting_escalas_8x8} explicitly shows this procedure for the scales \(s=1\), \(s=2\), \(s=4\), and \(s=8\). In each panel, the shaded regions indicate the boxes occupied by the set of interest.\\

\paragraph{Scale \(s=1\).}
For \(s=1\), each box coincides with a single lattice site. In this case, the number of occupied boxes is simply equal to the total number of \(+1\) spins in the configuration. Counting the dark sites in Figure~\ref{fig:rede_8x8_boxcounting}, we obtain
\begin{equation}
    N(1)=24.
\end{equation}

\paragraph{Scale \(s=2\).}
For \(s=2\), the lattice is divided into \(2\times2\) blocks. Since the lattice has linear size \(L=8\), this produces
\begin{equation}
    \left(\frac{8}{2}\right)^2 = 16
\end{equation}
boxes in total. A box is considered occupied if it contains at least one \(+1\) spin. In the present example, all \(16\) \(2\times2\) boxes are found to contain at least one element of the set of interest. Therefore,
\begin{equation}
    N(2)=16.
\end{equation}

\paragraph{Scale \(s=4\).}
For \(s=4\), the lattice is partitioned into \(4\times4\) blocks, resulting in
\begin{equation}
    \left(\frac{8}{4}\right)^2 = 4
\end{equation}
boxes. In this case, each of the four regions contains at least one \(+1\) spin, which implies
\begin{equation}
    N(4)=4.
\end{equation}

\paragraph{Scale \(s=8\).}
Finally, for \(s=8\), the entire lattice is treated as a single box. Since the configuration contains \(+1\) spins, this box is occupied. Therefore,
\begin{equation}
    N(8)=1.
\end{equation}

Thus, for the scales considered, the following data set is obtained
\begin{equation}
\begin{array}{c|cccc}
s      & 1 & 2 & 4 & 8 \\
\hline
N(s)   & 24 & 16 & 4 & 1
\end{array}
\label{eq:tabela_box_counting}
\end{equation}

Assuming that the set exhibits a scaling relation of the form
\begin{equation}
    N(s)\sim s^{-d_f},
\end{equation}
where \(d_f\) is the fractal dimension, we can take the logarithm of both sides and obtain
\begin{equation}
    \ln N(s) = -d_f \ln s + C,
\end{equation}
with \(C\) constant. In this way, the estimate of \(d_f\) can be obtained from the slope of the linear fit of the plot of \(\ln N(s)\) as a function of \(\ln s\). Calculating the logarithms of the values found, we obtain approximately
\begin{equation}
\begin{array}{c|cccc}
s & 1 & 2 & 4 & 8 \\
\hline
\ln s     & 0.000 & 0.693 & 1.386 & 2.079 \\
\ln N(s)  & 3.178 & 2.773 & 1.386 & 0.000
\end{array}
\end{equation}
Fitting these points with a straight line of the form
\begin{equation}
    \ln N(s)=a\,\ln s+b,
\end{equation}
for this example, a slope approximately equal to
\begin{equation}
    a \approx -1.58.
\end{equation}
Therefore, the estimate for the fractal dimension is
\begin{equation}
    d_f = -a \approx 1.58.
\end{equation}

It is important to emphasize that this value should be interpreted only as an illustration of the method. Since the configuration considered is small and was chosen only for didactic purposes, the estimate obtained should not be understood as a robust physical result. In real applications, the procedure is carried out on much larger lattices and, in general, after averaging over several independent samples, which makes it possible to identify more reliably the scaling region in which approximately linear behavior appears.\\

Even so, the example above makes the logic of the method clear. For each scale \(s\), the number of boxes occupied by the set of \(+1\) spins is counted; then, the dependence of \(N(s)\) on scale is analyzed. It is precisely this dependence that provides an estimate of the fractal dimension of the set. In this way, the \textit{box counting} method makes it possible to convert geometric information from the configuration into a numerical quantity directly comparable among different samples and different physical conditions of the system.\\

In summary, the method employed consists of identifying, at different scales, how many boxes are required to cover the set of spin-up sites present in the configuration, and using the corresponding scaling law to extract the fractal dimension. In this way, a quantitative measure of the degree of geometric irregularity of the analyzed set is obtained, allowing spatial properties of the configuration to be related to the physics of the system.

% =======================
% =======================

\chapter{Calculation of the spatial correlation function by spectral amplitude}
\label{ap:correlacao_fft}

\textbf{The images in this appendix were produced with the aid of artificial intelligence through the chatGPT® platform by providing the corresponding \textit{scripts} related to the content presented here}.\\

In this appendix, we describe the procedure used to calculate the spatial correlation function from the spin configuration of the Ising model. The method employed is based on the Wiener--Khinchin theorem, which makes it possible to obtain the autocorrelation of a function from its power spectrum, making the calculation numerically much more efficient than a direct sum over all pairs of lattice sites.\\

Consider a two-dimensional spin configuration represented by a discrete function \(S(x,y)\), defined on a square lattice of size \(L\times L\), with
\[
S(x,y)=\pm 1,
\]
where \(x\) and \(y\) identify the positions of the lattice sites. The central idea is to determine how the spin values at different points of the lattice are correlated as a function of spatial separation. The most direct way to define this correlation consists of introducing the two-dimensional autocorrelation function
\begin{equation}
C(dx,dy)=\left\langle S(x,y)\,S(x+dx,y+dy)\right\rangle,
\label{eq:autocorrelacao_bidimensional}
\end{equation}
where \((dx,dy)\) represents a displacement on the lattice and \(\langle\cdots\rangle\) indicates a spatial average over all sites compatible with this displacement. This quantity therefore measures the degree of correlation between spins separated by a vector \((dx,dy)\). However, directly calculating equation \eqref{eq:autocorrelacao_bidimensional} for all possible displacements can become computationally costly, especially for large systems. To overcome this difficulty, we use the Wiener--Khinchin theorem, according to which the spatial autocorrelation can be obtained from the inverse Fourier transform of the squared modulus of the direct transform of the configuration. In simple notation,
\begin{equation}
C(dx,dy)=\mathcal{F}^{-1}\left(|\hat{S}(k_x,k_y)|^2\right),
\label{eq:wk}
\end{equation}
where \(\hat{S}(k_x,k_y)=\mathcal{F}[S(x,y)]\) is the two-dimensional discrete Fourier transform of the spin lattice.\\

In practice, the procedure is carried out in three steps. First, the two-dimensional discrete Fourier transform is applied to the configuration \(S(x,y)\). Next, the power spectrum is constructed
\[
|\hat{S}(k_x,k_y)|^2.
\]
Finally, the inverse Fourier transform is applied to this quantity. The result is the spatial autocorrelation \(C(dx,dy)\), now obtained efficiently through the use of FFT algorithms. It is important to note that \(C(dx,dy)\) is still a two-dimensional function. This means that it explicitly depends on the direction of the displacement on the lattice, and not only on the distance between the spins. In other words, this function preserves the vector information of the separation. However, in many cases of physical interest, especially in isotropic systems, a correlation function that depends only on the scalar distance between sites is desired. For this purpose, the radial function \(G(r)\) is defined. The radial distance associated with a displacement \((dx,dy)\) is given by
\begin{equation}
r=\sqrt{dx^2+dy^2}.
\label{eq:distancia_radial}
\end{equation}
The function \(G(r)\) is then obtained from a radial average of \(C(dx,dy)\), grouping all lattice points whose displacements have approximately the same modulus \(r\). Conceptually,
\begin{equation}
G(r)=\langle C(dx,dy)\rangle_{r},
\label{eq:media_radial}
\end{equation}
where \(\langle\cdots\rangle_r\) indicates the average over all pairs \((dx,dy)\) such that \(\sqrt{dx^2+dy^2}\approx r\).\\

This step is fundamental, because it transforms the complete two-dimensional autocorrelation into a one-dimensional function, more appropriate for studying the spatial decay of correlations. Thus, while \(C(dx,dy)\) describes the correlation for each specific vector displacement, \(G(r)\) represents the average correlation between spins separated by a distance \(r\), independently of direction.\\

From a computational point of view, the spectral-amplitude method has an important advantage. Direct calculation of the correlation requires explicitly considering a large number of pairs of spins in the lattice, which can become prohibitive for large values of \(L\). In contrast, the use of FFT makes it possible to obtain the complete autocorrelation at a much lower computational cost, making the procedure particularly useful in simulations of large two-dimensional systems.\\

In summary, the method employed starts from the discrete spin configuration \(S(x,y)\), calculates its two-dimensional Fourier transform, constructs the corresponding power spectrum and, by inverse transform, obtains the autocorrelation \(C(dx,dy)\). Next, a radial average of this quantity provides the function \(G(r)\), which is the form used in the analyses presented in this work.

\section{Example of the calculation of \(G(r)\) by spectral amplitude on an \(8\times8\) lattice}
\label{ap:wk_8x8}

\paragraph{} In this section, we present an explicit example of calculating the radial correlation function \(G(r)\) from the spectral-amplitude method, using a discrete spin configuration \(S(x,y)\) on an \(8\times8\) lattice. The objective is to illustrate, step by step, how the two-dimensional autocorrelation \(C(dx,dy)\) is obtained via Fourier transform and then converted into a radial function depending only on the distance \(r\).\\

\begin{figure}[H]
    \centering
    \includegraphics[width=0.7\linewidth]{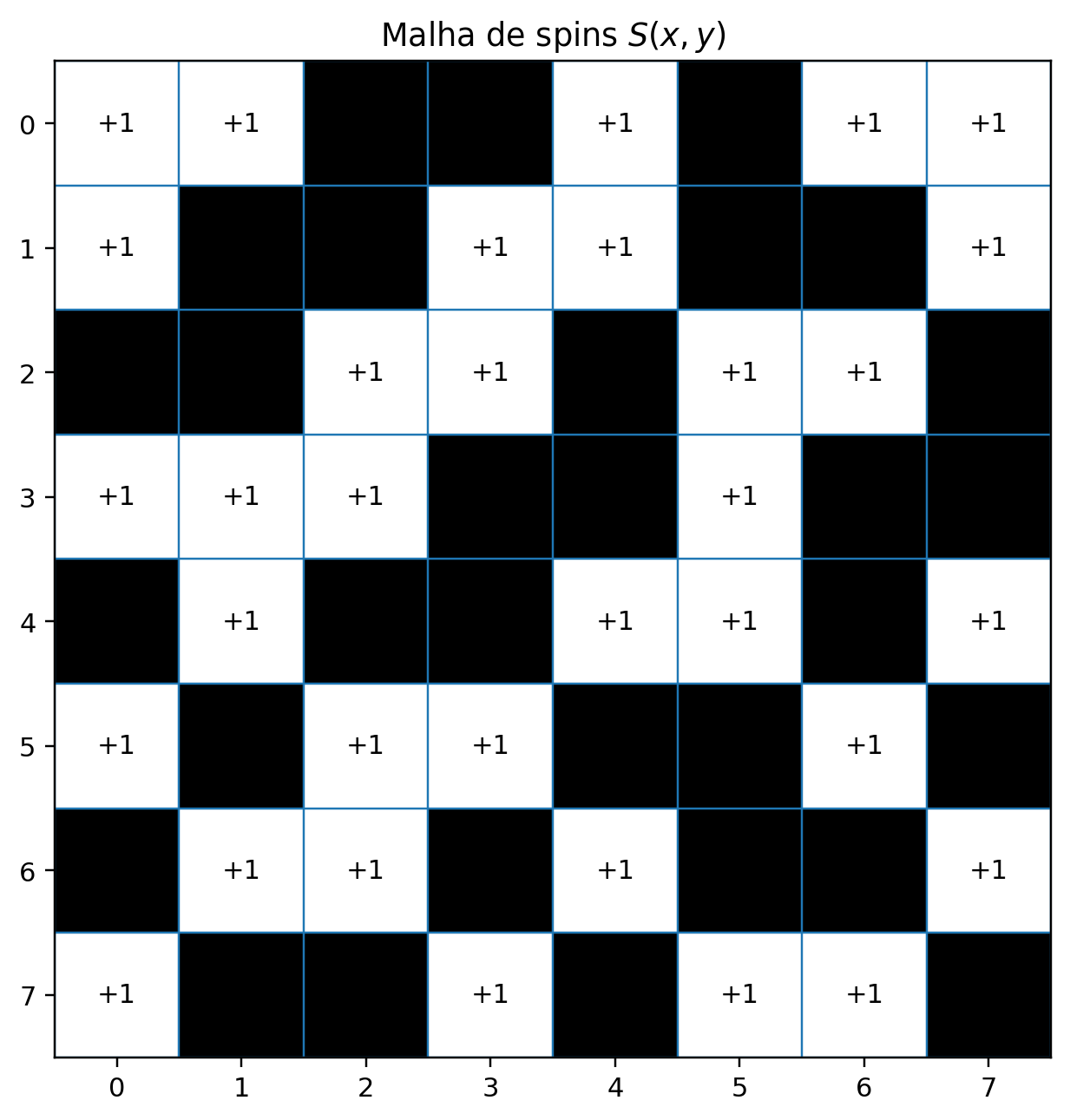}
    \caption{Spin configuration \(S(x,y)\) used in the numerical example. Figure produced with the aid of artificial intelligence.}
    \label{fig:wk8x8_lattice}
\end{figure}

Initially consider the spin configuration shown in Figure~\ref{fig:wk8x8_lattice}. Each lattice site takes values \(S(x,y)=\pm1\), corresponding to the two possible spin states.

\begin{figure}[ht]
    \centering
    \includegraphics[width=0.85\linewidth]{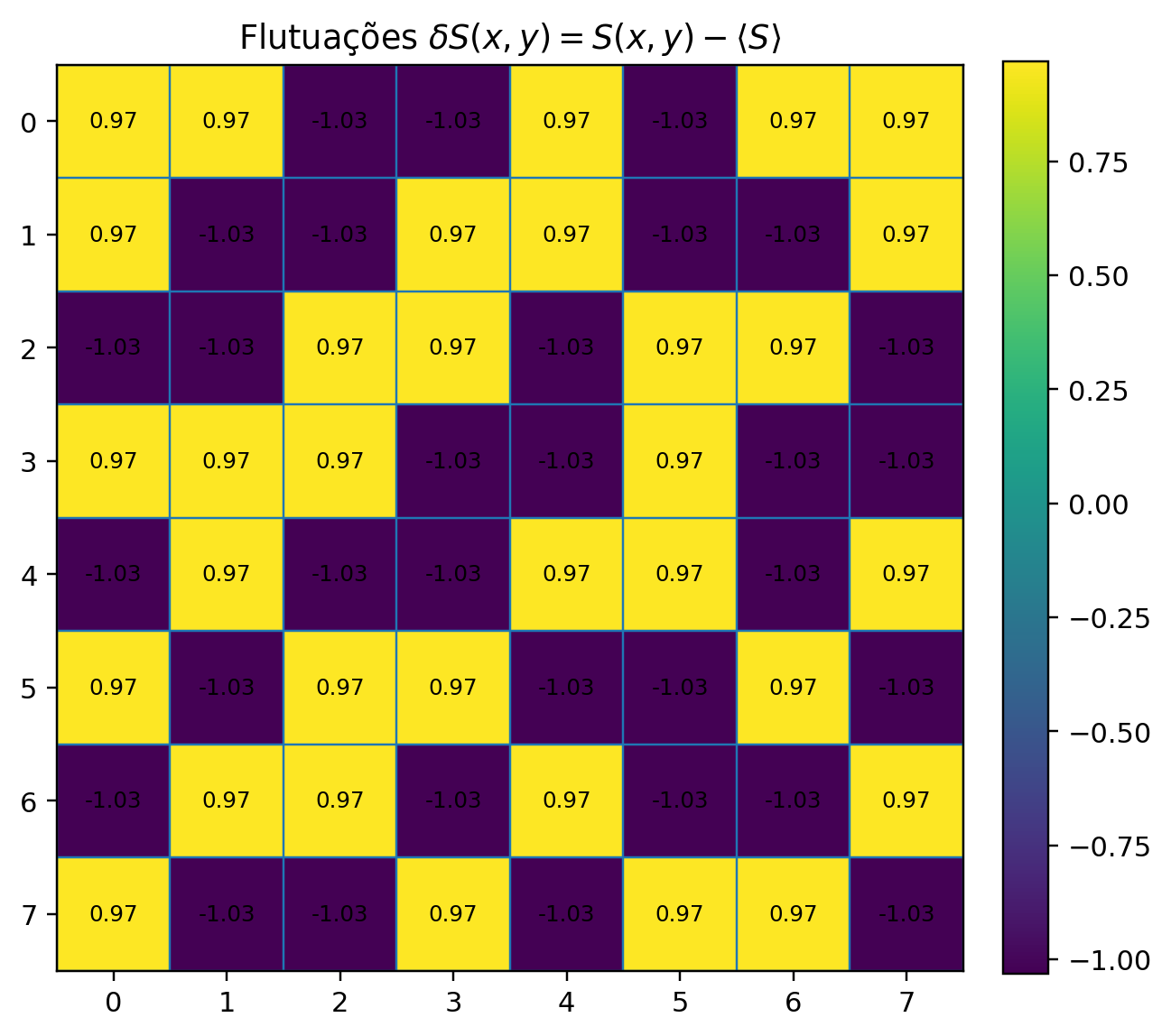}
    \caption{Lattice of fluctuations \(\delta S(x,y)=S(x,y)-\langle S\rangle\). Figure produced with the aid of artificial intelligence.}
    \label{fig:wk8x8_fluct}
\end{figure}

The spatial average of this configuration is
\[
\langle S \rangle = 0.031,
\]and from it, we define the fluctuations around the mean,
\[
\delta S(x,y)=S(x,y)-\langle S\rangle.
\]
In the present example, since \(\langle S\rangle\neq 0\), this subtraction is important so that the calculated correlation represents only the spatial fluctuations of the system. The fluctuation lattice is shown in Figure~\ref{fig:wk8x8_fluct}.\\

\begin{figure}[ht]
    \centering
    \includegraphics[width=0.75\linewidth]{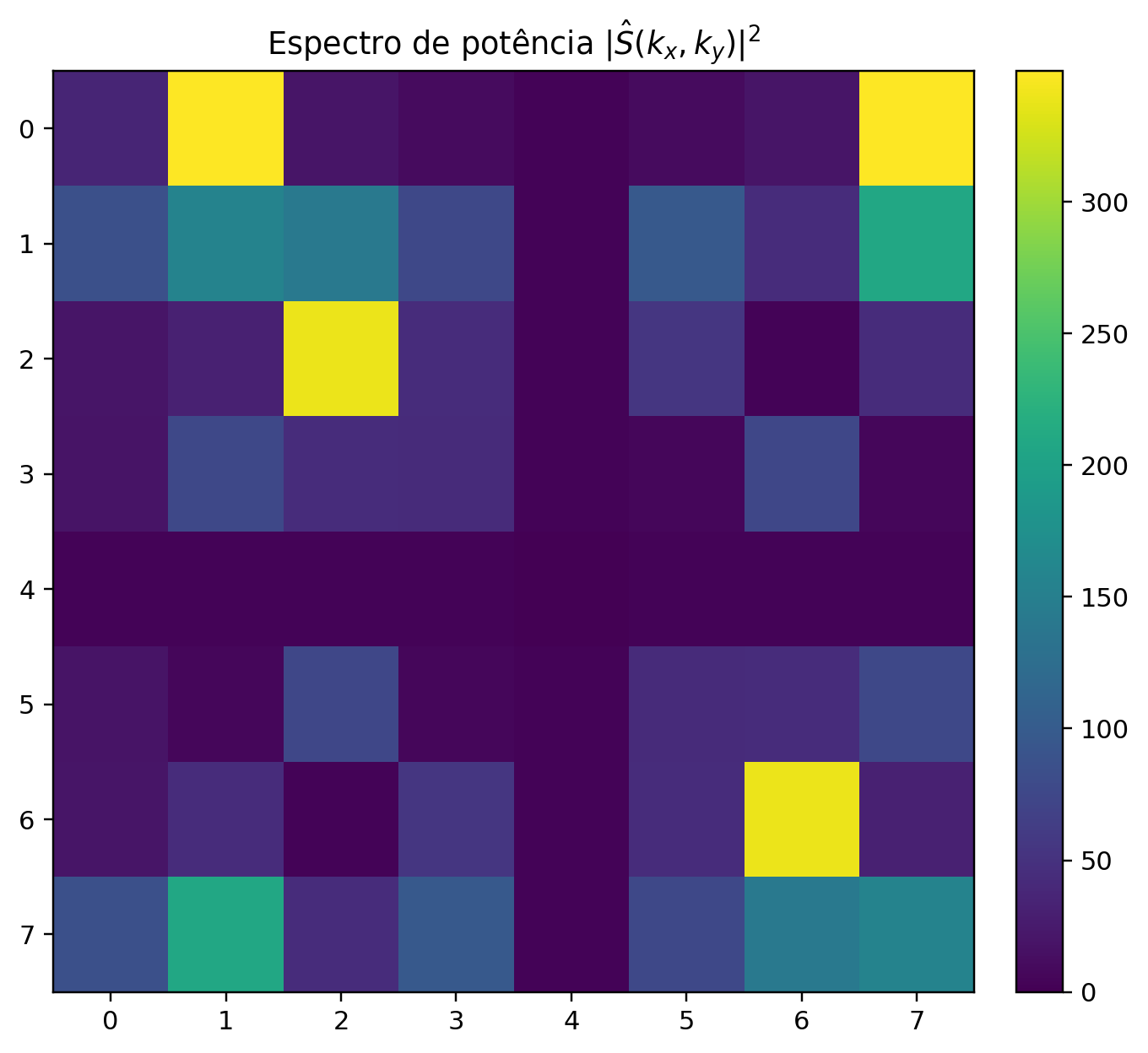}
    \caption{Power spectrum \( |\hat S(k_x,k_y)|^2 \) associated with the lattice configuration. Figure produced with the aid of artificial intelligence.}
    \label{fig:wk8x8_power}
\end{figure}

Next, the two-dimensional discrete Fourier transform is applied,
\[
\hat S(k_x,k_y)=\mathcal{F}[\delta S(x,y)].
\]
From this transform, the power spectrum is constructed
\[
|\hat S(k_x,k_y)|^2.
\]
Figure~\ref{fig:wk8x8_power} shows this spectrum for the chosen configuration.\\

\begin{figure}[ht]
    \centering
    \includegraphics[width=0.75\linewidth]{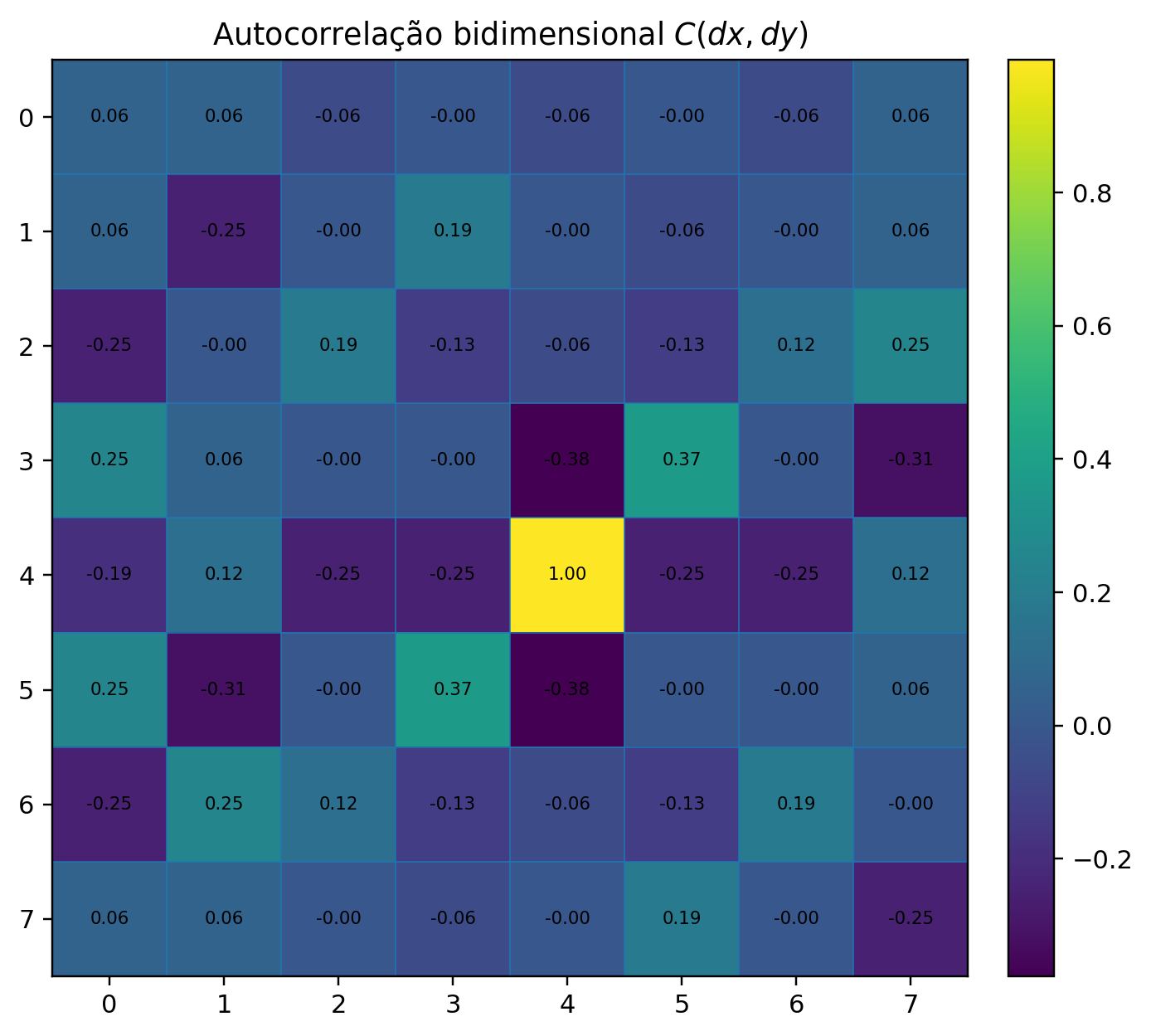}
    \caption{Two-dimensional autocorrelation \(C(dx,dy)\) obtained by the Wiener--Khinchin method. Figure produced with the aid of artificial intelligence.}
    \label{fig:wk8x8_corr}
\end{figure}

By the Wiener--Khinchin theorem, the two-dimensional autocorrelation can be obtained from the inverse transform of the power spectrum:
\[
C(dx,dy)=\mathcal{F}^{-1}\left(|\hat S(k_x,k_y)|^2\right).
\]
In the numerical calculation, after the inverse transform, the correlation matrix is shifted so that the point \((dx,dy)=(0,0)\) lies at the center of the figure, facilitating its interpretation. The result is shown in Figure~\ref{fig:wk8x8_corr}.\\

\begin{figure}[ht]
    \centering
    \includegraphics[width=0.75\linewidth]{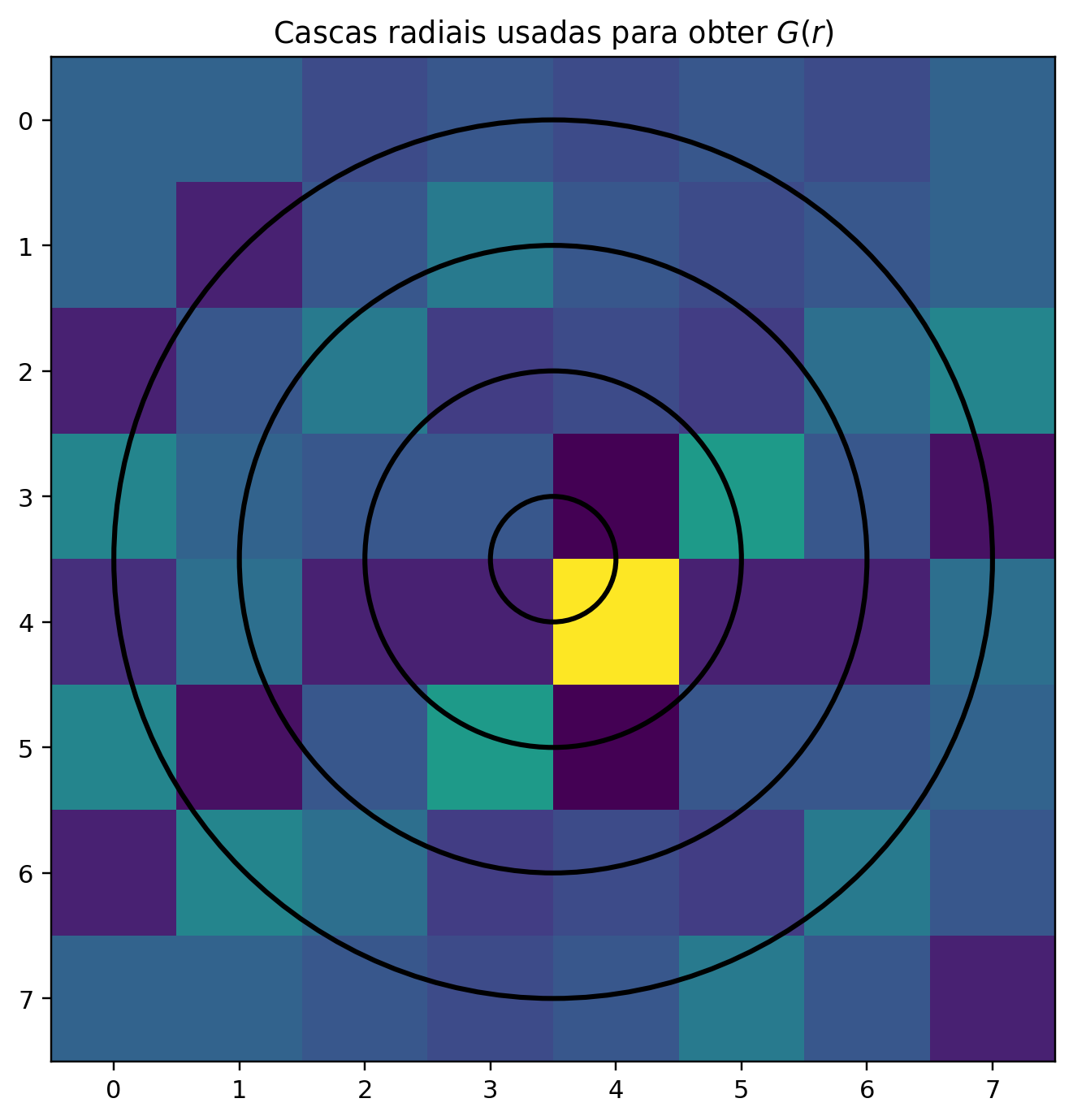}
    \caption{Radial shells used to group the values of \(C(dx,dy)\) with the same approximate distance from the center. Figure produced with the aid of artificial intelligence.}
    \label{fig:wk8x8_shells}
\end{figure}

This function still depends on the vector displacement \((dx,dy)\). However, for studying the spatial decay of the correlation, a quantity that depends only on the scalar distance is of interest
\[
r=\sqrt{dx^2+dy^2}.
\]
For this purpose, \(C(dx,dy)\) is grouped into radial shells centered at the origin, as schematically illustrated in Figure~\ref{fig:wk8x8_shells}.

\noindent The radial function is then defined by the average of the values of \(C(dx,dy)\) belonging to the same shell:
\[
G(r)=\langle C(dx,dy)\rangle_r.
\]
In the present example, we obtain:

\[
\begin{array}{c|c|c}
r & G(r) & \text{number of points in the average} \\
\hline
0 & 0.9990 & 1 \\
1 & -0.0635 & 8 \\
2 & -0.0947 & 12 \\
3 & 0.0381 & 16 \\
4 & -0.0123 & 22 \\
5 & 0.0615 & 4 \\
\end{array}
\]

\begin{figure}[ht]
    \centering
    \includegraphics[width=0.75\linewidth]{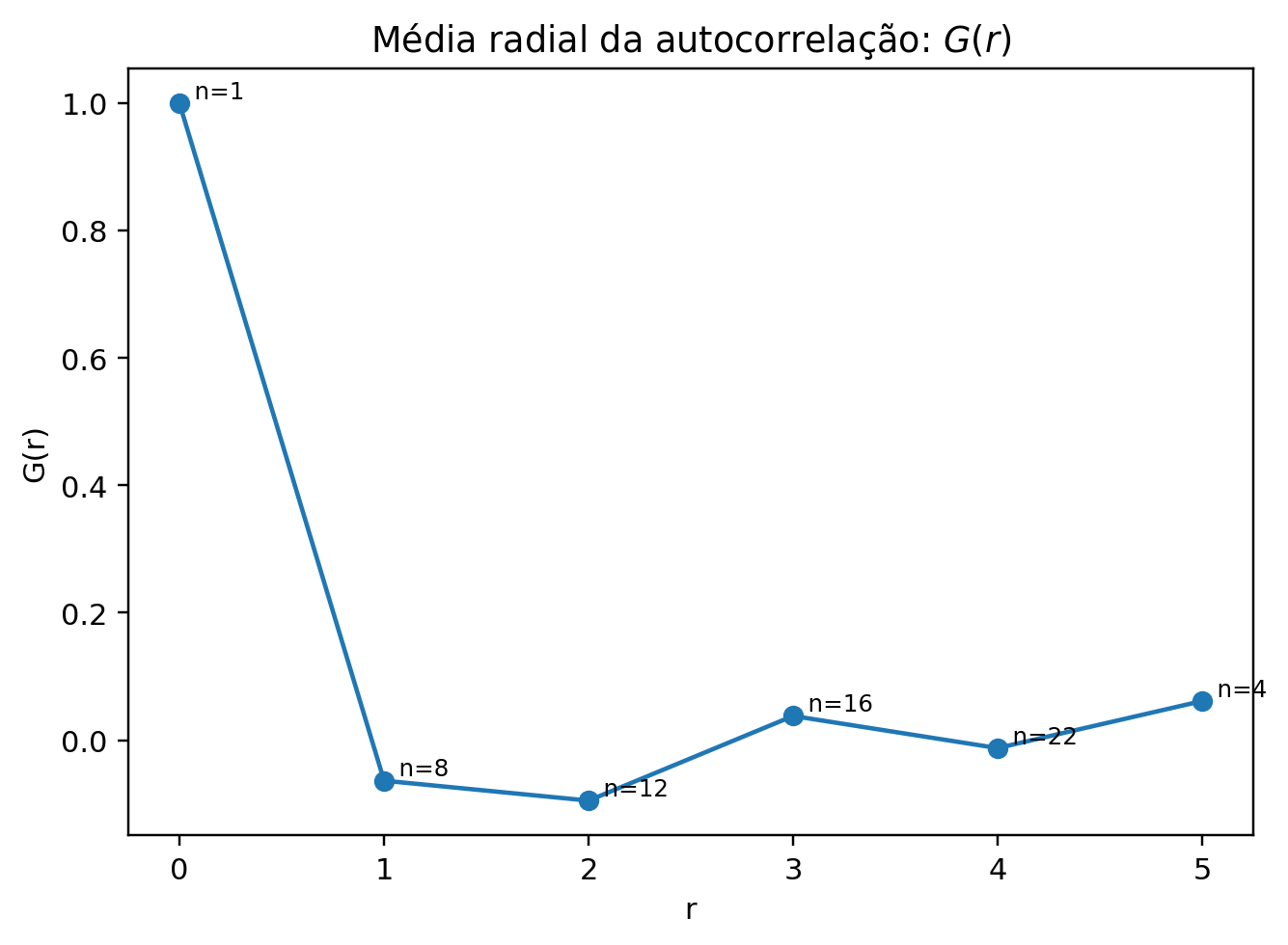}
    \caption{Radial correlation function \(G(r)\) obtained from the radial average of the two-dimensional autocorrelation. Figure produced with the aid of artificial intelligence.}
    \label{fig:wk8x8_gr}
\end{figure}

These values are shown graphically in Figure~\ref{fig:wk8x8_gr}.\\

This example concretely illustrates the procedure adopted throughout this work. Starting from a discrete spin configuration, the spatial mean is removed, the two-dimensional Fourier transform is calculated, the power spectrum is constructed, the inverse transform is applied to obtain \(C(dx,dy)\), and finally the radial average that defines \(G(r)\) is performed. Thus, the spectral-amplitude method provides a systematic and computationally efficient way to calculate spatial correlations in two-dimensional lattices.

\chapter{\textit{Checkerboard} update in the parallel Monte Carlo method}
\label{ap:checkerboard}

\textbf{The images in this appendix were produced with the aid of artificial intelligence through the chatGPT® platform by providing the corresponding \textit{scripts} related to the content presented here}.\\

\paragraph{} In the parallel implementation of the Monte Carlo method for the Ising model, it is not convenient to update neighboring spins at the same time, because the energy of each site depends directly on its nearest neighbors. To overcome this problem, the lattice is divided into two sublattices, as on a chessboard. This procedure is known as a \textit{checkerboard} update.\\

On a square lattice, each site \((i,j)\) is classified by the parity of \(i+j\):
\begin{equation}
(i+j)\bmod 2=
\begin{cases}
0, & \text{sublattice A},\\
1, & \text{sublattice B}.
\end{cases}
\end{equation}
Thus, all neighbors of a site in sublattice A belong to sublattice B, and vice versa.

The algorithm is then executed in two steps:
\begin{enumerate}
    \item all spins of one of the sublattices are updated in parallel, while the other is kept fixed;
    \item then, the spins of the other sublattice are updated.
\end{enumerate}

Thus, two immediate neighbors are prevented from being modified simultaneously. This scheme is especially useful on GPUs, because it allows many spins to be updated at the same time without local inconsistencies.\\

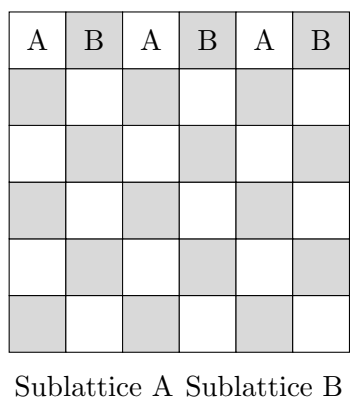
\begin{figure}[h]
\centering
\begin{tikzpicture}[scale=0.75]

% 6x6 grid
\foreach \i in {0,...,5}{
    \foreach \j in {0,...,5}{
        \pgfmathtruncatemacro{\cor}{mod(\i+\j,2)}
        \ifnum\cor=0
            \fill[gray!30] (\i,\j) rectangle (\i+1,\j+1);
        \else
            \fill[white] (\i,\j) rectangle (\i+1,\j+1);
        \fi
        \draw (\i,\j) rectangle (\i+1,\j+1);
    }
}

% labels
\node at (1.5,-0.6) {\small Sublattice A};
\node at (4.5,-0.6) {\small Sublattice B};

% example points
\node at (0.5,5.5) {\small A};
\node at (1.5,5.5) {\small B};
\node at (2.5,5.5) {\small A};
\node at (3.5,5.5) {\small B};
\node at (4.5,5.5) {\small A};
\node at (5.5,5.5) {\small B};

\end{tikzpicture}
\caption[Decomposition of a square lattice into two sublattices in the \textit{checkerboard} scheme.]{Decomposition of a square lattice into two sublattices in the \textit{checkerboard} scheme. The sites of one sublattice can be updated simultaneously, because their neighbors always belong to the complementary sublattice. Figure produced with the aid of artificial intelligence.}
\label{fig:checkerboard}
\end{figure}

Figure \ref{fig:checkerboard} illustrates this decomposition for a \(6\times 6\) square lattice.

\begin{thebibliography}{100}

\bibitem{Boyle1662}
R.~Boyle, ``A defence of the doctrine touching the spring and weight of the
  air,'' 1662.
\newblock London.

\bibitem{Black1761}
J.~Black, ``Experiments upon magnesia alba, quicklime, and some other alcaline
  substances,'' {\em Philosophical Transactions}, vol.~52, pp.~353--365, 1761.

\bibitem{Andrews1869}
T.~Andrews, ``The bakerian lecture: On the continuity of the gaseous and liquid
  states of matter,'' {\em Philosophical Transactions of the Royal Society of
  London}, vol.~159, pp.~575--590, 1869.

\bibitem{vdW1873}
J.~D. van~der Waals, {\em On the Continuity of the Gaseous and Liquid States}.
\newblock PhD thesis, Leiden University, Leiden, 1873.

\bibitem{fisher1964}
M.~E. Fisher, ``{Correlation Functions and the Critical Region of Simple
  Fluids},'' {\em Journal of Mathematical Physics}, vol.~5, pp.~944--962, 07
  1964.

\bibitem{Lima24}
H.~A. Lima, E.~E.~M. Luis, I.~S.~S. Carrasco, A.~Hansen, and F.~A. Oliveira,
  ``Geometrical interpretation of critical exponents,'' {\em Phys. Rev. E},
  vol.~110, p.~L062107, Dec 2024.

\bibitem{Lima25}
H.~A. de~Lima, I.~S.~S. Carrasco, M.~Santos, and F.~A. Oliveira, ``Scaling,
  fractal dynamics, and critical exponents: Application in a
  noninteger-dimensional ising model,'' {\em Phys. Rev. E}, vol.~112,
  p.~044109, Oct 2025.

\bibitem{Carrasco26}
I.~S. Carrasco, H.~A. de~Lima, and F.~A. Oliveira, ``A new fractal mean-field
  analysis in phase transition,'' {\em The European Physical Journal Special
  Topics}, p.~1, 2026.

\bibitem{lima26}
H.~A. Lima, K.~Hermann, I.~S. Carrasco, J.~R. de~Almeida, and F.~A. Oliveira,
  ``Strong universality class in disordered systems,'' {\em arXiv preprint
  arXiv:2605.15441}, 2026.

\bibitem{Muslih10}
S.~I. Muslih and O.~P. Agrawal, ``Riesz fractional derivatives and fractional
  dimensional space,'' {\em International Journal of Theoretical Physics},
  vol.~49, pp.~270--275, 2010.

\bibitem{Muslih10b}
S.~I. Muslih, ``Solutions of a particle with fractional $\delta$-potential in
  a fractional dimensional space,'' {\em International Journal of Theoretical
  Physics}, vol.~49, no.~9, pp.~2095--2104, 2010.

\bibitem{CudaSite}
``Cuda toolkit.'' \url{https://docs.nvidia.com/cuda/doc/index.html}.
\newblock Acessado: 13-02-2024.

\bibitem{Barbosa11}
M.~A.~A. Barbosa, F.~V. Barbosa, and F.~A. Oliveira, ``Thermodynamic and
  dynamic anomalies in a one-dimensional lattice model of liquid water,'' {\em
  J. Chem. Phys.}, vol.~134, no.~2, p.~024511, 2011.

\bibitem{Silva15}
F.~B.~V. da~Silva, F.~A. Oliveira, and M.~A.~A. Barbosa, ``Residual entropy and
  waterlike anomalies in the repulsive one dimensional lattice gas,'' {\em J.
  Chem. Phys.}, vol.~142, no.~14, p.~144506, 2015.

\bibitem{Braz25}
F.~Braz, S.~de~Souza, M.~Lyra, and O.~Rojas, ``Thermodynamic constraints and
  pseudotransition behavior in a one-dimensional waterlike system,'' {\em
  Physical Review E}, vol.~112, no.~4, p.~044144, 2025.

\bibitem{Habitzreuter25}
M.~A. Habitzreuter, ``Thermodynamic anomalies in spin-1/2 fermions,'' 2025.

\bibitem{belitz94}
D.~Belitz and T.~Kirkpatrick, ``The anderson-mott transition,'' {\em Reviews of
  modern physics}, vol.~66, no.~2, p.~261, 1994.

\bibitem{Garavelli91}
S.~L. Garavelli and F.~A. Oliveira, ``Analytical solution for a yukawa-type
  potential,'' {\em Phys. Rev. Lett.}, vol.~66, pp.~1310--1313, Mar 1991.

\bibitem{Garavelli92}
S.~L. GARAVELLI and F.~A. OLIVEIRA, ``Screening properties of hydrogen under
  high pressure,'' {\em Modern Physics Letters B}, vol.~06, no.~13,
  pp.~811--816, 1992.

\bibitem{Yukalov98}
V.~I. Yukalov, E.~P. Yukalova, and F.~A. Oliveira, ``Renormalization-group
  solutions for yukawa potential,'' {\em Journal of Physics A: Mathematical and
  General}, vol.~31, p.~4337, may 1998.

\bibitem{Patil01}
S.~H. Patil, ``Simple wavefunctions for yukawa- and hulth{\'e}n-type potentials,''
  {\em Journal of Physics A: Mathematical and General}, vol.~34, p.~3153, apr
  2001.

\bibitem{Penna09}
A.~L.~A. Penna, J.~B. Diniz, and F.~A. Oliveira, ``Charge degeneracy removal in
  the screened hydrogen atom,'' {\em Journal of Physics B: Atomic, Molecular
  and Optical Physics}, vol.~42, p.~215001, oct 2009.

\bibitem{Diniz08}
J.~B. Diniz, A.~L. Penna, S.~L. Garavelli, and F.~A. Oliveira, ``Charge
  degeneracy for a thomas--fermi hydrogen molecule: Bound--unbound
  transition,'' {\em Solid State Communications}, vol.~146, no.~3,
  pp.~169--174, 2008.

\bibitem{Penna10}
A.~Penna, J.~Diniz, and F.~Oliveira, ``Landau description for an atomic
  screened system: the charge symmetry breaking effect,'' {\em The European
  Physical Journal D}, vol.~56, no.~1, pp.~27--31, 2010.

\bibitem{salinas}
S.~R.~A. Salinas, {\em Introdu{\c c}{\~a}o \`a F{\'i}sica Estat{\'i}stica}.
\newblock 2~ed., 2005.

\bibitem{landau1937theory}
L.~Landau, ``On the theory of phase transitions,'' {\em Phys. Z. Sowjetunion},
  vol.~11, pp.~26--35, 1937.

\bibitem{onsager44}
L.~Onsager, ``Crystal statistics. i. a two-dimensional model with an
  order-disorder transition,'' {\em Physical Review}, vol.~65, no.~3 e 4,
  p.~117, 1944.

\bibitem{kaufman_1949}
B.~Kaufman and L.~Onsager, ``Crystal statistics. iii. short-range order in a
  binary ising lattice,'' {\em Phys. Rev.}, vol.~76, pp.~1244--1252, Oct 1949.

\bibitem{ornstein_zernike_1914}
L.~S. Ornstein and F.~Zernike, ``Accidental deviations of density and
  opalescence at the critical point of a simple substance,'' {\em Proceedings
  of the Section of Sciences of the Royal Academy of Arts and Sciences of the
  Netherlands}, vol.~17, pp.~793--806, 1914.
\newblock Originally published in Dutch as: "Accidentele Dichtheids- en
  Lichtverschijnselen bij het Kritieke Punt eener Eenvoudige Stoff",
  Proceedings of the Koninklijke Akademie van Wetenschappen te Amsterdam, 17,
  793--806 (1914).

\bibitem{Nowak22}
E.~Gudowska-Nowak, F.~A. Oliveira, and H.~S. Wio, ``Editorial: The
  fluctuation-dissipation theorem today,'' {\em Frontiers in Physics}, vol.~10,
  2022.

\bibitem{Oliveira19}
F.~Oliveira, R.~Ferreira, L.~Lapas, and M.~Vainstein, ``Anomalous diffusion: A
  basic mechanism for the evolution of inhomogeneous systems,'' {\em Frontiers
  in Physics}, vol.~7, p.~18, 2019.

\bibitem{GomesFilho25}
M.~S. Gomes-Filho, L.~C. Lapas, E.~Gudowska-Nowak, and F.~A. Oliveira, ``The
  fluctuation--dissipation relations: Growth, diffusion, and beyond,'' {\em
  Physics Reports}, vol.~1141, pp.~1--43, 2025.

\bibitem{Goldenfeld18}
N.~Goldenfeld, {\em Lectures on phase transitions and the renormalization
  group}.
\newblock CRC Press, 2018.

\bibitem{Costa03}
I.~V.~L. Costa, R.~Morgado, M.~V. B.~T. Lima, and F.~A. Oliveira, ``The
  fluctuation-dissipation theorem fails for fast superdiffusion,'' {\em
  Europhysics Letters}, vol.~63, p.~173, jul 2003.

\bibitem{Costa06}
I.~Costa, M.~Vainstein, L.~Lapas, A.~Batista, and F.~Oliveira, ``Mixing,
  ergodicity and slow relaxation phenomena,'' {\em Physica A: Statistical
  Mechanics and its Applications}, vol.~371, no.~1, pp.~130--134, 2006.

\bibitem{Lapas08}
L.~C. Lapas, R.~Morgado, M.~H. Vainstein, J.~M. Rub\'i, and F.~A. Oliveira,
  ``Khinchin theorem and anomalous diffusion,'' {\em Phys. Rev. Lett.},
  vol.~101, p.~230602, Dec 2008.

\bibitem{Wen23}
B.~Wen, M.-G. Li, J.~Liu, and J.-D. Bao, ``Ergodic measure and potential
  control of anomalous diffusion,'' {\em Entropy}, vol.~25, no.~7, p.~1012,
  2023.

\bibitem{Ricci-Tersenghi00}
F.~Ricci-Tersenghi, D.~A. Stariolo, and J.~J. Arenzon, ``Two time scales and
  violation of the fluctuation-dissipation theorem in a finite dimensional
  model for structural glasses,'' {\em Phys. Rev. Lett.}, vol.~84,
  pp.~4473--4476, May 2000.

\bibitem{Crisanti03}
A.~Crisanti and F.~Ritort, ``Violation of the
  fluctuation--dissipation theorem in glassy systems: basic notions
  and the numerical evidence,'' {\em Journal of Physics A: Mathematical and
  General}, vol.~36, p.~R181, may 2003.

\bibitem{Grigera99}
T.~S. Grigera and N.~E. Israeloff, ``Observation of
  fluctuation-dissipation-theorem violations in a structural glass,'' {\em
  Phys. Rev. Lett.}, vol.~83, pp.~5038--5041, Dec 1999.

\bibitem{Barrat98}
A.~Barrat, ``Monte carlo simulations of the violation of the
  fluctuation-dissipation theorem in domain growth processes,'' {\em Phys. Rev.
  E}, vol.~57, pp.~3629--3632, Mar 1998.

\bibitem{Bellon02}
L.~Bellon and S.~Ciliberto, ``Experimental study of the fluctuation dissipation
  relation during an aging process,'' {\em Physica D: Nonlinear Phenomena},
  vol.~168, pp.~325--335, 2002.

\bibitem{Bellon06}
L.~Bellon, L.~Buisson, M.~Ciccotti, S.~Ciliberto, and F.~Douarche, ``Thermal
  noise properties of two aging materials,'' in {\em Jamming, Yielding, and
  Irreversible Deformation in Condensed Matter}, pp.~23--52, Springer, 2006.

\bibitem{Vainstein05}
M.~Vainstein, R.~Morgado, F.~Oliveira, F.~{de Moura}, and M.~Coutinho-Filho,
  ``Stochastic description of the dynamics of a random-exchange heisenberg
  chain,'' {\em Physics Letters A}, vol.~339, no.~1, pp.~33--38, 2005.

\bibitem{Hayashi07}
K.~Hayashi and M.~Takano, ``Violation of the fluctuation-dissipation theorem in
  a protein system,'' {\em Biophysical journal}, vol.~93, no.~3, pp.~895--901,
  2007.

\bibitem{Kardar86}
M.~Kardar, G.~Parisi, and Y.-C. Zhang, ``Dynamic scaling of growing
  interfaces,'' {\em Phys. Rev. Lett.}, vol.~56, pp.~889--892, Mar 1986.

\bibitem{Cordeiro01}
J.~A. Cordeiro, M.~V. B.~T. Lima, R.~M. Dias, and F.~A. Oliveira, ``Morphology
  of growth by random walk deposition,'' {\em Physica A}, vol.~295, p.~209,
  2001.

\bibitem{Rodrigues15}
E.~A. Rodrigues, B.~A. Mello, and F.~A. Oliveira, ``Growth exponents of the
  etching model in high dimensions,'' {\em Journal of Physics A: Mathematical
  and Theoretical}, vol.~48, p.~035001, dec 2014.

\bibitem{Henkel26}
M.~Henkel and S.~Stoimenov, ``Schr{\"o}dinger-invariance in the voter model,''
  {\em International Journal of Theoretical Physics}, vol.~65, no.~2, p.~48,
  2026.

\bibitem{Gomes19}
W.~P. Gomes, A.~L. Penna, and F.~A. Oliveira, ``From cellular automata to
  growth dynamics: The kardar-parisi-zhang universality class,'' {\em Physical
  Review E}, vol.~100, no.~2, p.~020101, 2019.

\bibitem{GomesFilho20}
M.~S. Gomes-Filho, M.~A.~A. Barbosa, and F.~A. Oliveira, ``A statistical
  mechanical model for drug release: Relations between release parameters and
  porosity,'' {\em Physica A: Statistical Mechanics and its Applications},
  vol.~540, p.~123165, 2020.

\bibitem{GomesFilho21b}
M.~S. Gomes-Filho, A.~L. Penna, and F.~A. Oliveira, ``The kardar-parisi-zhang
  exponents for the 2+1 dimensions,'' {\em Results in Physics}, vol.~26,
  p.~104435, 2021.

\bibitem{Anjos21}
P.~H.~R. dos Anjos, M.~S. Gomes-Filho, W.~S. Alves, D.~L. Azevedo, and F.~A.
  Oliveira, ``The fractal geometry of growth: Fluctuation dissipation theorem
  and hidden symmetry,'' {\em Frontiers in Physics}, vol.~9, 2021.

\bibitem{GomesFilho24}
M.~S. Gomes-Filho, P.~de~Castro, D.~B. Liarte, and F.~A. Oliveira, ``Restoring
  the fluctuation--dissipation theorem in kardar--parisi--zhang universality
  class through a new emergent fractal dimension,'' {\em Entropy}, vol.~26,
  no.~3, 2024.

\bibitem{Perez-Madrid09}
A.~P{\'e}rez-Madrid, L.~C. Lapas, and J.~M. Rubi, ``Heat exchange between two
  interacting nanoparticles beyond the fluctuation-dissipation regime,'' {\em
  Phys. Rev. Lett.}, vol.~103, p.~048301, Jul 2009.

\bibitem{Averin10}
D.~V. Averin and J.~P. Pekola, ``Violation of the fluctuation-dissipation
  theorem in time-dependent mesoscopic heat transport,'' {\em Phys. Rev.
  Lett.}, vol.~104, p.~220601, Jun 2010.

\bibitem{Oliveira01}
F.~A. Oliveira, R.~Morgado, C.~Dias, G.~G. Batrouni, and A.~Hansen, ``{Comment
  on ``Nonstationarity Induced by Long-Time Noise Correlations in the Langevin
  Equation''},'' {\em Phys. Rev. Lett.}, vol.~86, p.~5839, 2001.

\bibitem{Morgado02}
R.~Morgado, F.~A. Oliveira, G.~G. Batrouni, and A.~Hansen, ``Relation between
  anomalous and normal diffusion in systems with memory,'' {\em Physical review
  letters}, vol.~89, no.~10, p.~100601, 2002.

\bibitem{Morgado04}
R.~Morgado, I.~V.~L. Costa, and F.~A. Oliveira, ``Normal and anomalous
  diffusion: Ergodicity and fluctuation-dissipation theorem,'' {\em Acta Phys.
  Polon. B}, vol.~35, p.~1359, 2004.

\bibitem{Bao03}
J.-D. Bao and Y.-Z. Zhuo, ``Ballistic diffusion induced by a thermal broadband
  noise,'' {\em Phys. Rev. Lett.}, vol.~91, p.~138104, Sep 2003.

\bibitem{Bao06}
J.~D. Bao, Y.~Z. Zhuo, F.~A. Oliveira, and P.~H\"{a}nggi, ``Intermediate
  dynamics between {N}ewton and {L}angevin,'' {\em Phys. Rev. E}, vol.~74,
  p.~061111, 2006.

\bibitem{Lapas07}
L.~C. Lapas, I.~V.~L. Costa, M.~H. Vainstein, and F.~A. Oliveira, ``Entropy,
  non-ergodicity and non-gaussian behaviour in ballistic transport,'' {\em
  Europhysics Letters}, vol.~77, p.~37004, jan 2007.

\bibitem{Vainstein06}
M.~H. Vainstein, Costa, I.~VL, and F.~Oliveira, ``Mixing, ergodicity and the
  fluctuation-dissipation theorem in complex systems,'' in {\em Jamming,
  Yielding, and Irreversible Deformation in Condensed Matter}, pp.~159--188,
  Springer, 2006.

\bibitem{Vainstein05a}
M.~H. Vainstein, R.~Morgado, and F.~A. Oliveira, ``Spatio-temporal conjecture
  for diffusion,'' {\em Physica A}, vol.~357, pp.~109--114, 2005.

\bibitem{Ferreira12}
R.~M.~S. Ferreira, M.~V.~S. Santos, C.~C. Donato, J.~S. Andrade, and F.~A.
  Oliveira, ``Analytical results for long-time behavior in anomalous
  diffusion,'' {\em Phys. Rev. E}, vol.~86, p.~021121, Aug 2012.

\bibitem{Ferreira22}
R.~M. Ferreira, ``From generalized {L}angevin stochastic dynamics to anomalous
  diffusion,'' {\em Physical Review E}, vol.~106, no.~5, p.~054157, 2022.

\bibitem{Barabasi95}
A.-L. Barabasi, H.~E. Stanley, {\em et~al.}, {\em Fractal concepts in surface
  growth}.
\newblock Cambridge university press, 1995.

\bibitem{Mello01}
B.~A. Mello, A.~S. Chaves, and F.~A. Oliveira, ``Discrete atomistic model to
  simulate etching of a crystalline solid,'' {\em Phys. Rev. E}, vol.~63,
  p.~041113, Mar 2001.

\bibitem{Rodrigues14}
E.~A. Rodrigues, B.~A. Mello, and F.~A. Oliveira, ``Growth exponents of the
  etching model in high dimensions,'' {\em Journal of Physics A: Mathematical
  and Theoretical}, vol.~48, no.~3, p.~035001, 2014.

\bibitem{Feder22}
J.~Feder, E.~G. Flekk{\o}y, and A.~Hansen, {\em Physics of flow in porous
  media}.
\newblock Cambridge University Press, 2022.

\bibitem{Luis22}
E.~E.~M. Luis, T.~A. de~Assis, and F.~A. Oliveira, ``Unveiling the connection
  between the global roughness exponent and interface fractal dimension in ew
  and kpz lattice models,'' {\em Journal of Statistical Mechanics: Theory and
  Experiment}, vol.~2022, p.~083202, aug 2022.

\bibitem{Luis23}
E.~E. Mozo~Luis, F.~A. Oliveira, and T.~A. de~Assis, ``Accessibility of the
  surface fractal dimension during film growth,'' {\em Phys. Rev. E}, vol.~107,
  p.~034802, Mar 2023.

\bibitem{Suzuki8}
M.~Suzuki, ``Phase transition and fractals,'' {\em Progress of Theoretical
  Physics}, vol.~69, no.~1, pp.~65--76, 1983.

\bibitem{Grimmett06}
G.~Grimmett, {\em The random-cluster model}, vol.~333.
\newblock Springer, 2006.

\bibitem{Cruz23}
M.-{\'A}.~M. Cruz, J.~P. Ortiz, M.~P. Ortiz, and A.~Balankin, ``Percolation on
  fractal networks: A survey,'' {\em Fractal and Fractional}, vol.~7, no.~3,
  p.~231, 2023.

\bibitem{Kroger00}
H.~Kr{\"o}ger, ``Fractal geometry in quantum mechanics, field theory and spin
  systems,'' {\em Physics Reports}, vol.~323, no.~2, pp.~81--181, 2000.

\bibitem{Family85}
F.~Family and T.~Vicsek, ``Scaling of the active zone in the eden process on
  percolation networks and the ballistic deposition model,'' {\em Journal of
  Physics A: Mathematical and General}, vol.~18, no.~2, p.~L75, 1985.

\bibitem{Rodrigues24}
E.~A. Rodrigues, E.~E.~M. Luis, T.~A. de~Assis, and F.~A. Oliveira, ``Universal
  scaling relations for growth phenomena,'' {\em Journal of Statistical
  Mechanics: Theory and Experiment}, vol.~2024, p.~013209, jan 2024.

\bibitem{Voller25}
V.~R. Voller and F.~D.~A. Reis, ``Anomalous infiltration in partially saturated
  porous media,'' {\em International Journal of Heat and Mass Transfer},
  vol.~240, p.~126611, 2025.

\bibitem{Chen17}
J.~P. Chen, L.~G. Rogers, L.~Anderson, U.~Andrews, A.~Brzoska, A.~Coffey,
  H.~Davis, L.~Fisher, M.~Hansalik, S.~Loew, {\em et~al.}, ``Power dissipation
  in fractal ac circuits,'' {\em Journal of Physics A: Mathematical and
  Theoretical}, vol.~50, no.~32, p.~325205, 2017.

\bibitem{Boyle07}
B.~Boyle, K.~Cekala, D.~Ferrone, N.~Rifkin, and A.~Teplyaev, ``Electrical
  resistance of n-gasket fractal networks,'' {\em Pacific Journal of
  Mathematics}, vol.~233, no.~1, pp.~15--40, 2007.

\bibitem{Anjos24}
P.~H. dos Anjos, F.~A. Oliveira, and D.~L. Azevedo, ``Fractality in resistive
  circuits: the fibonacci resistor networks,'' {\em The European Physical
  Journal B}, vol.~97, no.~8, p.~121, 2024.

\bibitem{Angulo00}
J.~Angulo, M.~Ruiz-Medina, V.~Anh, and W.~Grecksch, ``Fractional diffusion and
  fractional heat equation,'' {\em Advances in Applied Probability}, vol.~32,
  no.~4, pp.~1077--1099, 2000.

\bibitem{Metzler99}
R.~Metzler, E.~Barkai, and J.~Klafter, ``Anomalous diffusion and relaxation
  close to thermal equilibrium: A fractional {F}okker-{P}lanck equation
  approach,'' {\em Phys. Rev. Lett.}, vol.~82, no.~18, p.~3563, 1999.

\bibitem{Barkai01}
E.~Barkai, ``Fractional fokker-planck equation, solution, and application,''
  {\em Physical Review E}, vol.~63, no.~4, p.~046118, 2001.

\bibitem{Sokolov01}
I.~Sokolov, ``Thermodynamics and fractional fokker-planck equations,'' {\em
  Physical Review E}, vol.~63, no.~5, p.~056111, 2001.

\bibitem{Laskin07}
N.~Laskin, ``Levy flights over quantum paths,'' {\em Communications in
  Nonlinear Science and Numerical Simulation}, vol.~12, no.~1, pp.~2--18, 2007.

\bibitem{Coniglio89}
A.~Coniglio, ``Fractal structure of ising and potts clusters: Exact results,''
  {\em Phys. Rev. Lett.}, vol.~62, pp.~3054--3057, Jun 1989.

\bibitem{Pelissetto02}
A.~Pelissetto and E.~Vicari, ``Critical phenomena and renormalization-group
  theory,'' {\em Physics Reports}, vol.~368, no.~6, pp.~549--727, 2002.

\bibitem{sitecuda2}
``High-performance python -- gpus.''
  \url{https://www.admin-magazine.com/HPC/Articles/High-Performance-Python-3}.
\newblock Acessado 21/11/2025.

\bibitem{sitecuda3}
``Gpu-accelerated computing with python.''
  \url{https://developer.nvidia.com/how-to-cuda-python}.
\newblock Acessado 21/11/2025.

\bibitem{sitecuda1}
``A complete introduction to gpu programming with practical examples in cuda
  and python.''
  \url{https://www.cherryservers.com/blog/introduction-to-gpu-programming-with-cuda-and-python}.
\newblock Acessado 21/11/2025.

\bibitem{Guillou85}
J.~Le~Guillou and J.~Zinn-Justin, ``Accurate critical exponents from the
  $\varepsilon$-expansion,'' {\em Journal de Physique Lettres}, vol.~46, no.~4,
  pp.~137--141, 1985.

\bibitem{Guillou87}
J.~C. Le~Guillou and J.~Zinn-Justin, ``Accurate critical exponents for ising
  like systems in non-integer dimensions,'' {\em Journal de physique}, vol.~48,
  no.~1, pp.~19--24, 1987.

\bibitem{isingnoninteger}
J.~{Le Guillou} and J.~Z. Justin, ``Accurate critical exponents for ising like
  systems in non-integer dimensions,'' vol.~7, pp.~559--564, 1990.

\bibitem{Cardy96}
J.~Cardy, {\em Scaling and renormalization in statistical physics}, vol.~5.
\newblock Cambridge university press, 1996.

\bibitem{reis96}
F.~Aar{\~a}o~Reis, d.~SL, and R.~dos Santos, ``Weak versus strong universality in
  the two-dimensional random-bond ising ferromagnet,'' {\em Physical review. B,
  Condensed matter}, vol.~54, pp.~R9616--R9619, 11 1996.

\bibitem{Potts52}
R.~B. Potts, ``Some generalized order-disorder transformations,'' in {\em
  Mathematical proceedings of the cambridge philosophical society}, vol.~48,
  pp.~106--109, Cambridge University Press, 1952.

\bibitem{f.y.wu}
F.~Y. Wu, ``The potts model,'' {\em Rev. Mod. Phys.}, vol.~54, pp.~235--268,
  Jan 1982.

\bibitem{Xu25}
Y.~Xu, T.~Chen, Z.~Zhou, J.~Salas, and Y.~Deng, ``Correction-to-scaling
  exponent for percolation and the fortuin-kasteleyn potts model in two
  dimensions,'' {\em Physical Review E}, vol.~111, no.~3, p.~034108, 2025.

\bibitem{Kosterlitz74}
J.~M. Kosterlitz, ``The critical properties of the two-dimensional xy model,''
  {\em Journal of Physics C: Solid State Physics}, vol.~7, no.~6,
  pp.~1046--1060, 1974.

\bibitem{PhysRevB.43.6087}
P.~Peczak, A.~M. Ferrenberg, and D.~P. Landau, ``High-accuracy monte carlo
  study of the three-dimensional classical heisenberg ferromagnet,'' {\em Phys.
  Rev. B}, vol.~43, pp.~6087--6093, Mar 1991.

\bibitem{Guillou80}
J.~Le~Guillou and J.~Zinn-Justin, ``Critical exponents from field theory,''
  {\em Physical Review B}, vol.~21, no.~9, p.~3976, 1980.

\bibitem{Brezin85}
E.~Br{\'e}zin and J.~Zinn-Justin, ``Finite size effects in phase transitions,''
  {\em Nuclear Physics B}, vol.~257, pp.~867--893, 1985.

\bibitem{Holm93}
C.~Holm and W.~Janke, ``Critical exponents of the classical three-dimensional
  heisenberg model: A single-cluster monte carlo study,'' {\em Physical Review
  B}, vol.~48, no.~2, p.~936, 1993.

\bibitem{Campostrini02}
M.~Campostrini, M.~Hasenbusch, A.~Pelissetto, P.~Rossi, and E.~Vicari,
  ``Critical exponents and equation of state of the three-dimensional
  heisenberg universality class,'' {\em Physical Review B}, vol.~65, no.~14,
  p.~144520, 2002.

\bibitem{Pecora91}
L.~M. Pecora and T.~L. Carroll, ``Driving systems with chaotic signals,'' {\em
  Physical review A}, vol.~44, no.~4, p.~2374, 1991.

\bibitem{Longa96}
L.~Longa, E.~M.~F. Curado, and F.~A. Oliveira, ``Roundoff-induced coalescence
  of chaotic trajectories,'' {\em Phys. Rev. E}, vol.~54, p.~R2201, 1996.

\bibitem{Ciesla01}
M.~Cie\'sla, S.~P. Dias, L.~Longa, and F.~A. Oliveira, ``Synchronization
  induced by {L}angevin dynamics,'' {\em Phys. Rev. E}, vol.~63, p.~065202, May
  2001.

\bibitem{Morgado07}
R.~Morgado, M.~Cie\'sla, L.~Longa, and F.~A. Oliveira, ``Synchronization in the
  presence of memory,'' {\em Europhys. Lett.}, vol.~79, p.~10002, 2007.

\bibitem{Turing52}
A.~M. Turing, ``The chemical basis of morphogenesis,'' {\em Philosophical
  Transactions of the Royal Society of London. Series B, Biological Sciences},
  vol.~237, no.~641, pp.~37--72, 1952.

\bibitem{Cross93}
M.~C. Cross and P.~C. Hohenberg, ``Pattern formation outside of equilibrium,''
  {\em Rev. Mod. Phys.}, vol.~65, pp.~851--1112, Jul 1993.

\bibitem{DaCunha09}
J.~Da~Cunha, A.~Penna, M.~Vainstein, R.~Morgado, and F.~Oliveira,
  ``Self-organization analysis for a nonlocal convective fisher equation,''
  {\em Physics Letters A}, vol.~373, no.~6, pp.~661--667, 2009.

\bibitem{DaCunha11}
J.~A. da~Cunha, A.~L. Penna, and F.~A. Oliveira, ``Pattern formation and
  coexistence domains for a nonlocal population dynamics,'' {\em Physical
  Review E}, vol.~83, no.~1, p.~015201, 2011.

\bibitem{Barbosa17}
F.~V. Barbosa, A.~A. Penna, R.~M. Ferreira, K.~L. Novais, J.~A. da~Cunha, and
  F.~A. Oliveira, ``Pattern transitions and complexity for a nonlocal logistic
  map,'' {\em Physica A: Statistical Mechanics and its Applications}, vol.~473,
  pp.~301--312, 2017.

\bibitem{Fuentes03}
M.~Fuentes, M.~Kuperman, and V.~Kenkre, ``Nonlocal interaction effects on
  pattern formation in population dynamics,'' {\em Physical Review Letters},
  vol.~91, no.~15, p.~158104, 2003.

\bibitem{Aranda20-1}
O.~T. Aranda and F.~A. Oliveira, ``Analytical and numerical solutions of the
  riccati equation using the method of variation of parameters. application to
  population dynamics,'' {\em Journal of Computational and Nonlinear Dynamics},
  2020.

\bibitem{Aranda21}
O.~T. Aranda, A.~L. Penna, and F.~A. Oliveira, ``Nonlocal pattern formation
  effects in evolutionary population dynamics,'' {\em Physica A: Statistical
  Mechanics and its Applications}, vol.~572, p.~125865, 2021.

\bibitem{Aranda20-2}
O.~T. Aranda, A.~L. Penna, and F.~A. Oliveira, ``Nonlinear self-organized
  population dynamics induced by external selective nonlocal processes,'' {\em
  Communications in Nonlinear Science and Numerical Simulation}, vol.~93,
  p.~105512, 2021.

\bibitem{Fuentes04}
M.~Fuentes, M.~Kuperman, and V.~Kenkre, ``Analytical considerations in the
  study of spatial patterns arising from nonlocal interaction effects,'' {\em
  The Journal of Physical Chemistry B}, vol.~108, no.~29, pp.~10505--10508,
  2004.

\bibitem{Carwood26}
O.~A. Carwood and E.~J. Carr, ``Functionally-graded drug delivery systems with
  binding reactions: Analytical and stochastic approaches for the fraction of
  drug released,'' {\em International Journal of Heat and Mass Transfer},
  vol.~260, p.~128455, 2026.

\bibitem{GomesFilho16}
M.~S. Gomes~Filho, F.~A. Oliveira, and M.~A.~A. Barbosa, ``A statistical
  mechanical model for drug release: Investigations on size and porosity
  dependence,'' {\em Physica A}, vol.~460, no.~C, pp.~29--37, 2016.

\bibitem{GomesFilho22}
M.~S. Gomes-Filho, F.~A. Oliveira, and M.~A.~A. Barbosa, ``Modeling the
  diffusion-erosion crossover dynamics in drug release,'' {\em Phys. Rev. E},
  vol.~105, p.~044110, Apr 2022.

\bibitem{Ziff86}
R.~M. Ziff, E.~Gulari, and Y.~Barshad, ``Kinetic phase transitions in an
  irreversible surface-reaction model,'' {\em Physical review letters},
  vol.~56, no.~24, p.~2553, 1986.

\bibitem{Fernandes18}
H.~A. Fernandes, R.~da~Silva, and A.~B. Bernardi, ``Two universality classes of
  the ziff-gulari-barshad model with co desorption via time-dependent monte
  carlo simulations,'' {\em Physical Review E}, vol.~98, no.~3, p.~032113,
  2018.

\bibitem{Santos24}
M.~Santos and F.~A. Oliveira, ``Phase transitions in the ziff-gulari-barshad
  model operating on periodic conditions,'' {\em Phys. Rev. E}, vol.~110,
  p.~044122, Oct 2024.

\bibitem{Pinto16}
P.~D. Pinto, F.~A. Oliveira, and A.~L.~A. Penna, ``Thermodynamics aspects of
  noise-induced phase synchronization,'' {\em Phys. Rev. E}, vol.~93,
  p.~052220, May 2016.

\bibitem{Pinto17}
P.~D. Pinto, A.~L. Penna, and F.~A. Oliveira, ``Critical behavior of
  noise-induced phase synchronization,'' {\em EPL (Europhysics Letters)},
  vol.~117, no.~5, p.~50009, 2017.

\bibitem{Kadar85}
M.~Kardar, ``Roughening by impurities at finite temperatures,'' {\em Phys. Rev.
  Lett.}, vol.~55, no.~26, p.~2923, 1985.

\bibitem{Alves16}
W.~S. Alves, E.~A. Rodrigues, H.~A. Fernandes, B.~A. Mello, F.~A. Oliveira, and
  I.~V.~L. Costa, ``Analysis of etching at a solid-solid interface,'' {\em
  Phys. Rev. E}, vol.~94, p.~042119, Oct 2016.

\bibitem{Krug92}
J.~Krug, P.~Meakin, and T.~Halpin-Healy, ``Amplitude universality for driven
  interfaces and directed polymers in random media,'' {\em Phys. Rev. A},
  vol.~45, pp.~638--653, Jan 1992.

\bibitem{Krug97}
J.~Krug, ``Origins of scale invariance in growth processes,'' {\em Advances in
  Physics}, vol.~46, no.~2, pp.~139--282, 1997.

\bibitem{feder2013fractals}
J.~Feder, {\em Fractals}.
\newblock Springer Science \& Business Media, 2013.

\bibitem{Derrida98}
B.~Derrida and J.~L. Lebowitz, ``Exact large deviation function in the
  asymmetric exclusion process,'' {\em Phys. Rev. Lett.}, vol.~80,
  pp.~209--213, Jan 1998.

\bibitem{Meakin86}
P.~Meakin, P.~Ramanlal, L.~M. Sander, and R.~C. Ball, ``Ballistic deposition on
  surfaces,'' {\em Phys. Rev. A}, vol.~34, pp.~5091--5103, Dec 1986.

\bibitem{Daryaei20}
E.~Daryaei, ``Universality and crossover behavior of single-step growth models
  in $1+1$ and $2+1$ dimensions,'' {\em Phys. Rev. E}, vol.~101, p.~062108, Jun
  2020.

\bibitem{Edwards82}
S.~F. Edwards and D.~Wilkinson, ``The surface statistics of a granular
  aggregate,'' {\em Proceedings of the Royal Society of London. A. Mathematical
  and Physical Sciences}, vol.~381, no.~1780, pp.~17--31, 1982.

\bibitem{Hansen00}
A.~Hansen, J.~Schmittbuhl, G.~G. Batrouni, and F.~A. de~Oliveira, ``Normal
  stress distribution of rough surfaces in contact,'' {\em Geophysical research
  letters}, vol.~27, no.~22, pp.~3639--3642, 2000.

\bibitem{Wolf}
W.~P. Wolf, ``The ising model and real magnetics materials,'' {\em Brazilian
  Journal of Physics}, vol.~30, no.~4, p.~794, 2000.

\bibitem{Merikoski03}
J.~Merikoski, J.~Maunuksela, M.~Myllys, J.~Timonen, and M.~J. Alava, ``Temporal
  and spatial persistence of combustion fronts in paper,'' {\em Phys. Rev.
  Lett.}, vol.~90, p.~024501, Jan 2003.

\bibitem{Odor10}
G.~Odor, B.~Liedke, and K.-H. Heinig, ``Directed $d$-mer diffusion describing
  the kardar-parisi-zhang-type surface growth,'' {\em Phys. Rev. E}, vol.~81,
  p.~031112, Mar 2010.

\bibitem{Takeuchi13}
K.~A. Takeuchi, ``Crossover from growing to stationary interfaces in the
  kardar-parisi-zhang class,'' {\em Phys. Rev. Lett.}, vol.~110, p.~210604, May
  2013.

\bibitem{Gwa92}
L.-H. Gwa and H.~Spohn, ``Six-vertex model, roughened surfaces, and an
  asymmetric spin hamiltonian,'' {\em Physical review letters}, vol.~68, no.~6,
  p.~725, 1992.

\bibitem{Vega85}
H.~De~Vega and F.~Woynarovich, ``Method for calculating finite size corrections
  in bethe ansatz systems: Heisenberg chain and six-vertex model,'' {\em
  Nuclear Physics B}, vol.~251, pp.~439--456, 1985.

\bibitem{Plischke87}
M.~Plischke, Z.~R{a}cz, and D.~Liu, ``Time-reversal invariance and universality
  of two-dimensional growth models,'' {\em Physical Review B}, vol.~35, no.~7,
  p.~3485, 1987.

\bibitem{Corwin18}
I.~Corwin, P.~Ghosal, A.~Krajenbrink, P.~Le~Doussal, and L.-C. Tsai,
  ``Coulomb-gas electrostatics controls large fluctuations of the
  kardar-parisi-zhang equation,'' {\em Phys. Rev. Lett.}, vol.~121, p.~060201,
  Aug 2018.

\bibitem{nahum17}
A.~Nahum, J.~Ruhman, S.~Vijay, and J.~Haah, ``Quantum entanglement growth under
  random unitary dynamics,'' {\em Phys. Rev. X}, vol.~7, p.~031016, Jul 2017.

\bibitem{ljubotina19}
M.~Ljubotina, M.~{\v Z}nidari{\v c}, and T.~Prosen,
  ``Kardar-parisi-zhang physics in the quantum heisenberg magnet,'' {\em Phys.
  Rev. Lett.}, vol.~122, p.~210602, May 2019.

\bibitem{DeNardis19}
J.~De~Nardis, M.~Medenjak, C.~Karrasch, and E.~Ilievski, ``Anomalous spin
  diffusion in one-dimensional antiferromagnets,'' {\em Phys. Rev. Lett.},
  vol.~123, p.~186601, Oct 2019.

\bibitem{Moca23}
C.~P. Moca, M.~A. Werner, A.~Valli, G.~Zar{\'a}nd, and T.~Prosen,
  ``Kardar-parisi-zhang scaling in the hubbard model,'' {\em arXiv preprint
  arXiv:2306.11540}, 2023.

\bibitem{Rodriguez19}
M.~A. Rodriguez and H.~S. Wio, ``Stochastic entropies and fluctuation theorems
  for a discrete one-dimensional kardar-parisi-zhang system,'' {\em Phys. Rev.
  E}, vol.~100, p.~032111, Sep 2019.

\bibitem{Amorim23}
P.~M. Amorim, E.~E. Mozo~Luis, F.~F. Dall'Agnol, and T.~A. de~Assis, ``{Role
  of finite probe size in measuring growth exponent in film deposition},'' {\em
  Journal of Applied Physics}, vol.~133, p.~235304, 06 2023.

\bibitem{bhattacharyya06}
P.~Bhattacharyya and B.~K. Chakrabarti, {\em Modelling critical and
  catastrophic phenomena in geoscience: a statistical physics approach},
  vol.~705.
\newblock Springer, 2006.

\bibitem{Zhang12}
Z.~Zhang and N.~March, ``Proposed connection between critical exponents and
  fractal dimensions in the ising model,'' {\em Journal of Mathematical
  Chemistry}, vol.~50, pp.~920--925, 2012.

\bibitem{Dumouchel05}
C.~Dumouchel, J.~Cousin, and K.~Triballier, ``Experimental analysis of
  liquid--gas interface at low weber number: interface length and fractal
  dimension,'' {\em Experiments in fluids}, vol.~39, no.~4, pp.~651--666, 2005.

\bibitem{Kuramoto84}
Y.~Kuramoto and Y.~Kuramoto, {\em Chemical turbulence}.
\newblock Springer, 1984.

\bibitem{Carrasco18}
I.~S.~S. Carrasco and T.~J. Oliveira, ``Kardar-parisi-zhang growth on
  one-dimensional decreasing substrates,'' {\em Phys. Rev. E}, vol.~98,
  p.~010102, Jul 2018.

\bibitem{Pham23}
D.~T. Pham and Z.~E. Musielak, ``Spectra of reduced fractals and their
  applications in biology,'' {\em Fractal and Fractional}, vol.~7, no.~1, 2023.

\bibitem{Reis04}
F.~{Aar\~ao Reis}, ``Universality in two-dimensional {Kardar-Parisi-Zhang}
  growth,'' {\em Phys. Rev. E.}, vol.~69, p.~021610, 2004.

\bibitem{Wallace79}
D.~Wallace and R.~Zia, ``Euclidean group as a dynamical symmetry of surface
  fluctuations: The planar interface and critical behavior,'' {\em Physical
  Review Letters}, vol.~43, no.~12, p.~808, 1979.

\bibitem{Kawabata19}
K.~Kawabata, K.~Shiozaki, M.~Ueda, and M.~Sato, ``Symmetry and topology in
  non-hermitian physics,'' {\em Physical Review X}, vol.~9, no.~4, p.~041015,
  2019.

\bibitem{salman24}
S.~Salman, F.~H. Shah, and S.~J. Kim, ``Role of statistical physics formalism
  in pharmaceutical science,'' {\em Letters in Drug Design \& Discovery},
  vol.~21, no.~14, pp.~2891--2902, 2024.

\bibitem{Sherrington75}
D.~Sherrington and S.~Kirkpatrick, ``Solvable model of a spin-glass,'' {\em
  Physical review letters}, vol.~35, no.~26, p.~1792, 1975.

\bibitem{Almeida78}
J.~R.~L. de~Almeida and D.~J. Thouless, ``Stability of the
  sherrington-kirkpatrick solution of a spin glass model,'' {\em Journal of
  Physics A: Mathematical and General}, vol.~11, p.~983, may 1978.

\bibitem{Kalosakas22}
G.~Kalosakas and E.~Panagopoulou, ``Lag time in diffusion-controlled release
  formulations containing a drug-free outer layer,'' {\em Processes}, vol.~10,
  no.~12, p.~2592, 2022.

\bibitem{Lin89}
M.~Lin, H.~Lindsay, D.~Weitz, R.~Ball, R.~Klein, and P.~Meakin, ``Universality
  of fractal aggregates as probed by light scattering,'' {\em Proceedings of
  the Royal Society of London. A. Mathematical and Physical Sciences},
  vol.~423, no.~1864, pp.~71--87, 1989.

\end{thebibliography}
\end{document}